\documentclass[11pt]{article}
\usepackage{epsfig} 
\usepackage{amssymb}
\newcommand{\sect}[1]{ \section{#1} \setcounter{equation}{0} } 

\newcommand{\half}{\mbox{\small{$\frac{1}{2}$}}}

\newcommand{\MSbar}{\overline{\mbox{MS}}}

\newcommand{\Nf}{N_{\!f}}

\begin{document}
\title{Six dimensional Landau-Ginzburg-Wilson theory}
\author{J.A. Gracey \& R.M. Simms, \\ Theoretical Physics Division, \\ 
Department of Mathematical Sciences, \\ University of Liverpool, \\ P.O. Box 
147, \\ Liverpool, \\ L69 3BX, \\ United Kingdom.} 
\date{}
\maketitle 

\vspace{5cm} 
\noindent 
{\bf Abstract.} We renormalize the six dimensional cubic theory with an
$O(N)$~$\times$~$O(m)$ symmetry at three loops in the modified minimal
subtraction ($\MSbar$) scheme. The theory lies in the same universality class
as the four dimensional Landau-Ginzburg-Wilson model. As a check we show that 
the critical exponents derived from the three loop renormalization group 
functions at the Wilson-Fisher fixed point are in agreement with the large $N$ 
$d$-dimensional critical exponents of the underlying universal theory. Having
established this connection we analyse the fixed point structure of the
perturbative renormalization group functions to estimate the location of the 
conformal window of the $O(N)$~$\times$~$O(2)$ model.  

\vspace{-17cm}
\hspace{13cm}
{\bf LTH 1118}

\newpage 

\sect{Introduction.}

Scalar quantum field theories have been the subject of intense interest in 
recent years in the context of trying to develop our understanding of conformal
field theories in dimensions greater than two using established concepts,
\cite{1,2,3,4,5,6,7}, in a modern application, \cite{8,9,10,11}. One of the 
main aspects of this activity is in finding the conformal window of a theory 
where there are non-trivial fixed points of the $\beta$-function. In this 
window one in principle has a theory where ideas for extending Zamolodchikov's 
$c$-theorem, \cite{12}, to higher dimensions can be explored, for example, as 
well as other properties of strictly two dimensional conformal field theories. 
One of the first examples of a quantum field theory with a conformal window was
that deriving from the Banks-Zaks fixed point in Quantum Chromodynamics (QCD),
\cite{13,14}. For a range of the number of quark flavours, $\Nf$, there is a 
non-trivial fixed point at a non-zero value of the strong coupling constant. 
This is established from two loop perturbation theory and the upper bound of 
the window follows from the one loop term. However, the lower bound has not 
been unambiguously resolved partly because near this lower end the value of the 
critical coupling constant is not small and the perturbative approximation 
appears to break down. Non-perturbative lattice studies have yet to provide 
definitive data which determine the location of the lower bound precisely. 
Until relatively recently these two methods of perturbative analysis of
$\beta$-functions and lattice studies of the breakdown of the chiral limit were
generally the main techniques to access conformal windows of a theory. A third 
approach was developed several years ago which falls under the banner of the 
conformal bootstrap technique, \cite{8,9,10,11}. It is a numerical analysis of 
the operator content of a scalar field theory and exploits the decoupling of 
operators from the spectrum in the limit to a fixed point. This allows one
to obtain accurate values for scaling dimensions, \cite{10}. To date one main 
application has been to study scalar field theories across different spacetime 
dimensions. Recent examples include trying to find the conformal window for 
$O(N)$ scalar $\phi^3$ theory in five dimensions, \cite{15,16,17,18,19}. 
Originally Ma found that the conformal window was located at $N$~$=$~$1038$, 
\cite{20}, in strictly six dimensions but the higher order terms in the
$\epsilon$ in $d$~$=$~$6$~$-$~$2\epsilon$ dimensions were computed to three, 
\cite{21}, and four, \cite{22}, loops. Using summation techniques the bound in 
five dimensions was reduced but not to the low values indicated by, for 
example, conformal bootstrap analyses, \cite{17}. While such agreement between 
techniques is yet to be resolved for other quantities in the conformal window, 
such as estimates of critical exponents, there is very strong overlap in the 
values which suggest these complementary methods do provide a solid insight 
into the properties of these scalar quantum field theories. 

One upshot of the reopening of studies in higher dimensional theories has been 
that higher order perturbative results have been computed beyond a few loops. 
For instance, the three loop result of \cite{23,24} from thirty years ago for 
six dimensional $\phi^3$ theory were only extended to four loops recently,
\cite{22}. Equally the five loop renormalization group function of four 
dimensional $\phi^4$ theory from the mid-1990's have now been extended to six 
loops in \cite{25,26} and to seven loops for the field anomalous dimension, 
\cite{27}. Given the interest in the conformal bootstrap and its potential 
application to non-scalar theories or to scalar theories with symmetry other 
than $O(N)$ it is worthwhile providing higher loop perturbative results to 
complement recent, \cite{28}, and future bootstrap studies in various 
dimensions. Therefore the aim of this article is to renormalize the six 
dimensional extension of the Landau-Wilson-Ginzburg (LGW) model to three loops.
In effect this is a $\phi^3$ type theory but endowed with an 
$O(N)$~$\times$~$O(m)$ symmetry. It has applications in condensed matter 
problems such as randomly dilute spin models, \cite{29,30}. A conformal 
bootstrap analysis has recently been provided from the conformal bootstrap 
technology, \cite{28}, and one aim is to provide data to complement similar 
bootstrap analyses in the future. In addition analysis for the 
$O(4)$~$\times$~$O(2)$ theory, which describes the chiral phase transition in 
two flavour QCD, has been discussed in \cite{31}. A second motivation is to 
continue exploring the tower of theories across the dimensions which are in
the same universality class at the Wilson-Fisher fixed point, \cite{32}. The 
LGW model with an $O(N)$~$\times$~$O(m)$ is a new example to continue this
investigation which we will carry out in depth here. The extension of the four 
dimensional $O(N)$~$\times$~$O(m)$ symmetric theory of \cite{30} to six 
dimensions is termed the ultraviolet completion of the theory. Once a theory 
has been constructed in a fixed dimension with the same symmetries as its lower
dimensional counterpart the $\epsilon$ expansion of the critical exponents at 
the fixed point can be used to access non-perturbative fixed point properties 
in the lower dimensional partner. This ultraviolet-infrared connection was 
recognized earlier in \cite{33,34} but its power is being exploited in present 
analyses. To connect theories across dimensions requires a technique beyond 
perturbation theory since coupling constants become dimensionful outside their 
critical dimension. The interpolating expansion parameter has to be 
dimensionless and in the context of theories with an $O(N)$ symmetry the 
parameter is $1/N$ where $N$ is regarded as large. Earlier work by Vasil'ev's 
group, \cite{35,36,37}, provided critical exponents to three orders in $1/N$ 
for the universal theory in $d$-dimensions with an $O(N)$ symmetry. In such 
exponents the origin of the $d$ dependence is not from dimensional 
regularization. Rather large $N$ Feynman integrals within the formalism of 
\cite{35,36,37} are computed with analytic regularization. So the $d$ 
dependence in the exponents is a true reflection of the properties and 
structure of the universal theory in any dimension. Indeed any perturbative 
expansion of a critically evaluated renormalization group function in the large
$N$ expansion agrees with the exponents of \cite{35,36,37}. For the 
$O(N)$~$\times$~$O(m)$ extension this is also the case due to the computations 
of \cite{38} in the large $N$ expansion and explicit four dimensional 
perturbation theory, \cite{39,40,41,42,43,44,45}. Moreover, such a check will 
also be important in establishing the correctness of our three loop $\MSbar$ 
results for the six dimensional theory in the same universality class prior to 
analysing the renormalization group functions at criticality for a variety of 
values of $N$ and $m$. We will do this both in the fixed dimension of six as 
well as within the $\epsilon$ expansion. Moreover, we will draw similar 
conclusions to others, \cite{28}, where finding the specific location of the 
conformal window is not straightforward.

The paper is organized as follows. We construct the six dimensional version of
the theory with an $O(N)$~$\times$~$O(m)$ symmetry which is in the same 
universality class as the four dimensional Landau-Ginzburg-Wilson model in
section $2$. The necessary large $N$ analysis which will allow us to confirm
this is also reviewed in that section. The main three loop renormalization 
group functions are given in section $3$ together with a summary of the 
technology required to determine them. From these results we derive the large
$N$ critical exponents in section $4$ in order to compare with known $O(1/N^2)$
exponents. Having verified agreement with known results we discuss the search 
for a conformal window in section $5$. A detailed fixed point analysis at three
loops for a variety of values of $N$ in provided in section $6$. Concluding 
observations are given in section $7$. Two appendices are provided. The first
gives the remaining renormalization group functions which were not displayed in
the main text for space reasons including the mass mixing matrix. While the
other appendix provides the full spectrum of fixed points for $N$~$=$~$1000$ as
a complete example of the rich structure of this model.

\sect{Background.}

As we will be considering the six dimensional model with $O(N)$~$\times$~$O(m)$
symmetry which is in the same universality class as the four dimensional
theory with the same symmetry we begin by recalling the relevant aspects of the 
latter theory. In this case the Lagrangian involves a quartic interaction for
a scalar field $\phi^{ia}$ where $1$~$\leq$~$i$~$\leq$~$N$ and 
$1$~$\leq$~$a$~$\leq$~$m$. Consequently the Lagrangian is, \cite{38},  
\begin{equation}
L^{(4)} ~=~ \frac{1}{2} \partial^\mu \phi^{i a} \partial_\mu \phi^{i a} ~+~
\frac{\bar{g}_1}{4!} \left( \phi^{i a} \phi^{i a} \right)^2 ~+~
\frac{\bar{g}_2}{4!} \left[ \left( \phi^{i a} \phi^{i b} \right)^2 ~-~
\left( \phi^{i a} \phi^{i a} \right)^2 \right]
\end{equation}
where $\bar{g}_i$ are the couplings of the respective interactions. This 
version of the Landau-Ginzburg-Wilson theory is not the most useful for 
developing the large $N$ expansion or indeed for seeing the connection with
lower and higher dimensional theories. Instead it is better to reformulate
$L^{(4)}$ in terms of cubic interactions by introducing a set of auxiliary
fields $\tilde{\sigma}$ and $\tilde{T}^{ab}$ where the latter is antisymmetric 
and traceless in its $O(m)$ indices. Then $L^{(4)}$ becomes, \cite{38}, 
\begin{equation}
L^{(4)} ~=~ \frac{1}{2} \partial^\mu \phi^{i a} \partial_\mu \phi^{i a} ~+~
\frac{1}{2} \tilde{\sigma} \phi^{i a} \phi^{i a} ~+~
\frac{1}{2} \tilde{T}^{a b} \phi^{i a} \phi^{i b} ~-~
\frac{3\tilde{\sigma}^2}{2\tilde{g}_1} ~-~ 
\frac{3\tilde{T}^{a b} \tilde{T}^{a b}}{2\tilde{g}_2}
\label{laglgw4g}
\end{equation}
where we have introduced $\tilde{g}_1$~$=$~$\bar{g}_1$~$+$~$(m-1) 
\tilde{g}_2/m$ and $\bar{g}_2$~$=$~$\tilde{g}_2$, \cite{29,30}. Here the 
coupling constants appear within the quadratic part of the Lagrangian which is 
the first step in constructing the critical exponents using the large $N$ 
methods of \cite{35,36}. However, for perturbative computations it is more 
appropriate for the couplings to appear with the actual interactions. So using 
a simple rescaling $L^{(4)}$ becomes 
\begin{eqnarray}
L^{(4)} &=& \frac{1}{2} \partial^\mu \phi^{i a} \partial_\mu \phi^{i a} ~+~
\frac{1}{2} \sigma^2 ~+~
\frac{1}{2} T^{a b} T^{a b} ~+~
\frac{1}{2} g_1 \sigma \phi^{i a} \phi^{i a} ~+~
\frac{1}{2} g_2 T^{a b} \phi^{i a} \phi^{i b} ~. 
\label{laglgw4}
\end{eqnarray}
In this formulation one can build the equivalent six dimensional theory based
on the dimensionalities of the fields and ensuring that the Lagrangian is
renormalizable. In $d$-dimensions the $\phi^{ia}$ field has dimensions
$(\half d - 1)$ while $\sigma$ and $T^{ab}$ are both dimension $2$. Clearly
(\ref{laglgw4}) is renormalizable in four dimensions. The key to constructing
the six dimensional extension is the retention of the two basic interactions of
$\phi^{ia}$ with the auxiliary fields. This means that the dimensionalities
of all three fields are preserved at the connecting Wilson-Fisher fixed point
in $d$-dimensions. However, the equivalent Lagrangian in six dimensions,
$L^{(6)}$, has to have dimension $6$ in order to retain a dimensionless action.
Therefore the auxiliary field sector of (\ref{laglgw4}) has to be replaced. 
This leads to 
\begin{eqnarray}
L^{(6)} &=& \frac{1}{2} \partial^\mu \phi^{i a} \partial_\mu \phi^{i a} ~+~
\frac{1}{2} \partial_\mu \sigma \partial^\mu \sigma ~+~
\frac{1}{2} \partial^\mu T^{a b} \partial_\mu T^{a b} ~+~
\frac{1}{2} g_1 \sigma \phi^{i a} \phi^{i a} ~+~
\frac{1}{6} g_2 \sigma^3 \nonumber \\
&& +~ \frac{1}{2} g_3 T^{a b} \phi^{i a} \phi^{i b} ~+~ 
\frac{1}{2} g_4 \sigma T^{a b} T^{a b} ~+~ 
\frac{1}{6} g_5 T^{a b} T^{a c} T^{b c} 
\label{laglgw6}
\end{eqnarray}
as the ultraviolet completion which is renormalizable in six dimensions and 
which should be equivalent to (\ref{laglgw4}) in four dimensions at the 
Wilson-Fisher fixed point. As in previous extensions there are more 
interactions but also the $\sigma$ and $T^{ab}$ fields now cease being 
auxiliary fields and become propagating with fundamental propagators. The 
additional interactions which depend solely on $\sigma$ and $T^{ab}$ are 
referred to as spectators since they are only present in the critical 
dimension. The interactions with couplings $g_1$ and $g_3$ are the core ones 
which are present at the Wilson-Fisher fixed point through all the dimensions. 
They seed the universal theory in the sense that they determine the canonical 
dimensions of the fields. Thereby they induce the structure of the spectator 
interactions in each critical dimension by requiring renormalizability. 
Although our focus is primarily in relation to critical theories one could 
include masses for the three basic fields which would produce 
\begin{equation}
L_m^{(6)} ~=~ L^{(6)} - \frac{1}{2} m_1^2 \phi^{i a} \phi^{i a} ~-~
\frac{1}{2} m_2^2 \sigma^2 ~-~ \frac{1}{2} m_3^2 T^{a b} T^{a b} 
\label{laglgw6m}
\end{equation}
where $m_i$ are the masses. Similar terms can be added to $L^{(4)}$.

To appreciate properties of these scalar models with an $O(N)$~$\times$~$O(m)$
symmetry we recall relevant properties of the $\beta$-functions of 
(\ref{laglgw4}) which have been computed to several loops orders, \cite{38,35}.
Although for the present purposes it is sufficient to quote the results to two 
loops which are 
\begin{eqnarray}
\beta_1(\bar{g}_1,\bar{g}_2) &=& \frac{1}{2}(d-4)\bar{g}_1 ~+~ 
\frac{(mN+8)}{6}\bar{g}_1^2 ~-~ \frac{1}{3}(m-1)(N-1) \bar{g}_2 
\left(\bar{g}_1-\frac{\bar{g}_2}{2} \right) \nonumber \\
&& -~ \frac{1}{6} (3mN+14) \bar{g}_1^3 ~+~ 
(m-1)(N-1) \left( \frac{11}{9}\bar{g}_1^2 - \frac{13}{12} \bar{g}_1\bar{g}_2 
+ \frac{5}{18}\bar{g}_2^2 \right) \bar{g}_2 \nonumber \\
&& +~ O \left( \bar{g}_i^4 \right)                     
\end{eqnarray}
and
\begin{eqnarray}
\beta_2(\bar{g}_1,\bar{g}_2) &=& \frac{1}{2}(d-4)\bar{g}_2 ~+~ 
2\bar{g}_1\bar{g}_2 ~+~ \frac{1}{6}(m+N-8)\bar{g}_2^2 ~-~ 
\frac{1}{18}(5mN+82)\bar{g}_1^2\bar{g}_2 \nonumber \\
&& +~ \frac{1}{9}[5mN - 11(m+N) + 53]\bar{g}_1\bar{g}_2^2 ~-~ 
\frac{1}{36}[13mN - 35(m+N) + 99] \bar{g}_2^3 \nonumber \\
&& +~ O \left( \bar{g}_i^4 \right)             
\end{eqnarray}
where the order symbol is understood to mean any combinations of the two
coupling constants. For (\ref{laglgw4}), \cite{38,35}, there are several fixed 
points which are the free field Gaussian fixed point, that corresponding to the
Heisenberg model and two where both critical couplings are non-zero. The fixed 
point corresponding to the Heisenberg case corresponds to 
$\bar{g}_1$~$\neq$~$0$ and $\bar{g}_2$~$=$~$0$ irrespective of whether $m$ is 
set to unity or not. In the case when $m$~$\neq$~$1$ the parameter $m$ always 
appears as a multiplier of $N$. For the two fixed points where both critical 
couplings are non-zero one is known as the chiral stable (CS) fixed point and 
the other as the anti-chiral unstable (AU) one. For the Heisenberg fixed point 
in the context of the $(\bar{g}_1,\bar{g}_2)$-plane it is actually a 
saddle-point and so is unstable to perturbations in the $\bar{g}_2$ direction. 
In the reduction to the single coupling $O(N)$ scalar theory the Heisenberg 
fixed point would be stable.

As we will be using the large $N$ results, \cite{38,46}, with which to compare 
our six dimensional perturbative results it is worthwhile recalling some of 
those results as well as giving a perspective on the fixed point structure. In 
the large $N$ method of \cite{35,36} the critical exponents such as $\eta$ and
$\omega$ are computed by analysing the skeleton Schwinger-Dyson equations at
criticality. At that point the propagators and Green's functions obey scaling
law type forms where the powers are in effect the critical exponents. If one
expands the exponent $\eta$, for example, in powers of $1/N$ where $N$ is 
large,
\begin{equation}
\eta ~=~ \sum_{i=1}^\infty \frac{\eta_i}{N^i}
\end{equation}
then each term, $\eta_i$, can be deduced from evaluating the relevant Feynman
diagrams at each order of the $1/N$ expansion. While such diagrams are 
divergent they are {\em analytically} regularized which means that the solution
for each $\eta_i$ and the other exponents are determined as functions of the
spacetime dimension $d$. Therefore these exponents, to as many orders in $1/N$
as they can be computed, correspond to the exponents of the universal quantum
field theory which underlies the Wilson-Fisher fixed point in $d$-dimensions.
Thus when the exponents are expanded in powers of $\epsilon$, where
$d$~$=$~$D$~$-$~$2\epsilon$ and $D$ is the critical dimension of a specific
theory, then the $\epsilon$ expansion will agree with the same expansion of the
corresponding renormalization group function at the same fixed point. For
theories such as (\ref{laglgw4}) and (\ref{laglgw6}) which are in the same
universality class the large $N$ critical exponents computed in \cite{38,46}
reflect the three non-trivial fixed points noted above. The different solutions
for the Heisenberg, AU and CS cases emerge from simple conditions which are 
best seen in the Lagrangian formulation involving the fields $\sigma$ and 
$T^{ab}$. These can be summarized by the vector $(\sigma, T^{ab})$ so that the 
Heisenberg fixed point is $(\sigma,0)$, AU is $(0,T^{ab})$ and CS is
$(\sigma,T^{ab})$ where a zero entry in the vector means the corresponding
field is absent at that fixed point. In other words in the large $N$ 
construction the critical exponents for a particular fixed point are determined
by including only those non-zero fields in the vector in the skeleton
Schwinger-Dyson expansion. 

If we define the scaling dimensions of the fields $\phi^{ia}$, $\sigma$ and
$T^{ab}$ by $\alpha$, $\beta$ and $\gamma$ respectively then 
\begin{equation} 
\alpha ~=~ \mu ~-~ 1 ~+~ \half \eta ~~~,~~~ \beta ~=~ 2 ~-~ \eta ~-~ 
\chi ~~~,~~~ \gamma ~=~ 2 ~-~ \eta ~-~ \chi_T 
\end{equation}
where $d$~$=$~$2\mu$, define the respective anomalous dimensions with $\eta$
corresponding to that of $\phi^{ia}$. The exponents $\chi$ and $\chi_T$ 
correspond to the respective vertex anomalous dimensions of $\sigma$ and 
$T^{ab}$ with $\phi^{ia}$. These interactions are present in the universal 
theory. By contrast the spectator interactions, which involve only these two 
fields with themselves and not $\phi^{ia}$, are present in the different forms 
in the theory in each critical spacetime dimension $D$. For completeness it is 
worth noting the leading large $N$ critical exponent expressions,
\cite{35,36,38},  
\begin{eqnarray} 
\eta^{\mbox{\footnotesize{H}}}_1 &=& -~ 
\frac{4\Gamma(2\mu-2)}{\Gamma(2-\mu)\Gamma(\mu-1)\Gamma(\mu-2) \Gamma(\mu+1)m} 
\nonumber \\
\eta^{\mbox{\footnotesize{CS}}}_1 &=& -~ \frac{2(m+1)\Gamma(2\mu-2)} 
{\Gamma(\mu+1)\Gamma(\mu-1) \Gamma(\mu-2)\Gamma(2-\mu)} \nonumber \\ 
\eta^{\mbox{\footnotesize{AU}}}_1 &=& -~ \frac{2(m-1)(m+2)\Gamma(2\mu-2)} 
{m\Gamma(\mu+1)\Gamma(\mu-1) \Gamma(\mu-2)\Gamma(2-\mu)} \nonumber \\
\chi^{\mbox{\footnotesize{H}}}_1 &=& -~ \frac{\mu(4\mu-5) 
\eta^{\mbox{\footnotesize{H}}}_1}{(\mu-2)} ~~~,~~~ 
\chi^{\mbox{\footnotesize{CS}}}_1 ~=~ -~ \frac{\mu(4\mu-5) 
\eta^{\mbox{\footnotesize{CS}}}_1}{(\mu-2)} \nonumber \\
\chi^{\mbox{\footnotesize{CS}}}_{T,1} &=& -~ \frac{\mu[(2\mu-3)m + (4\mu-5)] 
\eta^{\mbox{\footnotesize{CS}}}_1}{(\mu-2)(m+1)} \nonumber \\
\chi^{\mbox{\footnotesize{AU}}}_{T,1} &=& -~ \frac{\mu(m-2)[(m+4)(2\mu-3) + 1] 
\eta^{\mbox{\footnotesize{AU}}}_1}{(m-1)(m+2)(\mu-2)} ~.  
\end{eqnarray} 
Higher order corrections are available in \cite{35,36,37,38,46}. For the four 
dimensional theory the exponents corresponding to the critical slope of the 
$\beta$-functions have also been determined, \cite{46}, which also give an 
insight into the stability of each fixed point. With
\begin{equation}
\omega ~=~ (\mu - 2) ~+~ \sum_{i=1}^\infty \frac{\omega_i}{N^i} 
\end{equation}
then, \cite{46}, 
\begin{eqnarray}  
\omega^{\mbox{\footnotesize{Heis}}}_{+\,1} &=& -~ \frac{4(2\mu-1)^2 
\Gamma(2\mu-2)}{\Gamma(2-\mu)\Gamma(\mu-1)\Gamma(\mu-2) \Gamma(\mu+1)mN} 
\nonumber \\
\omega^{\mbox{\footnotesize{AU}}}_{+\,1} &=& -~ \left[ 2\mu^2 - 3\mu 
- 1 ~+~ \frac{\mu (m-2) [ 2\mu-5 - 2(m+4)(2\mu-3) ]}{(m-1)(m+2)} \right] 
\frac{\eta^{\mbox{\footnotesize{AU}}}_1}{N} \nonumber \\
\omega^{\mbox{\footnotesize{CS}}}_{\pm\,1} &=& 
\frac{(2\mu-1)\eta^{\mbox{\footnotesize{CS}}}_1}{2(m+1)(\mu-2)N} \left[ 
m(\mu-1)(\mu-4) + (2\mu^2 - 7\mu + 4) \right. \nonumber \\ 
&& \left. \pm~ \mu \left[ (m^2-1)(\mu-1)^2 + 2(m-1)(2\mu-3)(\mu-1) 
+ (5\mu-8)^2 \right]^{\half} \right]
\end{eqnarray}
where the $\pm$ sign corresponds to two solutions in the CS case due to the
presence of the two fields $\sigma$ and $T^{ab}$. For the other two fixed 
points there is only one solution since there is in effect only one coupling 
constant relevant at these respective points. These large $N$ exponents in 
essence appear to provide a more fundamental insight into the critical point 
structure of the underlying universal theory in the large $N$ expansion. 
Although it is worth emphasising that these results are useful for checking 
explicit perturbative expressions it is the critical point structure of the 
$O(N)$~$\times$~$O(m)$ theory for finite $N$ which is our main focus. 

In addition to the wave function and coupling constant renormalization we will
also consider the renormalization of the three masses in (\ref{laglgw6m}) and
determine the mixing matrix of anomalous dimensions to three loops. However,
the comparison of the exponents derived from this matrix to the corresponding
large $N$ critical exponents is rather subtle which derives from the underlying
operator. This is apparent in comparing the structure of the quadratic terms in
$\sigma$ and $T^{ab}$ in (\ref{laglgw4g}), (\ref{laglgw4}) and 
(\ref{laglgw6m}). In four dimensions the quadratic terms are present to
implement the auxiliary field formaulation of the quartic interaction. By
contrast in six dimensions these fields have no auxiliary interpretation and
the quadratic parts have to appear with a mass in order to have a consistent
dimensionality. In terms of (\ref{laglgw4g}) the couplings $g_1$ and $g_2$ are
not dimensionless away from four dimensions and can be interpreted as mass
scales in higher dimensions. In other words at criticality the exponents
$\omega$ at each of the three fixed points will be related to the mass
anomalous dimensions of $\sigma$ and $T^{ab}$ computed in perturbation theory 
using (\ref{laglgw6}) and then evaluated at criticality. In reality it is not a
direct relation since one is dealing with a mass mixing matrix. Instead one
compares the appropriate exponent $\omega$ with the eigen-anomalous dimension
of the mixing matrix at criticality. The situation for the mass exponent of 
$\phi^{ia}$ is slightly different. In perturbation theory the three mass
operators have the same canonical dimension and hence the mixing matrix is 
$3$~$\times$~$3$. In the large $N$ expansion the canonical dimension of $\half
\phi^{ia} \phi^{ia}$ differs from the other two mass operators and the critical
exponent associated with the $\phi^{ia}$ mass operator is not related to an
$\omega$ exponent. Instead like in the $O(N)$ scalar theory the $\phi^{ia}$
mass dimension is given by the anomalous dimension of the $\sigma$ field. In
other words it is proportional to the sum of $\eta$ and $\chi$.  

\sect{Results.}

We now turn to the derivation of the various renormalization group functions
for (\ref{laglgw6}) which builds essentially on the method developed in 
\cite{22} for the $O(N)$ case but with a minor caveat in the computation of 
several $\beta$-functions. The general procedure is to use an automated
Feynman diagram approach where all the graphs are generated electronically
using the {\sc Qgraf} package, \cite{47}. With indices appended to this output 
and Feynman rules substituted, the resulting scalar integrals are integrated by
applying the integration by parts algorithm developed by Laporta, \cite{48}.
This reduces all the integrals to a basic set of what is termed master 
integrals whose $\epsilon$ expansion is substituted at the final step. To
implement the Laporta algorithm we have used the {\sc Reduze} version,
\cite{49,50}, and used the masters given in \cite{51}. In \cite{22} we checked 
that the three loop masters were consistent with the known four dimensional 
masters by applying the Tarasov method, \cite{52,53}. This is a way of relating
$d$-dimensional Feynman integrals to $(d+2)$-dimensional ones. For all the 
renormalization group functions we determine we use the method of \cite{54} to 
implement the renormalization in an automatic way. The Feynman diagrams are all
evaluated as functions of the bare parameters, such as the coupling constants, 
and then these are replaced by their renormalized counterparts which involved 
the as yet undetermined counterterms. The counterterms are then chosen to 
render the appropriate Green's function finite with reference to a particular 
scheme which throughout will be the $\MSbar$ scheme. All the computations we 
carry out could not be possible without the use of the symbolic manipulation 
language {\sc Form} and its threaded version {\sc Tform}, \cite{55,56}. 

With the $O(N)$~$\times$~$O(m)$ symmetry the Feynman rules for the propagators 
and vertices involving the field $T^{ab}$ have an associated colour tensor.
In other words the $T^{ab}$ propagator will involve the tensor, \cite{38},
\begin{equation}
P^{abcd} ~=~ \frac{1}{2} \left[ \delta^{ac} \delta^{bd} 
+ \delta^{ad} \delta^{bc} - \frac{2}{m} \delta^{ab} \delta^{cd} \right]
\end{equation}
which satisfies the trace properties
\begin{equation}
P^{abcc} ~=~ P^{aacd} ~=~ 0 ~~,~~ 
P^{abcb} ~=~ \frac{(m-1)(m+2)}{2m} \delta^{ac} ~. 
\end{equation}
It also satisfies the projector relations 
\begin{equation}
P^{abpq} P^{pqcd} ~=~ P^{abcd} ~~,~~ 
P^{abpq} P^{cpdq} ~=~ \frac{(m-2)}{2m} P^{abcd} ~.
\end{equation}
Equipped with this the Feynman rule for the triple $T^{ab}$ vertex involves
the rank $6$ colour tensor
\begin{equation}
P_3^{abcdef} ~=~ P^{abpq} P^{cdpr} P^{efqr} ~.
\end{equation}
Consequently
\begin{eqnarray}
P_3^{abcdpq} P^{efpq} &=& P_3^{abcdef} ~~,~~ 
P_3^{abcded} ~=~ \frac{(m-2)(m+4)}{4m} P^{abce} \nonumber \\ 
P_3^{abpqrs} P_3^{cdpqrs} &=& \frac{(m-2)(m+4)}{4m} P^{abcd} 
\end{eqnarray}
for instance. Encoding these within a {\sc Form} module allows the group theory 
evaluation of the higher loop graphs to proceed more efficiently. 

{\begin{table}[ht]
\begin{center}
\begin{tabular}{|c||c|c|c|c|}
\hline
Green's function & One loop & Two loop & Three loop & Total \\
\hline
$\phi^{ia} \phi^{jb}$ & $~2$ & $~\,23$ & $~~~514$ & $~~~539$ \\
$\sigma \sigma$ & $~3$ & $~\,19$ & $~~~343$ & $~~~365$ \\
$T^{ab} T^{cd}$ & $~3$ & $~\,27$ & $~~~589$ & $~~~619$ \\
$T^{ab} \phi^{ia} \phi^{jb}$ & $~5$ & $137$ & $~\,4984$ & $~\,5126$ \\
$T^{ab} T^{cd} T^{ef}$ & $~5$ & $155$ & $~\,5857$ & $~\,6017$ \\
\hline
Total & $18$ & $361$ & $12287$ & $12666$ \\
\hline
\end{tabular}
\end{center}
\begin{center}
{Table 1. Number of Feynman diagrams for each $2$- and $3$-point function.}
\end{center}
\end{table}}

The procedure we used to compute the large number of Feynman diagrams for the
most part follows that described in \cite{22} to which we refer the reader for
the more technical aspects and focus on the amendments we made here to
renormalize (\ref{laglgw6m}) at three loops. One useful technique which was 
exploited in \cite{22} was that in addition to the wave function 
renormalization constants, the coupling constant and mass renormalization 
constants could be determined by purely evaluating the $2$-point functions of 
each field. This was because each basic scalar propagator $1/k^2$ could be 
replaced by
\begin{equation}
\frac{\delta^{ij}}{k^2} ~\mapsto~ \frac{\delta^{ij}}{k^2} ~+~
\frac{\lambda_1\delta^{ij}}{(k^2)^2} ~+~ \frac{\lambda_2 g d^{ijk_e}}{(k^2)^2}
\end{equation}
where the parameters $\lambda_1$ and $\lambda_2$ tag the mass operator
insertion and $3$-point vertex insertion both at zero momentum. The group
structure of the general cubic theory is included on the final term and $k_e$ 
is a fixed index corresponding to the external leg of that vertex. In 
performing this replacement one truncates the expansion at the linear term in
$\lambda_i$ as this reproduces all the relevant graphs for the respective
mass operator and vertex renormalizations with one nullified external leg.
This expansion does not lead to any problems in six dimensions as a $1/(k^2)^2$ 
propagator is infrared safe unlike in four dimensions. We have recalled this
procedure partly because we have exploited it to minimize the amount of
computations we need to perform. Equally because it is not fully applicable to 
renormalizing (\ref{laglgw6}) since it misses out certain graphs which involve 
the $\sigma T^{ab} T^{ab}$ vertex. While we used it for the mass mixing matrix 
for (\ref{laglgw6m}) the replacement does not generate all the vertex graphs 
for the renormalization of the couplings $g_3$ and $g_5$. Instead for the 
associated $3$-point Green's functions we had to generate all the Feynman 
diagrams separately using {\sc Qgraf} and evaluate them with one nullified 
external vertex. While tedious there were no major difficulties. To gauge the 
size of the overall renormalization which was carried out, the number of graphs 
we computed for each Green's function is given in Table $1$. 
 
The results of our computations are the renormalization group functions. As we
will mainly focus our analysis on the $O(N)$~$\times$~$O(2)$ theory we record
these, partly because of that but also due to space consideration, but note 
that the full $O(N)$~$\times$~$O(m)$ expressions are provided in the associated
data file. First, the anomalous dimensions for the three fields are 
\begin{eqnarray}
\left. \gamma_\phi(g_i) \right|_{m=2} &=& 
-~ \frac{1}{6} \left[ g_1^2 + g_3^2 \right] \nonumber \\
&& +~ \frac{1}{432} \left[ -~ 22 N g_1^4 + 26 g_1^4 + 48 g_1^3 g_2 
- 11 g_1^2 g_2^2 + 52 g_1^2 g_3^2 - 22 g_1^2 g_4^2 + 144 g_1 g_3^2 g_4
\right. \nonumber \\
&& \left. ~~~~~~~~~~
-~ 11 N g_3^4 - 22 g_3^4 - 22 g_3^2 g_4^2 \right] \nonumber \\
&& +~ \frac{1}{31104} \left[ 52 N^2 g_1^6 - 464 N g_1^6 + 5184 \zeta_3 g_1^6 
- 9064 g_1^6 + 5292 N g_1^5 g_2 - 3264 g_1^5 g_2
\right. \nonumber \\
&& \left. ~~~~~~~~~~~~~
-~ 772 N g_1^4 g_2^2 + 5184 \zeta_3 g_1^4 g_2^2 - 11762 g_1^4 g_2^2 
+ 40 N g_1^4 g_3^2 + 15552 \zeta_3 g_1^4 g_3^2 
\right. \nonumber \\
&& \left. ~~~~~~~~~~~~~
-~ 27192 g_1^4 g_3^2 + 104 N g_1^4 g_4^2 + 236 g_1^4 g_4^2 + 942 g_1^3 g_2^3 
- 3264 g_1^3 g_2 g_3^2 
\right. \nonumber \\
&& \left. ~~~~~~~~~~~~~
+~ 2388 g_1^3 g_2 g_4^2 + 5292 N g_1^3 g_3^2 g_4 - 9792 g_1^3 g_3^2 g_4 
- 504 g_1^3 g_4^3 + 327 g_1^2 g_2^4 
\right. \nonumber \\
&& \left. ~~~~~~~~~~~~~
+ 118 g_1^2 g_2^2 g_3^2 - 772 g_1^2 g_2^2 g_4^2 
+ 10368 \zeta_3 g_1^2 g_2 g_3^2 g_4 - 23760 g_1^2 g_2 g_3^2 g_4 
\right. \nonumber \\
&& \left. ~~~~~~~~~~~~~
+ 2904 g_1^2 g_2 g_4^3 - 736 N g_1^2 g_3^4 + 2304 g_1^2 g_3^4 
- 1648 N g_1^2 g_3^2 g_4^2 
\right. \nonumber \\
&& \left. ~~~~~~~~~~~~~
+ 20736 \zeta_3 g_1^2 g_3^2 g_4^2 - 47048 g_1^2 g_3^2 g_4^2 - 144 g_1^2 g_4^4 
+ 1194 g_1 g_2^2 g_3^2 g_4 
\right. \nonumber \\
&& \left. ~~~~~~~~~~~~~
- 756 g_1 g_2 g_3^2 g_4^2 + 5292 N g_1 g_3^4 g_4 + 1944 g_1 g_3^4 g_4 
+ 6408 g_1 g_3^2 g_4^3 
\right. \nonumber \\
&& \left. ~~~~~~~~~~~~~
- 412 g_2^2 g_3^2 g_4^2 + 1452 g_2 g_3^2 g_4^3 + 13 N^2 g_3^6 - 1282 N g_3^6 
+ 5184 \zeta_3 g_3^6 
\right. \nonumber \\
&& \left. ~~~~~~~~~~~~~
- 9844 g_3^6 - 360 N g_3^4 g_4^2 - 3724 g_3^4 g_4^2 - 144 g_3^2 g_4^4 
\right] ~+~ O(g_i^8) \nonumber \\  
\left. \gamma_\sigma(g_i) \right|_{m=2} &=&
\frac{1}{12} \left[ - 2 N g_1^2 - g_2^2 - 2 g_4^2 \right]
\nonumber \\
&& +~ \frac{1}{432} \left[ 4 N g_1^4 + 96 N g_1^3 g_2 - 22 N g_1^2 g_2^2 
+ 4 N g_1^2 g_3^2 + 96 N g_1 g_3^2 g_4 + 13 g_2^4 - 22 g_2^2 g_4^2 
\right. \nonumber \\
&& \left. ~~~~~~~~~~
+ 96 g_2 g_4^3 - 22 N g_3^2 g_4^2 + 4 g_4^4 \right]
\nonumber \\
&& +~ \frac{1}{62208} \left[ - 11048 N^2 g_1^6 + 10368 \zeta_3 N g_1^6 
- 17120 N g_1^6 + 4608 N^2 g_1^5 g_2 + 2112 N g_1^5 g_2 
\right. \nonumber \\
&& \left. ~~~~~~~~~~~~~
+ 12 N^2 g_1^4 g_2^2 + 25920 \zeta_3 N g_1^4 g_2^2 - 53292 N g_1^4 g_2^2 
+ 20736 \zeta_3 N g_1^4 g_3^2 
\right. \nonumber \\
&& \left. ~~~~~~~~~~~~~
- 34240 N g_1^4 g_3^2 - 824 N g_1^4 g_4^2 - 3120 N g_1^3 g_2^3 
- 2688 N g_1^3 g_2 g_3^2 
\right. \nonumber \\
&& \left. ~~~~~~~~~~~~~
+ 4608 N g_1^3 g_2 g_4^2 + 11712 N g_1^3 g_3^2 g_4 - 20448 N g_1^3 g_4^3 
+ 1904 N g_1^2 g_2^4 
\right. \nonumber \\
&& \left. ~~~~~~~~~~~~~
- 392 N g_1^2 g_2^2 g_3^2 + 24 N g_1^2 g_2^2 g_4^2 
+ 31104 \zeta_3 N g_1^2 g_2 g_3^2 g_4 - 66672 N g_1^2 g_2 g_3^2 g_4 
\right. \nonumber \\
&& \left. ~~~~~~~~~~~~~
+ 4608 N g_1^2 g_2 g_4^3 - 5524 N^2 g_1^2 g_3^4 - 3776 N g_1^2 g_3^4 
+ 41472 \zeta_3 N g_1^2 g_3^2 g_4^2 
\right. \nonumber \\
&& \left. ~~~~~~~~~~~~~
- 77824 N g_1^2 g_3^2 g_4^2 - 824 N g_1^2 g_4^4 + 5808 N g_1 g_2^2 g_3^2 g_4 
- 12240 N g_1 g_2 g_3^2 g_4^2 
\right. \nonumber \\
&& \left. ~~~~~~~~~~~~~
+ 2304 N^2 g_1 g_3^4 g_4 - 672 N g_1 g_3^4 g_4 + 4992 N g_1 g_3^2 g_4^3 
+ 2592 \zeta_3 g_2^6 - 5195 g_2^6 
\right. \nonumber \\
&& \left. ~~~~~~~~~~~~~
+ 1904 g_2^4 g_4^2 - 3120 g_2^3 g_4^3 - 1648 N g_2^2 g_3^2 g_4^2 
+ 25920 \zeta_3 g_2^2 g_4^4- 53280 g_2^2 g_4^4 
\right. \nonumber \\
&& \left. ~~~~~~~~~~~~~
+ 4776 N g_2 g_3^2 g_4^3 + 6720 g_2 g_4^5 + 6 N^2 g_3^4 g_4^2 
- 8408 N g_3^4 g_4^2 + 680 N g_3^2 g_4^4 
\right. \nonumber \\
&& \left. ~~~~~~~~~~~~~
+ 10368 \zeta_3 g_4^6 - 28168 g_4^6 \right] ~+~ O(g_i^8) \nonumber \\
\left. \gamma_T(g_i) \right|_{m=2} &=&
\frac{1}{12} \left[ - N g_3^2 - 2 g_4^2 \right] \nonumber \\
&& +~ \frac{1}{432} \left[2 N g_1^2 g_3^2 - 22 N g_1^2 g_4^2 
+ 96 N g_1 g_3^2 g_4 - 11 g_2^2 g_4^2 + 48 g_2 g_4^3 - 22 N g_3^4 
\right. \nonumber \\
&& \left. ~~~~~~~~~~
- 11 N g_3^2 g_4^2 + 4 g_4^4 \right]
\nonumber \\
&& +~ \frac{1}{31104} \left[ - 206 N^2 g_1^4 g_3^2 + 2592 \zeta_3 N g_1^4 g_3^2
- 4280 N g_1^4 g_3^2 + 52 N^2 g_1^4 g_4^2 - 196 N g_1^4 g_4^2 
\right. \nonumber \\
&& \left. ~~~~~~~~~~~~~
+ 1200 N g_1^3 g_2 g_3^2 + 2904 N g_1^3 g_2 g_4^2 + 1152 N^2 g_1^3 g_3^2 g_4 
- 1344 N g_1^3 g_3^2 g_4 
\right. \nonumber \\
&& \left. ~~~~~~~~~~~~~
- 504 N g_1^3 g_4^3 - 103 N g_1^2 g_2^2 g_3^2 - 772 N g_1^2 g_2^2 g_4^2 
+ 5184 \zeta_3 N g_1^2 g_2 g_3^2 g_4 
\right. \nonumber \\
&& \left. ~~~~~~~~~~~~~
- 9576 N g_1^2 g_2 g_3^2 g_4 + 2388 N g_1^2 g_2 g_4^3 - 2556 N^2 g_1^2 g_3^4 
+ 2168 N g_1^2 g_3^4 
\right. \nonumber \\
&& \left. ~~~~~~~~~~~~~
- 46 N^2 g_1^2 g_3^2 g_4^2 + 15552 \zeta_3 N g_1^2 g_3^2 g_4^2 
- 33836 N g_1^2 g_3^2 g_4^2 + 340 N g_1^2 g_4^4 
\right. \nonumber \\
&& \left. ~~~~~~~~~~~~~
+ 576 N g_1 g_2^2 g_3^2 g_4 - 5364 N g_1 g_2 g_3^2 g_4^2 
+ 576 N^2 g_1 g_3^4 g_4 + 8376 N g_1 g_3^4 g_4 
\right. \nonumber \\
&& \left. ~~~~~~~~~~~~~
+ 2496 N g_1 g_3^2 g_4^3 + 327 g_2^4 g_4^2 + 942 g_2^3 g_4^3 
- 23 N g_2^2 g_3^2 g_4^2 + 5184 \zeta_3 g_2^2 g_4^4
\right. \nonumber \\
&& \left. ~~~~~~~~~~~~~
- 12534 g_2^2 g_4^4 + 576 N g_2 g_3^2 g_4^3 + 2028 g_2 g_4^5 - 412 N^2 g_3^6 
+ 2592 \zeta_3 N g_3^6 
\right. \nonumber \\
&& \left. ~~~~~~~~~~~~~
- 5354 N g_3^6 + 13 N^2 g_3^4 g_4^2 - 2152 N g_3^4 g_4^2 - 36 N g_3^2 g_4^4 
+ 5184 \zeta_3 g_4^6 
\right. \nonumber \\
&& \left. ~~~~~~~~~~~~~
- 9476 g_4^6 \right] ~+~ O(g_i^8) 
\end{eqnarray}
where $\zeta_z$ is the Riemann zeta function and the argument of the functions
represents all five coupling constants. The five $\beta$-functions are of
similar form and we note that  
\begin{eqnarray} 
\left. \beta_1(g_i) \right|_{m=2} &=& 
\frac{1}{24} \left[ - 2 N g_1^3 + 8 g_1^3 + 12 g_1^2 g_2 - g_1 g_2^2 
+ 8 g_1 g_3^2 - 2 g_1 g_4^2 + 12 g_3^2 g_4 \right] \nonumber \\
&& +~ \frac{1}{864} \left[ - 172 N g_1^5 
- 536 g_1^5 
+ 264 N g_1^4 g_2 
- 360 g_1^4 g_2 
- 22 N g_1^3 g_2^2 
- 628 g_1^3 g_2^2 
\right. \nonumber \\
&& \left. ~~~~~~~~~
+ 4 N g_1^3 g_3^2 
- 1072 g_1^3 g_3^2 
+ 40 g_1^3 g_4^2 
- 24 g_1^2 g_2^3 
- 240 g_1^2 g_2 g_3^2 
+ 168 g_1^2 g_2 g_4^2 
\right. \nonumber \\
&& \left. ~~~~~~~~~
+ 96 N g_1^2 g_3^2 g_4 
- 600 g_1^2 g_3^2 g_4 
- 216 g_1^2 g_4^3 
+ 13 g_1 g_2^4 
- 22 g_1 g_2^2 g_4^2 
\right. \nonumber \\
&& \left. ~~~~~~~~~
- 648 g_1 g_2 g_3^2 g_4 
+ 96 g_1 g_2 g_4^3 
- 88 N g_1 g_3^4 
+ 16 g_1 g_3^4 
- 22 N g_1 g_3^2 g_4^2 
\right. \nonumber \\
&& \left. ~~~~~~~~~
- 1256 g_1 g_3^2 g_4^2 
+ 4 g_1 g_4^4 
- 108 g_2 g_3^2 g_4^2 
+ 84 N g_3^4 g_4 
- 24 g_3^4 g_4 
+ 60 g_3^2 g_4^3 \right] 
\nonumber \\
&& +~ \frac{1}{124416} \left[ 14648 N^2 g_1^7 
+ 259200 \zeta_3 N g_1^7 
- 81376 N g_1^7 
+ 20736 \zeta_3 g_1^7 
+ 251360 g_1^7 
\right. \nonumber \\
&& \left. ~~~~~~~~~~~~~~
- 144 N^2 g_1^6 g_2 
- 311040 \zeta_3 N g_1^6 g_2 
+ 249408 N g_1^6 g_2 
+ 186624 \zeta_3 g_1^6 g_2 
\right. \nonumber \\
&& \left. ~~~~~~~~~~~~~~
+ 18000 g_1^6 g_2 
+ 12 N^2 g_1^5 g_2^2 
+ 25920 \zeta_3 N g_1^5 g_2^2 
- 107980 N g_1^5 g_2^2 
\right. \nonumber \\
&& \left. ~~~~~~~~~~~~~~
- 41472 \zeta_3 g_1^5 g_2^2 
+ 358480 g_1^5 g_2^2 
+ 20736 \zeta_3 N g_1^5 g_3^2 
- 106848 N g_1^5 g_3^2 
\right. \nonumber \\
&& \left. ~~~~~~~~~~~~~~
+ 62208 \zeta_3 g_1^5 g_3^2 
+ 754080 g_1^5 g_3^2 
+ 23496 N g_1^5 g_4^2 
- 15712 g_1^5 g_4^2 
\right. \nonumber \\
&& \left. ~~~~~~~~~~~~~~
- 9120 N g_1^4 g_2^3 
+ 124416 \zeta_3 g_1^4 g_2^3 
+ 97776 g_1^4 g_2^3 
+ 7488 N g_1^4 g_2 g_3^2 
\right. \nonumber \\
&& \left. ~~~~~~~~~~~~~~
+ 248832 \zeta_3 g_1^4 g_2 g_3^2 
+ 59712 g_1^4 g_2 g_3^2 
- 4896 N g_1^4 g_2 g_4^2 
- 20736 g_1^4 g_2 g_4^2 
\right. \nonumber \\
&& \left. ~~~~~~~~~~~~~~
- 186624 \zeta_3 N g_1^4 g_3^2 g_4 
+ 160704 N g_1^4 g_3^2 g_4 
+ 435456 \zeta_3 g_1^4 g_3^2 g_4 
\right. \nonumber \\
&& \left. ~~~~~~~~~~~~~~
- 29424 g_1^4 g_3^2 g_4 
+ 6624 N g_1^4 g_4^3 
- 50688 g_1^4 g_4^3 
+ 1904 N g_1^3 g_2^4 
\right. \nonumber \\
&& \left. ~~~~~~~~~~~~~~
+ 62208 \zeta_3 g_1^3 g_2^4 
+ 9960 g_1^3 g_2^4 
- 392 N g_1^3 g_2^2 g_3^2 
+ 158032 g_1^3 g_2^2 g_3^2 
\right. \nonumber \\
&& \left. ~~~~~~~~~~~~~~
+ 24 N g_1^3 g_2^2 g_4^2 
- 44032 g_1^3 g_2^2 g_4^2 
+ 31104 \zeta_3 N g_1^3 g_2 g_3^2 g_4 
\right. \nonumber \\
&& \left. ~~~~~~~~~~~~~~
- 98352 N g_1^3 g_2 g_3^2 g_4 
- 82944 \zeta_3 g_1^3 g_2 g_3^2 g_4 
+ 655776 g_1^3 g_2 g_3^2 g_4 
\right. \nonumber \\
&& \left. ~~~~~~~~~~~~~~
+ 4608 N g_1^3 g_2 g_4^3 
- 124416 \zeta_3 g_1^3 g_2 g_4^3 
+ 17664 g_1^3 g_2 g_4^3 
- 5524 N^2 g_1^3 g_3^4 
\right. \nonumber \\
&& \left. ~~~~~~~~~~~~~~
+ 373248 \zeta_3 N g_1^3 g_3^4 
- 27552 N g_1^3 g_3^4 
- 124416 \zeta_3 g_1^3 g_3^4 
+ 218016 g_1^3 g_3^4 
\right. \nonumber \\
&& \left. ~~~~~~~~~~~~~~
+ 41472 \zeta_3 N g_1^3 g_3^2 g_4^2 
- 62336 N g_1^3 g_3^2 g_4^2 
- 165888 \zeta_3 g_1^3 g_3^2 g_4^2 
\right. \nonumber \\
&& \left. ~~~~~~~~~~~~~~
+ 924160 g_1^3 g_3^2 g_4^2 
- 824 N g_1^3 g_4^4 
+ 248832 \zeta_3 g_1^3 g_4^4 
+ 43968 g_1^3 g_4^4 
\right. \nonumber \\
&& \left. ~~~~~~~~~~~~~~
- 31104 \zeta_3 g_1^2 g_2^5 
+ 33612 g_1^2 g_2^5 
- 8352 g_1^2 g_2^3 g_3^2 
- 6000 g_1^2 g_2^3 g_4^2 
\right. \nonumber \\
&& \left. ~~~~~~~~~~~~~~
+ 5808 N g_1^2 g_2^2 g_3^2 g_4 
+ 124416 \zeta_3 g_1^2 g_2^2 g_3^2 g_4 
+ 142128 g_1^2 g_2^2 g_3^2 g_4 
\right. \nonumber \\
&& \left. ~~~~~~~~~~~~~~
- 10656 g_1^2 g_2^2 g_4^3 
- 93312 \zeta_3 N g_1^2 g_2 g_3^4 
+ 100488 N g_1^2 g_2 g_3^4 
\right. \nonumber \\
&& \left. ~~~~~~~~~~~~~~
- 36480 g_1^2 g_2 g_3^4 
- 20688 N g_1^2 g_2 g_3^2 g_4^2 
+ 373248 \zeta_3 g_1^2 g_2 g_3^2 g_4^2 
\right. \nonumber \\
&& \left. ~~~~~~~~~~~~~~
+ 275568 g_1^2 g_2 g_3^2 g_4^2 
- 186624 \zeta_3 g_1^2 g_2 g_4^4 
+ 245616 g_1^2 g_2 g_4^4 
\right. \nonumber \\
&& \left. ~~~~~~~~~~~~~~
+ 2304 N^2 g_1^2 g_3^4 g_4 
- 248832 \zeta_3 N g_1^2 g_3^4 g_4 
+ 128688 N g_1^2 g_3^4 g_4 
\right. \nonumber \\
&& \left. ~~~~~~~~~~~~~~
+ 248832 \zeta_3 g_1^2 g_3^4 g_4 
- 201120 g_1^2 g_3^4 g_4 
- 16128 N g_1^2 g_3^2 g_4^3 
\right. \nonumber \\
&& \left. ~~~~~~~~~~~~~~
+ 248832 \zeta_3 g_1^2 g_3^2 g_4^3 
+ 183840 g_1^2 g_3^2 g_4^3 
- 16128 g_1^2 g_4^5 
+ 2592 \zeta_3 g_1 g_2^6 
\right. \nonumber \\
&& \left. ~~~~~~~~~~~~~~
- 5195 g_1 g_2^6 
+ 1904 g_1 g_2^4 g_4^2 
- 27072 g_1 g_2^3 g_3^2 g_4 
- 3120 g_1 g_2^3 g_4^3 
\right. \nonumber \\
&& \left. ~~~~~~~~~~~~~~
- 1648 N g_1 g_2^2 g_3^2 g_4^2 
+ 186624 \zeta_3 g_1 g_2^2 g_3^2 g_4^2 
+ 60176 g_1 g_2^2 g_3^2 g_4^2 
\right. \nonumber \\
&& \left. ~~~~~~~~~~~~~~
+ 25920 \zeta_3 g_1 g_2^2 g_4^4 
- 53280 g_1 g_2^2 g_4^4 
- 42912 N g_1 g_2 g_3^4 g_4 
\right. \nonumber \\
&& \left. ~~~~~~~~~~~~~~
- 124416 \zeta_3 g_1 g_2 g_3^4 g_4 
+ 36288 g_1 g_2 g_3^4 g_4 
+ 4776 N g_1 g_2 g_3^2 g_4^3 
\right. \nonumber \\
&& \left. ~~~~~~~~~~~~~~
+ 62208 \zeta_3 g_1 g_2 g_3^2 g_4^3 
+ 60672 g_1 g_2 g_3^2 g_4^3 
+ 6720 g_1 g_2 g_4^5 
+ 6424 N^2 g_1 g_3^6 
\right. \nonumber \\
&& \left. ~~~~~~~~~~~~~~
+ 32144 N g_1 g_3^6 
+ 82944 \zeta_3 g_1 g_3^6 
+ 72416 g_1 g_3^6 
+ 6 N^2 g_1 g_3^4 g_4^2 
\right. \nonumber \\
&& \left. ~~~~~~~~~~~~~~
- 45512 N g_1 g_3^4 g_4^2 
+ 209504 g_1 g_3^4 g_4^2 
+ 680 N g_1 g_3^2 g_4^4 
+ 124416 \zeta_3 g_1 g_3^2 g_4^4 
\right. \nonumber \\
&& \left. ~~~~~~~~~~~~~~
- 85632 g_1 g_3^2 g_4^4 
+ 10368 \zeta_3 g_1 g_4^6 
- 28168 g_1 g_4^6 
+ 11808 g_2^3 g_3^2 g_4^2 
\right. \nonumber \\
&& \left. ~~~~~~~~~~~~~~
- 62208 \zeta_3 g_2^2 g_3^2 g_4^3 
+ 63744 g_2^2 g_3^2 g_4^3 
+ 3600 N g_2 g_3^4 g_4^2 
- 62208 \zeta_3 g_2 g_3^4 g_4^2 
\right. \nonumber \\
&& \left. ~~~~~~~~~~~~~~
+ 70272 g_2 g_3^4 g_4^2 
- 43488 g_2 g_3^2 g_4^4 
- 1188 N^2 g_3^6 g_4 
+ 31848 N g_3^6 g_4 
\right. \nonumber \\
&& \left. ~~~~~~~~~~~~~~
- 62208 \zeta_3 g_3^6 g_4 
+ 165840 g_3^6 g_4 
- 2208 N g_3^4 g_4^3 
+ 124416 \zeta_3 g_3^4 g_4^3 
\right. \nonumber \\
&& \left. ~~~~~~~~~~~~~~
+ 23136 g_3^4 g_4^3 
- 62208 \zeta_3 g_3^2 g_4^5 
+ 118320 g_3^2 g_4^5 \right] ~+~ O(g_i^9) ~. 
\label{beta1}
\end{eqnarray} 
The remaining expressions are given in Appendix A where the mixing matrix of
mass anomalous dimensions is also provided. One test of the expressions we
have computed is that the double and triple poles of all the underlying 
renormalization constants correctly emerge as predicted by the renormalization
group formalism. Equally we have checked the limit back to the pure $O(N)$
theory where the $O(m)$ indices are completely passive and found agreement
with \cite{21}. The final checks which we have derive from the comparison with 
the large $N$ exponents which we devolve to the next section.

\sect{Large $N$ analysis.}

Equipped with the explicit forms of the renormalization group functions we are
in a position to check them against the large $N$ critical exponents for each
of the three fixed points. In order to do this we follow the prescription
introduced in \cite{21} and define scaled coupling constants by 
\begin{equation}
g_1 ~=~ i \sqrt{\frac{12\epsilon}{mN}} x ~~,~~
g_2 ~=~ i \sqrt{\frac{12\epsilon}{mN}} y ~~,~~
g_3 ~=~ i \sqrt{\frac{12\epsilon}{N}} z ~~,~~
g_4 ~=~ i \sqrt{\frac{12\epsilon}{mN}} t ~~,~~
g_5 ~=~ i \sqrt{\frac{12\epsilon}{N}} w ~.
\label{coupfac}
\end{equation}
With these we can deduce the location of each fixed point in a large $N$ 
expansion where each coefficient of the power of $1/N$ is a function of
$\epsilon$ having set $d$~$=$~$6$~$-$~$2\epsilon$. Each of the three fixed 
points is defined by different field content and therefore for the Heisenberg 
and AU only several of the coupling constants are non-zero. From the respective
$\beta$-functions we find  
\begin{eqnarray}
x &=& 1
+ \left(
22 
- \frac{155}{3} \epsilon 
+ \frac{1777}{36} \epsilon^2 
\right) \frac{1}{mN} \nonumber \\
&& + \left(
726 
- 3410 \epsilon 
+ \frac{29093}{9} \epsilon^2 
- 4680 \zeta_3 \epsilon^2
\right) \frac{1}{m^2N^2} ~+~ O\left(\epsilon^3;\frac{1}{N^3}\right) 
\nonumber \\
y &=& 6
+ \left(
972 
- 1290 \epsilon 
+ \frac{2781}{2} \epsilon^2 
\right) \frac{1}{mN} \nonumber \\
&& + \left(
 412596 
- 1036020 \epsilon 
+ 1083644 \epsilon^2 
- 628560 \zeta_3 \epsilon^2
\right) \frac{1}{m^2N^2} ~+~ O\left(\epsilon^3;\frac{1}{N^3}\right) 
\nonumber \\
z &=& t ~=~ w ~=~ 0
\end{eqnarray}
for the Heisenberg case which is consistent with \cite{21} where the order
symbol represents the truncation point for the two independent expansions. For 
AU we have 
\begin{eqnarray}
x &=& y ~=~ t ~=~ 0 \nonumber \\ 
z &=& 1
+ \left(
11
- 40 \frac{1}{m}
+ \frac{7}{2} m
+ \frac{299}{3} \frac{1}{m} \epsilon
- \frac{155}{6} \epsilon
- \frac{25}{3} m \epsilon
- \frac{3829}{36} \frac{1}{m} \epsilon^2
+ \frac{1777}{72} \epsilon^2
+ \frac{80}{9} m \epsilon^2
\right) \frac{1}{N} \nonumber \\
&& + \left(
- \frac{477}{2}
+ 2400 \frac{1}{m^2}
- 1320 \frac{1}{m}
+ \frac{231}{2} m
+ \frac{147}{8} m^2
- 14180 \frac{1}{m^2} \epsilon
+ 6959 \frac{1}{m} \epsilon
+ 1949 \epsilon
\right. \nonumber \\
&& \left. ~~~~~
- \frac{2555}{4} m \epsilon
- 150 m^2 \epsilon
+ \frac{38755}{18} \frac{1}{m^2} \epsilon^2
- 20664 \zeta_3 \frac{1}{m^2} \epsilon^2
- \frac{136469}{36} \frac{1}{m} \epsilon^2
+ 10296 \zeta_3 \frac{1}{m} \epsilon^2
\right. \nonumber \\
&& \left. ~~~~~
- \frac{19919}{24} \epsilon^2
+ 1242 \zeta_3 \epsilon^2
+ \frac{123919}{144} m \epsilon^2
- 576 \zeta_3 m \epsilon^2
+ \frac{23695}{72} m^2 \epsilon^2
\right) \frac{1}{N^2} ~+~ O\left(\epsilon^3;\frac{1}{N^3}\right) \nonumber \\
w &=& 6
+ \left(
486
- 3240 \frac{1}{m}
+ 81 m
+ 5178 \frac{1}{m} \epsilon
- 645 \epsilon
- 150 m \epsilon
- \frac{12105}{2} \frac{1}{m} \epsilon^2
+ \frac{2781}{4} \epsilon^2
\right. \nonumber \\
&& \left. ~~~~~
+ 180 m \epsilon^2
\right) \frac{1}{N} \nonumber \\ 
&& + \left(
- 118071
+ 4874400 \frac{1}{m^2}
- 1417320 \frac{1}{m}
+ 32283 m
+ \frac{10161}{4} m^2
- 14470680 \frac{1}{m^2} \epsilon
\right. \nonumber \\
&& \left. ~~~~~
+ 3945054 \frac{1}{m} \epsilon
+ 464454 \epsilon
- \frac{186915}{2} m \epsilon
- 10830 m^2 \epsilon
+ 16668989 \frac{1}{m^2} \epsilon^2
\right. \nonumber \\
&& \left. ~~~~~
- 7556976 \zeta_3 \frac{1}{m^2} \epsilon^2
- \frac{8635273}{2} \frac{1}{m} \epsilon^2
+ 2238192 \zeta_3 \frac{1}{m} \epsilon^2
- \frac{2355195}{4} \epsilon^2
+ 236196 \zeta_3 \epsilon^2
\right. \nonumber \\
&& \left. ~~~~~
+ \frac{915527}{8} m \epsilon^2
- 42120 \zeta_3 m \epsilon^2
+ \frac{76709}{4} m^2 \epsilon^2
\right) \frac{1}{N^2} ~+~ O\left(\epsilon^3;\frac{1}{N^3}\right) 
\end{eqnarray}
and, finally, 
\begin{eqnarray}
x &=& 1
+ \left(
11
+ 11 m
- \frac{155}{6} \epsilon
- \frac{155}{6} m \epsilon 
+ \frac{1777}{72} \epsilon^2
+ \frac{1777}{72} m \epsilon^2 
\right) \frac{1}{N} \nonumber \\ 
&& + \left(
 \frac{1563}{2}
+ 63 m
- \frac{237}{2} m^2
- \frac{6855}{2} \epsilon
- \frac{835}{2} m \epsilon 
+ 435 \epsilon m^2
+ \frac{35345}{9} \epsilon^2
- 2646 \zeta_3 \epsilon^2
\right. \nonumber \\
&& \left. ~~~~~
+ \frac{4085}{72} m \epsilon^2
- 1602 \zeta_3 m \epsilon^2
- \frac{54101}{72} m^2 \epsilon^2
- 432 \zeta_3 m^2 \epsilon^2
\right) \frac{1}{N^2} ~+~ O\left(\epsilon^3;\frac{1}{N^3}\right) \nonumber \\ 
y &=& 6
+ \left(
486
+ 486 m
- 645 \epsilon
- 645 m \epsilon
+ \frac{2781}{4} \epsilon^2
+ \frac{2781}{4} m \epsilon^2
\right) \frac{1}{N} \nonumber \\ 
&& + \left(
248949
+ 133398 m
+ 30249 m^2
- 660675 \epsilon
- 317175 m \epsilon
- 58170 m^2 \epsilon
+ \frac{1419565}{2} \epsilon^2
\right. \nonumber \\
&& \left. ~~~~~
- 354780 \zeta_3 \epsilon^2
+ \frac{1289545}{4} m \epsilon^2
- 215460 \zeta_3 m \epsilon^2
+ \frac{205901}{4} m^2 \epsilon^2
- 58320 \zeta_3 m^2 \epsilon^2
\right) \frac{1}{N^2} \nonumber \\
&& +~ O\left(\epsilon^3;\frac{1}{N^3}\right) \nonumber \\ 
z &=& 1 
+ \left(
11
+ \frac{7}{2} m
- \frac{155}{6} \epsilon
- \frac{25}{3} m \epsilon
+ \frac{1777}{72} \epsilon^2
+ \frac{80}{9} m \epsilon^2
\right) \frac{1}{N} \nonumber \\ 
&& + \left(
\frac{1563}{2}
+ \frac{231}{2} m
+ \frac{147}{8} m^2
- \frac{6855}{2} \epsilon
- \frac{2555}{4} m \epsilon
- 150 m^2 \epsilon
+ \frac{35345}{9} \epsilon^2
- 2646 \zeta_3 \epsilon^2
\right. \nonumber \\
&& \left. ~~~~~
+ \frac{123919}{144} m \epsilon^2
- 576 \zeta_3 m \epsilon^2
+ \frac{23695}{72} m^2 \epsilon^2
\right) \frac{1}{N^2} ~+~ O\left(\epsilon^3;\frac{1}{N^3}\right) \nonumber \\ 
t &=& 6
+ \left(
486
+ 216 m
- 645 \epsilon
- 315 m \epsilon
+ \frac{2781}{4} \epsilon^2
+ \frac{1407}{4} m \epsilon^2
\right) \frac{1}{N} \nonumber \\ 
&& + \left(
248949
+ 65988 m
+ 7389 m^2
- 660675 \epsilon
- 168030 m \epsilon
- 19545 m^2 \epsilon
+ \frac{1419565}{2} \epsilon^2
\right. \nonumber \\
&& \left. ~~~~~
- 354780 \zeta_3 \epsilon^2
+ 183756 m \epsilon^2
- 99900 \zeta_3 m \epsilon^2
+ 25357 m^2 \epsilon^2
- 11664 \zeta_3 m^2 \epsilon^2
\right) \frac{1}{N^2} \nonumber \\
&& +~ O\left(\epsilon^3;\frac{1}{N^3}\right) \nonumber \\ 
w &=& 6
+ \left(
486
+ 81 m
- 645 \epsilon
- 150 m \epsilon
+ \frac{2781}{4} \epsilon^2
+ 180 m \epsilon^2
\right) \frac{1}{N} \nonumber \\ 
&& + \left(
248949
+ 32283 m
+ \frac{10161}{4} m^2
- 660675 \epsilon
- \frac{186915}{2} m \epsilon
- 10830 m^2 \epsilon
+ \frac{1419565}{2} \epsilon^2
\right. \nonumber \\
&& \left. ~~~~~
- 354780 \zeta_3 \epsilon^2
+ \frac{915527}{8} m \epsilon^2
- 42120 \zeta_3 m \epsilon^2
+ \frac{76709}{4} m^2 \epsilon^2
\right) \frac{1}{N^2} ~+~ O\left(\epsilon^3;\frac{1}{N^3}\right) 
\end{eqnarray} 
for CS where all the couplings are active. With these particular values at 
each of the three fixed points we find agreement with the known large $N$ 
exponents \cite{35,36,37,38,46} out to $O(\epsilon^3)$. This includes the mass 
mixing matrix. However, the comparison with the mass dimension exponents is not
straightforward since one has to compare with the anomalous dimensions of the
eigenvalues of the mass mixing matrix $\gamma_{ij}(g_k)$ evaluated at each 
critical point. For instance, at AU the exponent 
$\omega^{\mbox{\footnotesize{AU}}}_{+\,1}$ is in precise agreement with the 
critical eigen-anomalous dimension. Equally at CS the exponents 
$\eta$~$+$~$\chi$ and the linear combination 
$\omega^{\mbox{\footnotesize{CS}}}_{+\,1}$~$+$~$
\omega^{\mbox{\footnotesize{CS}}}_{-\,1}$ are also in exact correspondence with
the $O(\epsilon^3)$ terms of the eigen-anomalous dimensions. These nontrivial
large $N$ checks at each of the three fixed points on the three loop $\MSbar$ 
renormalization group functions provide confidence that our perturbative
computation is correct.  

\sect{Conformal window search.} 

One of our aims is to find the conformal window for (\ref{laglgw6}). Given the
nature of the renormalization group equations computed at three loops it 
transpires that pinning down the actual range of the conformal window is not 
straightforward. A similar observation was made in \cite{28} for the four 
dimensional $O(N)$~$\times$~$O(3)$ case using the conformal bootstrap method.
For the pure $O(N)$ case, \cite{21}, which has two coupling constants unlike 
our five here the conformal window was determined by solving the equations 
\begin{equation}
\beta_1(g_i) ~=~ \beta_2(g_i) ~=~ 0 ~~~,~~~ 
\det \left( \frac{\partial \beta_i}{\partial g_j} \right) ~=~ 0 
\label{betazerohess}
\end{equation} 
where $i$~$=$~$1$ and $2$. As the generalization of these equations to five
couplings is
\begin{equation}
\beta_1(g_i) ~=~ \beta_2(g_i) ~=~ \ldots ~=~ \beta_5(g_i) ~=~ 0
\label{betazero}
\end{equation}
together with the Hessian it turned out our computer resources were not
sufficient to solve the complete system numerically in general. Instead we have
resorted to an
alternative strategy which could equally well have been applied to the pure 
$O(N)$ theory. One observation of \cite{21,22} in respect of the conformal
window in the $O(N)$ case was the nature of the fixed point spectrum above
and below a conformal window boundary. At leading order the main window
boundary is at $N_{cr}$~$=$~$1038$, \cite{20}, for $O(N)$. Above this value of 
$N_{cr}$ there are fixed points with real couplings. By contrast below this 
point there are no real fixed points. Given this distinguishing property we 
have solved the equations (\ref{betazero}) for fixed values of $N$ and then 
analysed the stability properties of the real solutions. The stability of a 
fixed point is determined by finding the eigenvalues of the stability matrix 
${\cal S}$ at each real fixed point in turn for the chosen value of $N$ where 
${\cal S}$ is defined by 
\begin{equation}
{\cal S} ~=~ \left( \frac{\partial \beta_i}{\partial g_j} \right) ~.
\end{equation} 
Specifically if all the eigenvalues are negative then this signifies 
ultraviolet (UV) stability, while if all eigenvalues are positive then that 
fixed point would be UV unstable and consequently infrared (IR) stable. 
Obtaining a mixed signature indicates that the fixed point is a saddle point. 
In the situation where the eigenvalues are zero, we can only conclude that the 
fixed point is marginal and beyond the linear approximation. We did not find 
any such cases for the values of $N$ analysed. While this may appear to be a 
tedious process for finding the conformal window boundary it turned out to be 
relatively quick since one can narrow the search area by a process of 
sectioning. 

To illustrate the process we focus for the moment on the $O(N)$~$\times$~$O(2)$
theory. First, given the fact that there are more couplings in (\ref{laglgw6})
the criteria defining the window boundary differs slightly from the properties
of the $O(N)$ case. In order to define this we need to introduce a descriptive
syntax which derives partly from the nature of the fixed points which emerge
and the structure of the four dimensional $O(N)$~$\times$~$O(m)$ coupling
constant plane. In (\ref{laglgw4g}) there were three non-trivial fixed points
designated Heisenberg, anti-chiral unstable and chiral stable and they were
associated with different combinations of the fields $\sigma$ and $T^{ab}$ that
were active or not at a fixed point. Moreover with fewer couplings in four 
dimensions each type of fixed point had a definite stability which led to the 
notation AU or CS aside from the Heisenberg solution which was necessarily a 
saddle point. In our conformal window analysis of (\ref{laglgw6}) we will 
retain our AU and CS syntax as well as Heisenberg but use it to represent the 
field content only. So, for instance, indicating an AU fixed point will mean 
that only interactions involving the $T^{ab}$ field are present while a CS type
of fixed point will correspond to all interactions of (\ref{laglgw6}) being 
active. This readjustment in syntax is necessary since, as will become clear, 
the fixed point structure is much richer than that of the six dimensional 
$O(N)$ $\phi^3$ theory and (\ref{laglgw4g}). So we will refer to Heisenberg, AU
and CS types of solutions. Illustrating this with the coupling vector
$(g_1,g_2,g_3,g_4,g_5)$ their characteristic critical coupling constant 
patterns respectively are $(x,y,0,0,0)$, $(0,0,z,0,w)$ and $(x,y,z,t,w)$ where 
we mean that $x$, $y$, $z$, $t$ and $w$ are non-zero in these patterns. For
simplicity we have omitted the constant of proportionality given in 
(\ref{coupfac}). It is important to appreciate that for the Heisenberg, AU and 
CS patterns the actual fixed point which is present could actually be stable or
unstable and not be related to the U or S of the label type. In one respect the 
emergence of these patterns within the perturbative context, where we are now 
working, should not be surprising as the fixed $N$ analysis has to at least 
contain the Heisenberg, AU and CS large $N$ solutions. With this syntax for 
(\ref{laglgw6}) we can now give our criteria for the conformal window boundary.
From the analysis we have carried out we regard a window boundary to be where 
there is a change in the number of a particular pattern of fixed point such as 
CS. We note that as in the $O(N)$ case various fixed point solutions are 
connected to each other via symmetries, \cite{21}, and so we focus on a 
representative fixed point of each such class in the discussion. We also find a
large number of fixed points with complex and purely imaginary values which may
indicate non-unitarity solutions or even that a limit cycle exists. In our 
discussions in this and the next section we will focus only on the real 
solutions for the critical couplings as they lead to clear stability 
properties.

As a first stage to our search strategy it is best to summarize the analysis 
for the upper boundary we found which was $N$~$=$~$1105$ when $m$~$=$~$2$. For 
the case of $N$~$=$~$1106$ we have three CS type fixed points. One of these is 
UV stable which is at
\begin{eqnarray}
x &=&  1.024331 ~+~ 0.602917\epsilon ~-~ 618.493720\epsilon^{2} ~+~
O(\epsilon^{3}) \nonumber \\
y &=& 10.027831 ~-~ 224.568795\epsilon ~+~ 204744.131100\epsilon^{2} ~+~
O(\epsilon^{3}) \nonumber \\
z &=& 1.014679 ~+~ 0.242004\epsilon ~-~ 259.254500\epsilon^{2} ~+~
O(\epsilon^{3}) \nonumber \\
t &=&  8.413935 ~-~ 122.062932\epsilon ~+~ 110001.339800\epsilon^{2} ~+~
O(\epsilon^{3}) \nonumber \\
w &=&  7.750728 ~-~ 86.093662\epsilon ~+~ 77109.596670\epsilon^{2} ~+~
O(\epsilon^{3}) ~.
\end{eqnarray}
The corresponding critical exponents are
\begin{eqnarray}
\gamma^\ast_{\phi} &=& 0.002810\epsilon ~-~ 0.003531\epsilon^{2} ~-~ 
2.095198\epsilon^{3} ~+~ O(\epsilon^{4}) \nonumber \\
\gamma^\ast_{\sigma} &=& 1.158724\epsilon ~-~ 2.828644\epsilon^{2} ~+~ 
2307.673939\epsilon^{3} ~+~ O(\epsilon^{4}) \nonumber \\
\gamma^\ast_{T} &=& 1.093583\epsilon ~-~ 1.472805\epsilon^{2} ~+~ 
1165.028293\epsilon^{3} ~+~ O(\epsilon^{4}) ~. 
\end{eqnarray}
The other two CS style fixed points are saddle points at 
\begin{eqnarray}
x &=& 1.023546 ~-~ 0.790738\epsilon ~+~ 618.557767\epsilon^{2} ~+~
O(\epsilon^{3}) \nonumber \\
y &=&  10.288220 ~+~ 238.034889\epsilon ~-~ 204695.170900\epsilon^{2} ~+~
O(\epsilon^{3}) \nonumber \\
z &=&  1.014350 ~-~ 0.341297\epsilon ~+~ 259.727356\epsilon^{2} ~+~
O(\epsilon^{3}) \nonumber \\
t &=& 8.553710 ~+~ 126.145941\epsilon ~-~ 109987.441000\epsilon^{2} ~+~
O(\epsilon^{3}) \nonumber \\
w &=& 7.848666 ~+~ 87.779203\epsilon ~-~ 77103.604170\epsilon^{2} ~+~
O(\epsilon^{3}) 
\end{eqnarray}
and
\begin{eqnarray}
x &=& -~ 0.869900  ~-~ 0.200484\epsilon ~-~ 0.868576\epsilon^{2} ~+~
O(\epsilon^{3}) \nonumber \\
y &=& 20.723963 ~+~ 8.470150\epsilon ~-~ 14.322290\epsilon^{2} ~+~
O(\epsilon^{3}) \nonumber \\
z &=& 1.011451 ~-~ 0.019282\epsilon ~+~ 0.058843\epsilon^{2} ~+~
O(\epsilon^{3}) \nonumber \\
t &=& -~ 4.381299 ~-~ 2.162646\epsilon ~-~ 6.897939\epsilon^{2} ~+~
O(\epsilon^{3}) \nonumber \\
w &=& 5.927808 ~+~ 0.692949\epsilon ~+~ 3.355853\epsilon^{2} ~+~
O(\epsilon^{3}) ~.
\end{eqnarray}
In addition there are three Heisenberg fixed points, one of which is UV stable 
at 
\begin{eqnarray}
x &=& 1.010040 ~-~  0.023705\epsilon  ~+~ 0.020596\epsilon^{2} ~+~ 
O(\epsilon^{3}) \nonumber \\
y &=& 6.557735 ~-~  0.940183\epsilon  ~+~ 0.810426\epsilon^{2} ~+~ 
O(\epsilon^{3}) \nonumber \\
z &=& 0 ~~~,~~~ t ~=~ 0 ~~~,~~~ w ~=~ 0
\end{eqnarray}
with critical exponents
\begin{eqnarray}
\gamma^\ast_{\phi} &=& 0.000922\epsilon ~-~ 0.001777\epsilon^{2} ~-~ 
0.000152\epsilon^{3} ~+~ O(\epsilon^{4}) \nonumber \\
\gamma^\ast_{\sigma} &=& 1.039622\epsilon ~-~  0.075355\epsilon^{2} ~-~ 
0.008779\epsilon^{3} ~+~ O(\epsilon^{4}) \nonumber \\
\gamma^\ast_{T} &=& 0 ~. 
\end{eqnarray}
The other two fixed points are saddle points and are located at
\begin{eqnarray}
x &=& 0.979414 ~-~ 0.003228\epsilon ~+~ 0.071572\epsilon^{2} ~+~ 
O(\epsilon^{3}) \nonumber \\
y &=& 17.380571 ~+~  10.947386\epsilon  ~+~ 21.645075\epsilon^{2} ~+~ 
O(\epsilon^{3}) \nonumber \\
z &=& 0 ~~~,~~~ t ~=~ 0 ~~~,~~~ w ~=~ 0
\end{eqnarray}
and
\begin{eqnarray}
x &=& -~ 0.857078 ~-~ 0.208350\epsilon ~-~ 0.632470\epsilon^{2} ~+~ 
O(\epsilon^{3}) \nonumber \\
y &=& 19.745752 ~+~ 9.661778\epsilon ~-~ 2.588019\epsilon^{2} ~+~ 
O(\epsilon^{3}) \nonumber \\
z &=& 0 ~~~,~~~ t ~=~ 0 ~~~,~~~ w ~=~ 0 ~. 
\end{eqnarray}
There was one AU fixed point which is UV stable at 
\begin{eqnarray}
x &=& 0 ~~~,~~~ y ~=~ 0 ~~~,~~~ t ~=~ 0 \nonumber \\
z &=& 0.998197 ~+~ 0.006635\epsilon ~-~ 0.008935\epsilon^{2} ~+~ 
O(\epsilon^{3}) \nonumber \\
w &=& 5.367450 ~+~ 0.851212\epsilon ~-~ 1.446454\epsilon^{2} ~+~ 
O(\epsilon^{3}) 
\end{eqnarray}
with critical exponents
\begin{eqnarray}
\gamma^\ast_{\phi} &=&  0.001802\epsilon ~-~  0.003273\epsilon^{2} ~-~ 
0.000708\epsilon^{3} ~+~ O(\epsilon^{4}) \nonumber \\
\gamma^\ast_{\sigma} &=& 0 \nonumber \\
\gamma^\ast_{T} &=& 0.996396\epsilon ~+~ 0.006664\epsilon^{2} ~+~ 
0.002605\epsilon^{3} ~+~ O(\epsilon^{4}) ~.
\end{eqnarray}

For values of $N$ above $1106$ the same pattern and number of Heisenberg, AU 
and CS fixed points emerge with the same stability structure. By contrast for 
$N$~$=$~$1105$ a different style of solution emerges. This is first seen in the
CS type of fixed points in that we have only {\em one} such fixed point which 
is at
\begin{eqnarray}
x &=& -~ 0.869887 ~-~ 0.200513\epsilon ~-~ 0.868979\epsilon^{2} ~+~ 
O(\epsilon^{3}) \nonumber \\
y &=& 20.715552 ~+~ 8.465518\epsilon ~-~ 14.330113\epsilon^{2} ~+~ 
O(\epsilon^{3}) \nonumber \\
z &=& 1.011461 ~-~ 0.019297\epsilon ~+~ 0.058911\epsilon^{2} ~+~ 
O(\epsilon^{3}) \nonumber \\
t &=& -~ 4.380955 ~-~ 2.163247\epsilon ~-~ 6.901395\epsilon^{2} ~+~ 
O(\epsilon^{3}) \nonumber \\
w &=& 5.927669 ~+~ 0.693620\epsilon ~+~ 3.359563\epsilon^{2} ~+~ 
O(\epsilon^{3}) ~.
\end{eqnarray}
More crucially it is a saddle point. In other words there is no stable CS fixed
point. So given this change in pattern we regard $N$~$=$~$1105$ as the bound 
for the conformal window in six dimensions. It is instructive to provide the 
picture for the other types of fixed points for $N$~$=$~$1105$. There are also 
three Heisenberg fixed points. The UV stable one is 
\begin{eqnarray}
x &=& 1.010049 ~-~ 0.023726\epsilon ~+~ 0.020611\epsilon^{2} ~+~ 
O(\epsilon^{3}) \nonumber \\
y &=& 6.558394 ~-~ 0.941587\epsilon ~+~ 0.811596\epsilon^{2} ~+~ 
O(\epsilon^{3}) \nonumber \\
z &=& 0 ~~~, ~~~ t ~=~ 0 ~~~, ~~~ w ~=~ 0
\end{eqnarray}
with critical exponents
\begin{eqnarray}
\gamma^\ast_{\phi} &=& 0.000923\epsilon ~-~ 0.001779\epsilon^{2} ~-~ 
0.000152\epsilon^{3} ~+~ O(\epsilon^{4}) \nonumber \\
\gamma^\ast_{\sigma} &=& 1.039662\epsilon ~-~ 0.075439\epsilon^{2} ~-~ 
0.008783\epsilon^{3} ~+~ O(\epsilon^{4}) \nonumber \\
\gamma^\ast_{T} &=& 0 
\end{eqnarray}
while the other two fixed points are saddle points at 
\begin{eqnarray}
x &=& -~ 0.857055 ~-~ 0.208383\epsilon ~-~ 0.632604\epsilon^{2} ~+~ 
O(\epsilon^{3}) \nonumber \\
y &=& 19.736951 ~+~ 9.657499\epsilon ~-~ 2.589415\epsilon^{2} ~+~ 
O(\epsilon^{3}) \nonumber \\
z &=& 0 ~~~, ~~~ t ~=~ 0 ~~~, ~~~ w ~=~ 0
\end{eqnarray}
and 
\begin{eqnarray}
x &=& 0.979447 ~-~ 0.003297\epsilon ~+~ 0.071496\epsilon^{2} ~+~ 
O(\epsilon^{3}) \nonumber \\
y &=& 17.371128 ~+~ 10.944494\epsilon ~+~ 21.644028\epsilon^{2} ~+~ 
O(\epsilon^{3}) \nonumber \\
z &=& 0 ~~~, ~~~ t ~=~ 0 ~~~, ~~~ w ~=~ 0 ~. 
\end{eqnarray}
The one AU fixed point is UV stable and is located at 
\begin{eqnarray}
x &=& 0 ~~~, ~~~ y ~=~ 0 ~~~, ~~~ t ~=~ 0 \nonumber \\
z &=& 0.998195 ~+~ 0.006641\epsilon ~-~ 0.008942\epsilon^{2} ~+~ 
O(\epsilon^{3}) \nonumber \\
w &=& 5.367025 ~+~ 0.851662\epsilon ~-~ 1.447623\epsilon^{2} ~+~ 
O(\epsilon^{3}) 
\end{eqnarray}
with critical exponents
\begin{eqnarray}
\gamma^\ast_{\phi} &=& 0.001803\epsilon ~-~ 0.003276\epsilon^{2} ~-~ 
0.000709\epsilon^{3} ~+~ O(\epsilon^{4}) \nonumber \\
\gamma^\ast_{\sigma} &=& 0 \nonumber \\
\gamma^\ast_{T} &=& 0.996393\epsilon ~+~ 0.006670\epsilon^{2} ~+~ 
0.002608\epsilon^{3} ~+~ O(\epsilon^{4}) ~. 
\end{eqnarray}
For $N$~$<$~$1105$ we applied our algorithm of section searching for changes in
fixed point patterns but found no further boundaries. However the structure for
certain fixed values of $N$ will be recorded later for completeness. One 
observation on our window analysis is that the boundary at $N$~$=$~$1105$ is 
not dissimilar to the leading order value of $N_{cr}$~$=$~$1038$, \cite{20}, 
for the $O(N)$ case. In \cite{21,22} the $O(\epsilon^3)$ corrections to 
$N_{cr}$ were computed and by using resummation methods a value of $N_{cr}$ 
around $400$ was found for the five dimensional theory. Clearly applying our 
section search method cannot be readily extended beyond the leading order which
is for the strictly six dimensional theory. Instead solving (\ref{betazero}) 
simultaneously with $\det({\cal S})$~$=$~$0$ would be the way to extract such 
corrections but was beyond the range of our computational tools.

We close this section by briefly discussing a different tack for gaining more
insight into the conformal window problem for $O(N)$~$\times$~$O(m)$. As is 
apparent from the $O(N)$~$\times$~$O(2)$ case the change in nature of the fixed
points indicates a boundary. Moreover different types of (real) solutions
emerge. Therefore, for the general $O(N)$~$\times$~$O(m)$ case we searched for
the conformal window for the AU pattern of couplings. In other words we set
$x$~$=$~$y$~$=$~$t$~$=0$ at the outset for a selection of values of $m$ and 
solved (\ref{betazerohess}). Included in this is the equation for the Hessian
which allows us to determine the critical value of $N$ defining the window
boundary, which we will denote by $N^{(m)}_{cr}$ for this AU pattern, without 
having to do a section search. To get a perspective on our results we have 
provided the leading order value for $N^{(m)}_{cr}$ for various $m$ in Table 
$2$. As $m$~$\to$~$\infty$ we found that $N^{(m)}_{cr}$ asymptotes to a 
straight line. While this is only a partial picture for the situation for 
$m$~$>$~$2$ one thing is evident which is that in six dimensions when 
$m$~$\geq$~$5$ there should be a change in pattern for AU type fixed points for
a fixed $N$ search akin to that illustrated in our section based search for 
$m$~$=$~$2$. This is in addition to the change in pattern for the other style 
of solutions. The solution given in Table $2$ for $m$~$=$~$4$ reflects that 
there was no solution rather than an exact value of zero. Although we have 
recorded $0$ in the Table for that reason it does appear to be consistent with 
the monotonic increase in $N^{(m)}_{cr}$ with $m$. Since we are able to solve 
(\ref{betazerohess}) the three loop corrections to the leading order values in 
Table $2$ have been determined. We found 
\begin{eqnarray}
N^{(1)}_{cr} &=& -~ 2946.134605 ~+~ 3951.961993\epsilon ~+~ 
2676.699839\epsilon^{2} ~+~ O(\epsilon^{3}) \nonumber \\
z &=& 1.006955 ~-~ 0.008027\epsilon ~+~ 0.012574\epsilon^{2} ~+~ 
O(\epsilon^{3}) \nonumber \\
w &=& 8.952176 ~-~ 0.933006\epsilon ~+~ 1.840946\epsilon^{2} ~+~ 
O(\epsilon^{3}) \nonumber \\
N^{(2)}_{cr} &=& -~ 1087.488959 ~+~ 1415.172128\epsilon ~+~ 
261.248651\epsilon^{2} ~+~ O(\epsilon^{3}) \nonumber \\
z &=& 1.001844 ~-~ 0.004332\epsilon ~+~ 0.005483\epsilon^{2} ~+~ 
O(\epsilon^{3}) \nonumber \\
w &=& 9.000046 ~-~ 1.261448\epsilon ~-~ 1.365084\epsilon^{2} ~+~ 
O(\epsilon^{3}) \nonumber \\
N^{(3)}_{cr} &=& -~ 410.145045 ~+~  439.505646\epsilon ~+~ 
1591.300276\epsilon^{2} ~+~ O(\epsilon^{3}) \nonumber \\
z &=& 0.988129 ~+~ 0.002686\epsilon ~+~ 0.093273\epsilon^{2} ~+~ 
O(\epsilon^{3}) \nonumber \\
w &=& 9.206805 ~-~ 2.755615\epsilon ~+~ 36.244925\epsilon^{2} ~+~ 
O(\epsilon^{3}) \nonumber \\
N^{(5)}_{cr} &=& 216.767170 ~-~  419.773422\epsilon ~+~ 
25581.601520\epsilon^{2} ~+~ O(\epsilon^{3}) \nonumber \\
z &=& 1.094548 ~-~ 0.073610\epsilon ~-~ 9.071267\epsilon^{2} ~+~ 
O(\epsilon^{3}) \nonumber \\
w &=& 8.708936 ~+~ 1.332420\epsilon ~-~ 267.527508\epsilon^{2} ~+~ 
O(\epsilon^{3}) \nonumber \\
N^{(6)}_{cr} &=& 421.682453 ~-~ 774.149504\epsilon ~+~ 
8084.140233\epsilon^{2} ~+~ O(\epsilon^{3}) \nonumber \\
z &=& 1.053874 ~-~ 0.038490\epsilon ~-~ 0.487837\epsilon^{2} ~+~ 
O(\epsilon^{3}) \nonumber \\
w &=& 8.724938 ~+~ 0.659976\epsilon ~-~ 46.386543\epsilon^{2} ~+~ 
O(\epsilon^{3}) \nonumber \\
N^{(10)}_{cr} &=& 992.309977 ~-~ 1796.450905\epsilon ~+~ 
1605.099447\epsilon^{2} ~+~ O(\epsilon^{3}) \nonumber \\
z &=& 1.035563 ~-~ 0.025317\epsilon ~+~ 0.018625\epsilon^{2} ~+~ 
O(\epsilon^{3}) \nonumber \\
w &=& 8.766117 ~+~ 0.281651\epsilon ~+~ 0.445942\epsilon^{2} ~+~ 
O(\epsilon^{3}) \nonumber \\
N^{(20)}_{cr} &=& 1999.619696 ~-~  3823.678958\epsilon ~-~ 
645.564678\epsilon^{2} ~+~ O(\epsilon^{3}) \nonumber \\
z &=& 1.032648 ~-~ 0.021711\epsilon ~+~ 0.036581\epsilon^{2} ~+~ 
O(\epsilon^{3}) \nonumber \\
w &=& 8.770214 ~+~ 0.294322\epsilon ~+~ 6.987749\epsilon^{2} ~+~ 
O(\epsilon^{3}) \nonumber \\
N^{(30)}_{cr} &=& 2887.855771 ~-~ 5724.541609\epsilon ~-~ 
1656.156005\epsilon^{2} ~+~ O(\epsilon^{3}) \nonumber \\
z &=& 1.032739 ~-~ 0.020634\epsilon ~+~ 0.038671\epsilon^{2} ~+~ 
O(\epsilon^{3}) \nonumber \\
w &=& 8.766733 ~+~ 0.353761\epsilon ~+~ 8.295335\epsilon^{2} ~+~ 
O(\epsilon^{3}) \nonumber \\
N^{(40)}_{cr} &=& 3746.323521 ~-~ 7599.677475\epsilon ~-~ 
2431.924312\epsilon^{2} ~+~ O(\epsilon^{3}) \nonumber \\
z &=& 1.032998 ~-~ 0.020013\epsilon ~+~ 0.039622\epsilon^{2} ~+~ 
O(\epsilon^{3}) \nonumber \\
w &=& 8.763914 ~+~ 0.398295\epsilon ~+~ 8.889004\epsilon^{2} ~+~ 
O(\epsilon^{3}) \nonumber \\
N^{(50)}_{cr} &=& 4592.876982 ~-~ 9466.056881\epsilon ~-~ 
3115.903624\epsilon^{2} ~+~ O(\epsilon^{3}) \nonumber \\
z &=& 1.033232 ~-~ 0.019588\epsilon ~+~ 0.040210\epsilon^{2} ~+~ 
O(\epsilon^{3}) \nonumber \\
w &=& 8.761832 ~+~ 0.430906\epsilon ~+~ 9.233526\epsilon^{2} ~+~ 
O(\epsilon^{3}) 
\label{ncrm}
\end{eqnarray}
for a selection of $m$ where the respective critical couplings have been 
displayed. This is an important point. While we have provided values for 
$N^{(m)}_{cr}$ in (\ref{ncrm}) other solutions were found for each $m$. In 
\cite{21,22} there were three solutions but the small $N^{(m)}_{cr}$ solutions 
were discarded because they were negative or had complex critical couplings. We
have followed the same reasoning here. The negative values for $N^{(m)}_{cr}$ 
are in keeping with similar negative solutions for the eight dimensional 
ultraviolet completion of the $O(N)$ sequence of theories, \cite{57}. We have 
also excluded from this AU analysis values of $N^{(m)}_{cr}$ which have large 
critical couplings as such values are clearly outside the perturbative 
approximation we are using.

\begin{table}[h]
\begin{center}
\label{my-label}
\begin{tabular}{|c||c|c|c|c|c|c|c|c|c|c|c|}
\hline
$m$ & 1 & 2 & 3 & 4 & 5 & 6 & 10 & 20 & 30 & 40 & 50 \\ \hline
$N^{(m)}_{cr}\!$ & -2946.1$\!$ & -1087.5$\!$ & -410.2$\!$ & 0$\!$ & 216.8$\!$ &
421.7$\!$ & 992.3$\!$ & 1999.6$\!$ & 2887.9$\!$ & 3746.3$\!$ & 4592.9$\!$ \\ 
\hline
\end{tabular}
\end{center}
\begin{center}
{Table 2. Leading order value of $N^{(m)}_{cr}$ for the conformal window for 
different values of $m$.}
\end{center}
\end{table}

\sect{Fixed point analysis.}

In this section we present a fixed point analysis for a variety of specific 
values of $N$. This includes the determination of those fixed points which are 
stable or otherwise, in order to give a flavour of the fixed point spectrum 
away from $N$~$=$~$1105$. In addition we will indicate the potential for 
another conformal window boundary for non-CS type fixed points. While we focus 
on a selection of values of $N$ for a reader interested in exploring the 
solution space further the complete set of renormalization group functions for 
arbitrary $m$ can be analysed which are available in the attached data file. We
will begin by looking at $N$~$=$~$1000$ and then proceed to lower values of 
$N$. For $N$~$=$~$1000$ we have one CS fixed type point as expected which is a 
saddle point
\begin{eqnarray}
x &=& -~ 0.868555 ~-~ 0.203744\epsilon ~-~ 0.915849\epsilon^{2} ~+~ 
O(\epsilon^{3}) \nonumber \\
y &=& 19.811433 ~+~ 7.966436\epsilon ~-~ 15.205442\epsilon^{2} ~+~ 
O(\epsilon^{3}) \nonumber \\
z &=& 1.012581 ~-~ 0.020942\epsilon ~+~ 0.066905\epsilon^{2} ~+~ 
O(\epsilon^{3}) \nonumber \\
t &=& -~ 4.342552 ~-~ 2.231269\epsilon ~-~ 7.303390\epsilon^{2} ~+~ 
O(\epsilon^{3}) \nonumber \\
w &=& 5.911324 ~+~ 0.770705\epsilon ~+~ 3.795188\epsilon^{2} ~+~ 
O(\epsilon^{3}) ~. 
\end{eqnarray}
We also have three Heisenberg fixed points, one of which is UV stable at
\begin{eqnarray}
x &=& 1.011102 ~-~ 0.026162\epsilon ~+~ 0.022238\epsilon^{2} ~+~ 
O(\epsilon^{3}) \nonumber \\
y &=& 6.637801 ~-~ 1.117476\epsilon ~+~ 0.962982\epsilon^{2} ~+~ 
O(\epsilon^{3}) \nonumber \\
z &=&  0 ~~~, ~~~ t ~=~ 0 ~~~, ~~~ w ~=~ 0
\end{eqnarray}
with the critical exponents
\begin{eqnarray}
\gamma^\ast_{\phi} &=& 0.001022\epsilon ~-~ 0.001981\epsilon^{2} ~-~ 
0.000145\epsilon^{3} ~+~ O(\epsilon^{4}) \nonumber \\
\gamma^\ast_{\sigma} &=& 1.044358\epsilon ~-~ 0.085562\epsilon^{2} ~-~ 
0.009090\epsilon^{3} ~+~ O(\epsilon^{4}) \nonumber \\
\gamma^\ast_{T} &=& 0 ~. 
\end{eqnarray}
The other two fixed points 
\begin{eqnarray}
x &=& -~ 0.854446 ~-~ 0.212078\epsilon ~-~ 0.647751\epsilon^{2} ~+~ 
O(\epsilon^{3}) \nonumber \\
y &=& 18.789145 ~+~ 9.197094\epsilon ~-~ 2.733818\epsilon^{2} ~+~ 
O(\epsilon^{3}) \nonumber \\
z &=&  0 ~~~, ~~~ t ~=~ 0 ~~~, ~~~ w ~=~ 0
\end{eqnarray}
and
\begin{eqnarray}
x &=& 0.983210 ~-~ 0.011253\epsilon ~+~ 0.063259\epsilon^{2} ~+~ 
O(\epsilon^{3}) \nonumber \\
y &=& 16.345805 ~+~ 10.658027\epsilon ~+~ 21.524495\epsilon^{2} ~+~ 
O(\epsilon^{3}) \nonumber \\
z &=&  0 ~~~, ~~~ t ~=~ 0 ~~~, ~~~ w ~=~ 0
\end{eqnarray}
are saddle points. Again we also have one AU fixed point which is UV stable
which is located at 
\begin{eqnarray}
x &=& 0 ~~~, ~~~ y ~=~ 0 ~~~, ~~~ t ~=~ 0 \nonumber \\
z &=& 0.998006 ~+~ 0.007339\epsilon ~-~ 0.009793\epsilon^{2} ~+~ 
O(\epsilon^{3}) \nonumber \\
w &=& 5.318846 ~+~ 0.901757\epsilon ~-~ 1.582315\epsilon^{2} ~+~ 
O(\epsilon^{3}) 
\end{eqnarray}
giving the critical exponents
\begin{eqnarray}
\gamma^\ast_{\phi} &=& 0.001992\epsilon ~-~  0.003615\epsilon^{2} ~-~ 
0.000789\epsilon^{3} ~+~ O(\epsilon^{4}) \nonumber \\
\gamma^\ast_{\sigma} &=& 0 \nonumber \\
\gamma^\ast_{T} &=& 0.996016\epsilon ~+~ 0.007374\epsilon^{2} ~+~ 
0.003032\epsilon^{3} ~+~ O(\epsilon^{4}) ~. 
\end{eqnarray}
To illustrate the full spectrum of fixed points for a particular value of $N$
we have provided the remaining fixed points for $N$~$=$~$1000$ in Appendix B. 
In addition to other real solutions which do not fit the Heisenberg, AU or CS
pattern we record the complex solutions for completeness there.

Next examining the value of $N$~$=$~$600$ in order to illustrate a change in 
the fixed point pattern, we have one CS saddle point solution at
\begin{eqnarray}
x &=& -~ 0.862204 ~-~ 0.222460\epsilon ~-~ 1.255729\epsilon^{2} ~+~ 
O(\epsilon^{3}) \nonumber \\
y &=& 15.871131 ~+~ 5.764114\epsilon ~-~ 20.128689\epsilon^{2} ~+~ 
O(\epsilon^{3}) \nonumber \\
z &=& 1.020205 ~-~ 0.031243\epsilon ~+~ 0.133767\epsilon^{2} ~+~ 
O(\epsilon^{3}) \nonumber \\
t &=& -~ 4.133094 ~-~ 2.641545\epsilon ~-~ 10.197488\epsilon^{2} ~+~ 
O(\epsilon^{3}) \nonumber \\
w &=& 5.794456 ~+~ 1.264133\epsilon ~+~ 7.071880\epsilon^{2} ~+~ 
O(\epsilon^{3}) ~. 
\end{eqnarray}
In addition there are three Heisenberg fixed points with the one at
\begin{eqnarray}
x &=& 1.018022 ~-~ 0.037843\epsilon ~+~ 0.001985\epsilon^{2} ~+~ 
O(\epsilon^{3}) \nonumber \\
y &=& 7.507506 ~-~ 4.490389\epsilon ~+~ 9.490485\epsilon^{2} ~+~ 
O(\epsilon^{3}) \nonumber \\
z &=& 0 ~~~, ~~~ t ~=~ 0 ~~~, ~~~ w ~=~ 0
\end{eqnarray}
being UV stable giving 
\begin{eqnarray}
\gamma^\ast_{\phi} &=& 0.001727\epsilon ~-~ 0.003465\epsilon^{2} ~-~ 
0.000010\epsilon^{3} ~+~ O(\epsilon^{4}) \nonumber \\
\gamma^\ast_{\sigma} &=& 1.083337\epsilon ~-~ 0.195953\epsilon^{2} ~+~ 
0.089727\epsilon^{3} ~+~ O(\epsilon^{4}) \nonumber \\
\gamma^\ast_{T} &=& 0 
\end{eqnarray}
while the other two fixed points are saddle points at
\begin{eqnarray}
x &=& -~ 0.839313 ~-~ 0.232926\epsilon ~-~ 0.736639\epsilon^{2} ~+~ 
O(\epsilon^{3}) \nonumber \\
y &=& 14.602366 ~+~ 7.174642\epsilon ~-~ 3.193826\epsilon^{2} ~+~ 
O(\epsilon^{3}) \nonumber \\
z &=& 0 ~~~, ~~~ t ~=~ 0 ~~~, ~~~ w ~=~ 0
\end{eqnarray}
and
\begin{eqnarray}
x &=& 1.007039 ~-~ 0.068389\epsilon ~+~ 0.063230\epsilon^{2} ~+~ 
O(\epsilon^{3}) \nonumber \\
y &=& 11.302398 ~+~ 11.995559\epsilon ~+~ 13.765855\epsilon^{2} ~+~ 
O(\epsilon^{3}) \nonumber \\
z &=& 0 ~~~, ~~~ t ~=~ 0 ~~~, ~~~ w ~=~ 0 ~. 
\end{eqnarray}
There is also have one AU fixed point which is UV stable at
\begin{eqnarray}
x &=& 0 ~~~, ~~~ y ~=~ 0 ~~~, ~~~ t ~=~ 0 \nonumber \\
z &=& 0.996683 ~+~ 0.012238\epsilon ~-~ 0.015305\epsilon^{2} ~+~ 
O(\epsilon^{3}) \nonumber \\
w &=& 5.033838 ~+~ 1.165363\epsilon ~-~ 2.476601\epsilon^{2} ~+~ 
O(\epsilon^{3}) 
\end{eqnarray}
with exponents
\begin{eqnarray}
\gamma^\ast_{\phi} &=& 0.003311\epsilon ~-~  0.005969\epsilon^{2} ~-~ 
0.001383\epsilon^{3} ~+~ O(\epsilon^{4}) \nonumber \\
\gamma^\ast_{\sigma} &=& 0 \nonumber \\
\gamma^\ast_{T} &=& 0.993377\epsilon ~+~ 0.012333\epsilon^{2} ~+~ 
0.006784\epsilon^{3} ~+~ O(\epsilon^{4}) ~. 
\end{eqnarray}
We have recorded this spectrum to contrast it with that for $N$~$=$~$519$. So
for this value we then have one CS fixed point which is a saddle point at
\begin{eqnarray}
x &=& -~ 0.860686 ~-~ 0.228611\epsilon ~-~ 1.395472\epsilon^{2} ~+~ 
O(\epsilon^{3}) \nonumber \\
y &=& 14.937476 ~+~ 5.238428\epsilon ~-~ 21.686769\epsilon^{2} ~+~ 
O(\epsilon^{3}) \nonumber \\
z &=& 1.023093 ~-~ 0.0347571\epsilon ~+~ 0.165366\epsilon^{2} ~+~ 
O(\epsilon^{3}) \nonumber \\
t &=& -~ 4.069923 ~-~ 2.780163\epsilon ~-~ 11.373918\epsilon^{2} ~+~ 
O(\epsilon^{3}) \nonumber \\
w &=& 5.750244 ~+~ 1.435664\epsilon ~+~ 8.435080\epsilon^{2} ~+~ 
O(\epsilon^{3}) ~. 
\end{eqnarray}
However, by contrast, we have one Heisenberg fixed point which is a saddle 
point at 
\begin{eqnarray}
x &=& -~ 0.834431 ~-~ 0.239444\epsilon ~-~ 0.765623\epsilon^{2} ~+~ 
O(\epsilon^{3}) \nonumber \\
y &=& 13.591847 ~+~ 6.689873\epsilon ~-~ 3.246649\epsilon^{2} ~+~ 
O(\epsilon^{3}) \nonumber \\
z &=& 0 ~~~, ~~~ t ~=~ 0 ~~~, ~~~ w ~=~ 0 ~.
\end{eqnarray}
In addition there is one UV stable AU type fixed point at
\begin{eqnarray}
x &=& 0 ~~~, ~~~ y ~=~ 0 ~~~, ~~~ t ~=~ 0 \nonumber \\
z &=& 0.996169 ~+~ 0.014150\epsilon ~-~ 0.017238\epsilon^{2} ~+~ 
O(\epsilon^{3}) \nonumber \\
w &=& 4.941667 ~+~ 1.239792\epsilon ~-~ 2.803418\epsilon^{2} ~+~ 
O(\epsilon^{3}) 
\end{eqnarray}
giving critical exponents
\begin{eqnarray}
\gamma^\ast_{\phi} &=& 0.003824\epsilon ~-~ 0.006875\epsilon^{2} ~-~ 
0.0016281\epsilon^{3} ~+~ O(\epsilon^{4}) \nonumber \\
\gamma^\ast_{\sigma} &=& 0 \nonumber \\
\gamma^\ast_{T} &=& 0.992352\epsilon ~+~ 0.014277\epsilon^{2} ~+~ 
0.008615\epsilon^{3} ~+~ O(\epsilon^{4}) ~. 
\end{eqnarray}
So between $N$~$=$~$519$ and $N$~$=$~$600$ the behaviour of a Heisenberg type 
fixed point changes. This seems to indicate that a conformal window type region
exists with respect to the Heisenberg structure and thus there is a new window
between $519$ and $600$. However, its actual location is not of  major
significance in the context of (\ref{laglgw6}) as this in effect corresponds to
the original Heisenberg model with no $T^{ab}$ field. 

The final case we consider in detail in our excursion through fixed values of 
$N$ is $N$~$=$~$2$. It is of potential interest since for this value in a 
variety of models a supersymmetric solution emerged, \cite{21,58,59}. We have 
three CS fixed points all of which are saddle points at 
\begin{eqnarray}
x &=& -~ 0.454392 ~-~ 1.128422\epsilon ~-~ 10.883437\epsilon^{2} ~+~ 
O(\epsilon^{3}) \nonumber \\
y &=& 0.673205 ~+~ 1.783387\epsilon ~+~ 15.854883\epsilon^{2} ~+~ 
O(\epsilon^{3}) \nonumber \\
z &=& 0.318954 ~+~ 0.395758\epsilon ~+~ 3.102196\epsilon^{2} ~+~ 
O(\epsilon^{3}) \nonumber \\
t &=& 0.379850 ~+~ 0.510247\epsilon ~+~ 4.361634\epsilon^{2} ~+~ 
O(\epsilon^{3}) ~. 
\end{eqnarray}
The value for the coupling $w$ has not been provided with the others as a novel
feature emerged for this set. It transpired that there were three fixed points 
with the same $x$, $y$, $z$ and $t$ values but differing only in the $w$
value. Therefore, we note these values separately as
\begin{eqnarray}
w &\in& \{ 0.717916 ~+~ 0.824313\epsilon ~+~ 5.193907\epsilon^{2} ~+~ 
O(\epsilon^{3}) , \nonumber \\
&& ~ -~ 0.267715 ~-~ 0.020190\epsilon ~+~ 0.685452\epsilon^{2} ~+~ 
O(\epsilon^{3}) , \nonumber \\
&& ~ -~ 0.450201 ~-~ 0.479646\epsilon ~-~ 3.632751\epsilon^{2} ~+~ 
O(\epsilon^{3}) \} ~. 
\end{eqnarray}
There was one Heisenberg fixed point, which is a saddle point
\begin{eqnarray}
x &=& -~ 0.470736 ~-~ 0.737444\epsilon ~-~ 5.708527\epsilon^{2} ~+~ 
O(\epsilon^{3}) \nonumber \\
y &=& 0.762184 ~+~ 0.999917\epsilon ~+~ 6.174478\epsilon^{2} ~+~ 
O(\epsilon^{3}) \nonumber \\
z &=& 0 ~~~, ~~~ t ~=~ 0 ~~~, ~~~ w ~=~ 0 ~. 
\end{eqnarray}
In addition we found one AU fixed point which is UV stable
\begin{eqnarray}
x &=& 0 ~~~, ~~~ y ~=~ 0 ~~~, ~~~ t ~=~ 0 \nonumber \\
z &=& 0.577350 ~+~ 1.507526\epsilon ~+~ 19.533564\epsilon^{2} ~+~ 
O(\epsilon^{3}) \nonumber \\
w &=& 0.800625 ~+~ 1.806817\epsilon ~+~ 27.377665\epsilon^{2} ~+~ 
O(\epsilon^{3})
\end{eqnarray}
with critical exponents
\begin{eqnarray}
\gamma^\ast_{\phi} &=& \gamma^\ast_{T} ~=~ 0.333333\epsilon ~+~ 
1.333333\epsilon^{2} ~+~ 22.148148\epsilon^{3} ~+~ O(\epsilon^{4}) \nonumber \\
\gamma^\ast_{\sigma} &=& 0 ~.
\end{eqnarray}
One property of the emergent supersymmetric solutions found in earlier work, 
\cite{21,58,59}, was that critical couplings were equivalent. For this AU 
solution a different feature is apparent which is that the exponents of 
$\phi^{ia}$ and $T^{ab}$ are equal.

\sect{Discussion.}

We have provided a comprehensive three loop analysis of the extension of the
four dimensional Landau-Ginzburg-Wilson $O(N)$~$\times$~$O(m)$ symmetric 
theory to six dimensions. One aspect of our study was to investigate the 
ultraviolet completion beyond four dimensions. The main interest previously had
been in $O(N)$ symmetric theories and our extension (\ref{laglgw6}) fits in
with the vision of how to proceed. Briefly this requires a common interaction
which seeds the theories in various even dimensions along the thread of
Wilson-Fisher fixed points in $d$-dimensions. Each fixed dimension Lagrangian
is required to be renormalizable in its critical dimension which requires the
addition of matter field independent extra interactions. With the increase in
dimension the number of these so-called spectator interactions increases. For
(\ref{laglgw6}) overall there are five interactions each with its own coupling.
One reassuring aspect of our computations is the verification that the three 
loop renormalization group functions are consistent with the large $N$ critical
exponents of \cite{38}. These exponents are determined in the underlying
universal theory and as $1/N$ is a dimensionless coupling constant it 
transcends a specific dimension. In other words the exponents contain
information on the respective renormalization group functions in {\em all} the
theories connected via the Wilson-Fisher fixed point thread. Indeed verifying
that our three loop perturbative results were consistent with the results of
\cite{38} was an important check. 

One consequence of the larger number of coupling constants is a richer spectrum
of fixed points for specific values of $N$ and $m$. While our analysis of this 
concentrated on $m$~$=$~$2$ we do not expect that the general picture of fixed 
points differs conceptually for higher values of $m$. Instead the boundary
values will be at different values of $N$ as our AU study for various values of
$m$ illustrated. Our $m$~$=$~$2$ analysis was similar to the $O(N)$ case of 
\cite{21} with real and complex critical couplings with the latter 
corresponding to non-unitary theories. However, for real solutions we were able
to isolate fixed points which had a structure in keeping with the phase plane 
in the four dimensional model. In other words there are Heisenberg, AU and CS 
type solutions which depend on which combination of $\sigma$ and $T^{ab}$ 
fields are active and their stability was studied for certain values of $N$. 
One of the main areas of interest in the $O(N)$ and $O(N)$~$\times$~$O(m)$ 
symmetric theories is whether there is a fixed point in the five dimensional 
theory and if so what is the conformal window. In \cite{15} a bootstrap study 
indicated that this was not an easy exercise from the lower dimensional point 
of view unless one was examining AU type coupling patterns. Our investigation 
left us with a similar point of view. Although we were able to narrow down the 
leading order value of the window for CS solutions for $m$~$=$~$2$. By contrast
we were able to solve the AU set of equations and found that a window exists 
above $m$~$=$~$5$. However, we have provided the full data from our 
renormalization group functions which can be mined for future studies for other
values of $m$.  

\vspace{1cm}
\noindent
{\bf Acknowledgements.} This work was carried out with the support from the 
STFC through a studentship and the Consolidated Grant ST/L000431/1. Part of
the work was completed at the Galileo Galilei Institute for Theoretical Physics
and both authors are grateful for its hospitality as well as the partial
support of the INFN.

\appendix

\sect{Remaining renormalization group functions.}

For completeness we record the remaining $\beta$-functions for the 
$O(N)$~$\times$~$O(2)$ theory for comparison with (\ref{beta1}) as well as
various other renormalization group functions. The $\beta$-functions are
\begin{eqnarray} 
\left. \beta_2(g_i) \right|_{m=2} &=& 
\frac{1}{8} \left[ 8 N g_1^3 
- 2 N g_1^2 g_2 + 3 g_2^3 
- 2 g_2 g_4^2 + 8 g_4^3 \right] \nonumber \\
&& +~ \frac{1}{288} \left[ 
- 48 N g_1^5 
- 644 N g_1^4 g_2 
- 120 N g_1^3 g_2^2 
- 48 N g_1^3 g_3^2 
+ 62 N g_1^2 g_2^3 
+ 4 N g_1^2 g_2 g_3^2 
\right. \nonumber \\
&& \left. ~~~~~~~~~
- 648 N g_1^2 g_3^2 g_4 
+ 96 N g_1 g_2 g_3^2 g_4 
- 216 N g_1 g_3^2 g_4^2 
- 125 g_2^5 
+ 62 g_2^3 g_4^2 
\right. \nonumber \\
&& \left. ~~~~~~~~~
- 120 g_2^2 g_4^3 
- 22 N g_2 g_3^2 g_4^2 
- 644 g_2 g_4^4 
+ 84 N g_3^2 g_4^3 
- 48 g_4^5 \right] \nonumber \\
&& +~ \frac{1}{41472} \left[ 110784 N^2 g_1^7 
+ 68448 N g_1^7 
- 153896 N^2 g_1^6 g_2 
+ 10368 \zeta_3 N g_1^6 g_2 
\right. \nonumber \\
&& \left. ~~~~~~~~~~~~
+ 118816 N g_1^6 g_2 
+ 45216 N^2 g_1^5 g_2^2 
+ 124416 \zeta_3 N g_1^5 g_2^2 
+ 50592 N g_1^5 g_2^2 
\right. \nonumber \\
&& \left. ~~~~~~~~~~~~
+ 136896 N g_1^5 g_3^2 
+ 1920 N g_1^5 g_4^2 
- 3156 N^2 g_1^4 g_2^3 
+ 88128 \zeta_3 N g_1^4 g_2^3 
\right. \nonumber \\
&& \left. ~~~~~~~~~~~~
+ 255780 N g_1^4 g_2^3 
+ 20736 \zeta_3 N g_1^4 g_2 g_3^2 
+ 111200 N g_1^4 g_2 g_3^2 
\right. \nonumber \\
&& \left. ~~~~~~~~~~~~
- 65912 N g_1^4 g_2 g_4^2 
+ 116928 N g_1^4 g_3^2 g_4 
+ 108864 N g_1^4 g_4^3 
- 41472 \zeta_3 N g_1^3 g_2^4 
\right. \nonumber \\
&& \left. ~~~~~~~~~~~~
- 17376 N g_1^3 g_2^4 
- 45888 N g_1^3 g_2^2 g_3^2 
+ 45216 N g_1^3 g_2^2 g_4^2 
\right. \nonumber \\
&& \left. ~~~~~~~~~~~~
+ 248832 \zeta_3 N g_1^3 g_2 g_3^2 g_4 
+ 37632 N g_1^3 g_2 g_3^2 g_4 
- 175968 N g_1^3 g_2 g_4^3 
\right. \nonumber \\
&& \left. ~~~~~~~~~~~~
+ 55392 N^2 g_1^3 g_3^4 
+ 10560 N g_1^3 g_3^4 
+ 269760 N g_1^3 g_3^2 g_4^2 
+ 108864 N g_1^3 g_4^4 
\right. \nonumber \\
&& \left. ~~~~~~~~~~~~
- 12544 N g_1^2 g_2^5 
+ 4264 N g_1^2 g_2^3 g_3^2 
- 6312 N g_1^2 g_2^3 g_4^2 
- 155520 \zeta_3 N g_1^2 g_2^2 g_3^2 g_4 
\right. \nonumber \\
&& \left. ~~~~~~~~~~~~
+ 174384 N g_1^2 g_2^2 g_3^2 g_4 
+ 45216 N g_1^2 g_2^2 g_4^3 
- 5524 N^2 g_1^2 g_2 g_3^4 
\right. \nonumber \\
&& \left. ~~~~~~~~~~~~
- 3776 N g_1^2 g_2 g_3^4 
+ 476928 \zeta_3 N g_1^2 g_2 g_3^2 g_4^2 
+ 181376 N g_1^2 g_2 g_3^2 g_4^2 
\right. \nonumber \\
&& \left. ~~~~~~~~~~~~
- 65912 N g_1^2 g_2 g_4^4 
- 71424 N^2 g_1^2 g_3^4 g_4 
- 25632 N g_1^2 g_3^4 g_4 
\right. \nonumber \\
&& \left. ~~~~~~~~~~~~
+ 273936 N g_1^2 g_3^2 g_4^3 
+ 1920 N g_1^2 g_4^5 
- 14064 N g_1 g_2^3 g_3^2 g_4 
+ 3312 N g_1 g_2^2 g_3^2 g_4^2 
\right. \nonumber \\
&& \left. ~~~~~~~~~~~~
+ 2304 N^2 g_1 g_2 g_3^4 g_4 
- 672 N g_1 g_2 g_3^4 g_4 
- 124416 \zeta_3 N g_1 g_2 g_3^2 g_4^3 
\right. \nonumber \\
&& \left. ~~~~~~~~~~~~
+ 30912 N g_1 g_2 g_3^2 g_4^3 
+ 20304 N^2 g_1 g_3^4 g_4^2 
- 35424 N g_1 g_3^4 g_4^2 
\right. \nonumber \\
&& \left. ~~~~~~~~~~~~
- 61344 N g_1 g_3^2 g_4^4 
+ 12960 \zeta_3 g_2^7 
+ 33085 g_2^7 
- 12544 g_2^5 g_4^2 
- 41472 \zeta_3 g_2^4 g_4^3 
\right. \nonumber \\
&& \left. ~~~~~~~~~~~~
- 17376 g_2^4 g_4^3 
+ 5648 N g_2^3 g_3^2 g_4^2 
+ 88128 \zeta_3 g_2^3 g_4^4 
+ 252624 g_2^3 g_4^4 
\right. \nonumber \\
&& \left. ~~~~~~~~~~~~
- 24600 N g_2^2 g_3^2 g_4^3 
+ 124416 \zeta_3 g_2^2 g_4^5 
+ 95808 g_2^2 g_4^5 
+ 6 N^2 g_2 g_3^4 g_4^2 
\right. \nonumber \\
&& \left. ~~~~~~~~~~~~
- 8408 N g_2 g_3^4 g_4^2 
- 4360 N g_2 g_3^2 g_4^4 
+ 10368 \zeta_3 g_2 g_4^6 
- 35080 g_2 g_4^6 
\right. \nonumber \\
&& \left. ~~~~~~~~~~~~
- 1584 N^2 g_3^4 g_4^3 
+ 61248 N g_3^4 g_4^3 
- 1776 N g_3^2 g_4^5 
+ 179232 g_4^7 \right] ~+~ O(g_i^9) \nonumber \\
\left. \beta_3(g_i) \right|_{m=2} &=& 
\frac{g_3}{24} \left[ 8 g_1^2 + 24 g_1 g_4 
- N g_3^2 
- 4 g_3^2 
- 2 g_4^2 \right] \nonumber \\
&& +~ \frac{g_3}{864} \left[ 40 N g_1^4 
- 536 g_1^4 
- 120 g_1^3 g_2 
+ 168 N g_1^3 g_4 
- 480 g_1^3 g_4 
+ 20 g_1^2 g_2^2 
- 648 g_1^2 g_2 g_4 
\right. \nonumber \\
&& \left. ~~~~~~~~~
- 214 N g_1^2 g_3^2 
+ 56 g_1^2 g_3^2 
- 22 N g_1^2 g_4^2 
- 1256 g_1^2 g_4^2 
+ 84 g_1 g_2^2 g_4 
- 216 g_1 g_2 g_4^2 
\right. \nonumber \\
&& \left. ~~~~~~~~~
+ 180 N g_1 g_3^2 g_4 
+ 24 g_1 g_3^2 g_4 
+ 120 g_1 g_4^3 
- 11 g_2^2 g_4^2 
+ 48 g_2 g_4^3 
- 44 N g_3^4 
\right. \nonumber \\
&& \left. ~~~~~~~~~
- 476 g_3^4 
- 11 N g_3^2 g_4^2 
- 260 g_3^2 g_4^2 
+ 4 g_4^4 \right] \nonumber \\
&& +~ \frac{g_3}{62208} \left[ 
- 688 N^2 g_1^6 
- 22480 N g_1^6 
+ 10368 \zeta_3 g_1^6 
+ 125680 g_1^6 
- 10440 N g_1^5 g_2 
\right. \nonumber \\
&& \left. ~~~~~~~~~~~~
+ 62208 \zeta_3 g_1^5 g_2 
- 11856 g_1^5 g_2 
- 1584 N^2 g_1^5 g_4 
- 62208 \zeta_3 N g_1^5 g_4 
\right. \nonumber \\
&& \left. ~~~~~~~~~~~~
+ 40176 N g_1^5 g_4 
+ 62208 \zeta_3 g_1^5 g_4 
+ 41712 g_1^5 g_4 
+ 1312 N g_1^4 g_2^2 
\right. \nonumber \\
&& \left. ~~~~~~~~~~~~
- 20736 \zeta_3 g_1^4 g_2^2 
+ 74048 g_1^4 g_2^2 
- 45216 N g_1^4 g_2 g_4 
+ 127440 g_1^4 g_2 g_4 
\right. \nonumber \\
&& \left. ~~~~~~~~~~~~
+ 6562 N^2 g_1^4 g_3^2 
+ 189216 \zeta_3 N g_1^4 g_3^2 
+ 4032 N g_1^4 g_3^2 
- 93312 \zeta_3 g_1^4 g_3^2 
\right. \nonumber \\
&& \left. ~~~~~~~~~~~~
+ 131280 g_1^4 g_3^2 
+ 52 N^2 g_1^4 g_4^2 
- 40452 N g_1^4 g_4^2 
+ 158032 g_1^4 g_4^2 
\right. \nonumber \\
&& \left. ~~~~~~~~~~~~
- 11436 g_1^3 g_2^3 
+ 5712 N g_1^3 g_2^2 g_4 
+ 62208 \zeta_3 g_1^3 g_2^2 g_4 
+ 52752 g_1^3 g_2^2 g_4 
\right. \nonumber \\
&& \left. ~~~~~~~~~~~~
- 31104 \zeta_3 N g_1^3 g_2 g_3^2 
+ 15456 N g_1^3 g_2 g_3^2 
- 16176 g_1^3 g_2 g_3^2 
- 8184 N g_1^3 g_2 g_4^2 
\right. \nonumber \\
&& \left. ~~~~~~~~~~~~
+ 62208 g_1^3 \zeta_3 g_2 g_4^2 
+ 80808 g_1^3 g_2 g_4^2 
+ 360 N^2 g_1^3 g_3^2 g_4 
- 186624 \zeta_3 N g_1^3 g_3^2 g_4 
\right. \nonumber \\
&& \left. ~~~~~~~~~~~~
+ 212856 N g_1^3 g_3^2 g_4 
+ 62208 \zeta_3 g_1^3 g_3^2 g_4 
+ 25680 g_1^3 g_3^2 g_4 
- 2712 N g_1^3 g_4^3 
\right. \nonumber \\
&& \left. ~~~~~~~~~~~~
+ 124416 \zeta_3 g_1^3 g_4^3 
+ 80592 g_1^3 g_4^3 
- 204 g_1^2 g_2^4 
- 31104 \zeta_3 g_1^2 g_2^3 g_4 
\right. \nonumber \\
&& \left. ~~~~~~~~~~~~
+ 9360 g_1^2 g_2^3 g_4 
+ 3281 N g_1^2 g_2^2 g_3^2 
+ 2828 g_1^2 g_2^2 g_3^2 
- 772 N g_1^2 g_2^2 g_4^2 
\right. \nonumber \\
&& \left. ~~~~~~~~~~~~
+ 124416 \zeta_3 g_1^2 g_2^2 g_4^2 
- 5168 g_1^2 g_2^2 g_4^2 
+ 5184 \zeta_3 N g_1^2 g_2 g_3^2 g_4 
\right. \nonumber \\
&& \left. ~~~~~~~~~~~~
+ 6984 N g_1^2 g_2 g_3^2 g_4 
- 41472 \zeta_3 g_1^2 g_2 g_3^2 g_4 
+ 25056 g_1^2 g_2 g_3^2 g_4 
\right. \nonumber \\
&& \left. ~~~~~~~~~~~~
+ 2388 N g_1^2 g_2 g_4^3 
+ 62208 \zeta_3 g_1^2 g_2 g_4^3 
- 9312 g_1^2 g_2 g_4^3 
+ 828 N^2 g_1^2 g_3^4 
\right. \nonumber \\
&& \left. ~~~~~~~~~~~~
- 59016 N g_1^2 g_3^4 
+ 186624 \zeta_3 g_1^2 g_3^4 
- 15744 g_1^2 g_3^4 
- 46 N^2 g_1^2 g_3^2 g_4^2 
\right. \nonumber \\
&& \left. ~~~~~~~~~~~~
+ 15552 \zeta_3 N g_1^2 g_3^2 g_4^2 
- 56812 N g_1^2 g_3^2 g_4^2 
- 20736 \zeta_3 g_1^2 g_3^2 g_4^2 
+ 280544 g_1^2 g_3^2 g_4^2 
\right. \nonumber \\
&& \left. ~~~~~~~~~~~~
+ 340 N g_1^2 g_4^4 
+ 32352 g_1^2 g_4^4 
- 1716 g_1 g_2^4 g_4 
- 13320 g_1 g_2^3 g_4^2 
\right. \nonumber \\
&& \left. ~~~~~~~~~~~~
+ 180 N g_1 g_2^2 g_3^2 g_4 
+ 8364 g_1 g_2^2 g_3^2 g_4 
- 62208 \zeta_3 g_1 g_2^2 g_4^3 
+ 96912 g_1 g_2^2 g_4^3 
\right. \nonumber \\
&& \left. ~~~~~~~~~~~~
- 7740 N g_1 g_2 g_3^2 g_4^2 
- 62208 \zeta_3 g_1 g_2 g_3^2 g_4^2 
+ 76536 g_1 g_2 g_3^2 g_4^2 
- 22320 g_1 g_2 g_4^4 
\right. \nonumber \\
&& \left. ~~~~~~~~~~~~
+ 180 N^2 g_1 g_3^4 g_4 
- 17184 N g_1 g_3^4 g_4 
+ 121056 g_1 g_3^4 g_4 
+ 5520 N g_1 g_3^2 g_4^3 
\right. \nonumber \\
&& \left. ~~~~~~~~~~~~
+ 62208 \zeta_3 g_1 g_3^2 g_4^3 
+ 129984 g_1 g_3^2 g_4^3 
- 62208 \zeta_3 g_1 g_4^5 
+ 56112 g_1 g_4^5 
\right. \nonumber \\
&& \left. ~~~~~~~~~~~~
+ 327 g_2^4 g_4^2 
+ 942 g_2^3 g_4^3 
- 23 N g_2^2 g_3^2 g_4^2 
+ 2560 g_2^2 g_3^2 g_4^2 
+ 5184 \zeta_3 g_2^2 g_4^4 
\right. \nonumber \\
&& \left. ~~~~~~~~~~~~
- 12534 g_2^2 g_4^4 
+ 576 N g_2 g_3^2 g_4^3 
- 24312 g_2 g_3^2 g_4^3 
+ 2028 g_2 g_4^5 
- 386 N^2 g_3^6 
\right. \nonumber \\
&& \left. ~~~~~~~~~~~~
+ 2592 \zeta_3 N g_3^6 
+ 18434 N g_3^6 
+ 10368 \zeta_3 g_3^6 
+ 23512 g_3^6 
+ 13 N^2 g_3^4 g_4^2 
\right. \nonumber \\
&& \left. ~~~~~~~~~~~~
- 8416 N g_3^4 g_4^2 
+ 93312 \zeta_3 g_3^4 g_4^2 
- 70304 g_3^4 g_4^2 
- 36 N g_3^2 g_4^4 
\right. \nonumber \\
&& \left. ~~~~~~~~~~~~
+ 62208 \zeta_3 g_3^2 g_4^4 
+ 9648 g_3^2 g_4^4 
+ 5184 \zeta_3 g_4^6 
- 9476 g_4^6 \right] ~+~ O(g_i^9) \nonumber \\
\left. \beta_4(g_i) \right|_{m=2} &=& 
\frac{1}{24} \left[ - 2 N g_1^2 g_4 + 12 N g_1 g_3^2 
- g_2^2 g_4 + 12 g_2 g_4^2 
- 2 N g_3^2 g_4 + 6 g_4^3 \right] \nonumber \\
&& + \frac{1}{864} \left[ 4 N g_1^4 g_4 
+ 96 N g_1^3 g_2 g_4 
- 72 N g_1^3 g_3^2 
- 216 N g_1^3 g_4^2 
- 22 N g_1^2 g_2^2 g_4 
\right. \nonumber \\
&& \left. ~~~~~~~~
- 324 N g_1^2 g_2 g_3^2 
+ 168 N g_1^2 g_2 g_4^2 
- 1288 N g_1^2 g_3^2 g_4 
+ 40 N g_1^2 g_4^3 
\right. \nonumber \\
&& \left. ~~~~~~~~
- 216 N g_1 g_2 g_3^2 g_4 
+ 144 N g_1 g_3^4 
- 36 N g_1 g_3^2 g_4^2 
+ 13 g_2^4 g_4 
- 24 g_2^3 g_4^2 
- 650 g_2^2 g_4^3 
\right. \nonumber \\
&& \left. ~~~~~~~~
+ 42 N g_2 g_3^2 g_4^2 
- 96 g_2 g_4^4 
- 152 N g_3^4 g_4 
+ 40 N g_3^2 g_4^3 
- 708 g_4^5 \right] \nonumber \\
&& +~ \frac{1}{124416} \left[ 
- 11048 N^2 g_1^6 g_4 
+ 10368 \zeta_3 N g_1^6 g_4 
- 17120 N g_1^6 g_4 
+ 4608 N^2 g_1^5 g_2 g_4 
\right. \nonumber \\
&& \left. ~~~~~~~~~~~~~~
+ 2112 N g_1^5 g_2 g_4 
+ 57312 N^2 g_1^5 g_3^2 
+ 102672 N g_1^5 g_3^2 
+ 27072 N^2 g_1^5 g_4^2 
\right. \nonumber \\
&& \left. ~~~~~~~~~~~~~~
- 43200 N g_1^5 g_4^2 
+ 12 N^2 g_1^4 g_2^2 g_4 
+ 25920 \zeta_3 N g_1^4 g_2^2 g_4 
- 53292 N g_1^4 g_2^2 g_4 
\right. \nonumber \\
&& \left. ~~~~~~~~~~~~~~
- 32544 N^2 g_1^4 g_2 g_3^2 
+ 58464 N g_1^4 g_2 g_3^2 
- 4752 N^2 g_1^4 g_2 g_4^2 
\right. \nonumber \\
&& \left. ~~~~~~~~~~~~~~
- 186624 \zeta_3 N g_1^4 g_2 g_4^2 
+ 250368 N g_1^4 g_2 g_4^2 
- 143672 N^2 g_1^4 g_3^2 g_4 
\right. \nonumber \\
&& \left. ~~~~~~~~~~~~~~
+ 31104 \zeta_3 N g_1^4 g_3^2 g_4 
+ 239520 N g_1^4 g_3^2 g_4 
- 1376 N^2 g_1^4 g_4^3 
\right. \nonumber \\
&& \left. ~~~~~~~~~~~~~~
+ 248832 \zeta_3 N g_1^4 g_4^3 
+ 44520 N g_1^4 g_4^3 
- 3120 N g_1^3 g_2^3 g_4 
+ 79200 N g_1^3 g_2^2 g_3^2 
\right. \nonumber \\
&& \left. ~~~~~~~~~~~~~~
- 10656 N g_1^3 g_2^2 g_4^2 
+ 27072 N^2 g_1^3 g_2 g_3^2 g_4 
+ 248832 \zeta_3 N g_1^3 g_2 g_3^2 g_4 
\right. \nonumber \\
&& \left. ~~~~~~~~~~~~~~
- 15168 N g_1^3 g_2 g_3^2 g_4 
- 124416 \zeta_3 N g_1^3 g_2 g_4^3 
+ 22272 N g_1^3 g_2 g_4^3 
\right. \nonumber \\
&& \left. ~~~~~~~~~~~~~~
+ 108864 N^2 g_1^3 g_3^4 
+ 8640 N g_1^3 g_3^4 
+ 31680 N^2 g_1^3 g_3^2 g_4^2 
\right. \nonumber \\
&& \left. ~~~~~~~~~~~~~~
+ 248832 \zeta_3 N g_1^3 g_3^2 g_4^2 
+ 82656 N g_1^3 g_3^2 g_4^2 
- 44064 N g_1^3 g_4^4 
+ 1904 N g_1^2 g_2^4 g_4 
\right. \nonumber \\
&& \left. ~~~~~~~~~~~~~~
+ 10944 N g_1^2 g_2^3 g_3^2 
- 6000 N g_1^2 g_2^3 g_4^2 
+ 186624 \zeta_3 N g_1^2 g_2^2 g_3^2 g_4 
\right. \nonumber \\
&& \left. ~~~~~~~~~~~~~~
+ 67740 N g_1^2 g_2^2 g_3^2 g_4 
- 44008 N g_1^2 g_2^2 g_4^3 
- 19440 N^2 g_1^2 g_2 g_3^4 
\right. \nonumber \\
&& \left. ~~~~~~~~~~~~~~
- 62208 \zeta_3 N g_1^2 g_2 g_3^4 
+ 62352 N g_1^2 g_2 g_3^4 
- 1584 N^2 g_1^2 g_2 g_3^2 g_4^2 
\right. \nonumber \\
&& \left. ~~~~~~~~~~~~~~
+ 82944 \zeta_3 N g_1^2 g_2 g_3^2 g_4^2 
+ 528600 N g_1^2 g_2 g_3^2 g_4^2 
- 25632 N g_1^2 g_2 g_4^4 
\right. \nonumber \\
&& \left. ~~~~~~~~~~~~~~
- 106612 N^2 g_1^2 g_3^4 g_4 
+ 33696 N g_1^2 g_3^4 g_4 
- 1768 N^2 g_1^2 g_3^2 g_4^3 
\right. \nonumber \\
&& \left. ~~~~~~~~~~~~~~
+ 41472 \zeta_3 N g_1^2 g_3^2 g_4^3 
+ 635376 N g_1^2 g_3^2 g_4^3 
+ 7784 N g_1^2 g_4^5 
\right. \nonumber \\
&& \left. ~~~~~~~~~~~~~~
- 25344 N g_1 g_2^3 g_3^2 g_4 
- 62208 \zeta_3 N g_1 g_2^2 g_3^2 g_4^2 
+ 85656 N g_1 g_2^2 g_3^2 g_4^2 
\right. \nonumber \\
&& \left. ~~~~~~~~~~~~~~
+ 6768 N^2 g_1 g_2 g_3^4 g_4 
- 124416 \zeta_3 N g_1 g_2 g_3^4 g_4 
+ 44928 N g_1 g_2 g_3^4 g_4 
\right. \nonumber \\
&& \left. ~~~~~~~~~~~~~~
- 177120 N g_1 g_2 g_3^2 g_4^3 
+ 31104 N^2 g_1 g_3^6 
- 62208 \zeta_3 N g_1 g_3^6 
\right. \nonumber \\
&& \left. ~~~~~~~~~~~~~~
+ 162576 N g_1 g_3^6 
+ 24912 N^2 g_1 g_3^4 g_4^2 
+ 124416 \zeta_3 N g_1 g_3^4 g_4^2 
\right. \nonumber \\
&& \left. ~~~~~~~~~~~~~~
- 11232 N g_1 g_3^4 g_4^2 
- 124416 \zeta_3 N g_1 g_3^2 g_4^4 
+ 3312 N g_1 g_3^2 g_4^4 
+ 2592 \zeta_3 g_2^6 g_4 
\right. \nonumber \\
&& \left. ~~~~~~~~~~~~~~
- 5195 g_2^6 g_4 
- 31104 \zeta_3 g_2^5 g_4^2 
+ 33612 g_2^5 g_4^2 
+ 62208 \zeta_3 g_2^4 g_4^3 
\right. \nonumber \\
&& \left. ~~~~~~~~~~~~~~
+ 11864 g_2^4 g_4^3 
+ 2592 N g_2^3 g_3^2 g_4^2 
+ 124416 \zeta_3 g_2^3 g_4^4 
+ 88656 g_2^3 g_4^4 
\right. \nonumber \\
&& \left. ~~~~~~~~~~~~~~
- 16788 N g_2^2 g_3^2 g_4^3 
- 15552 \zeta_3 g_2^2 g_4^5 
+ 250512 g_2^2 g_4^5 
- 396 N^2 g_2 g_3^4 g_4^2 
\right. \nonumber \\
&& \left. ~~~~~~~~~~~~~~
+ 15072 N g_2 g_3^4 g_4^2 
+ 16680 N g_2 g_3^2 g_4^4 
- 124416 \zeta_3 g_2 g_4^6 
+ 267264 g_2 g_4^6 
\right. \nonumber \\
&& \left. ~~~~~~~~~~~~~~
- 13168 N^2 g_3^6 g_4 
+ 72576 \zeta_3 N g_3^6 g_4 
- 4856 N g_3^6 g_4 
- 1130 N^2 g_3^4 g_4^3 
\right. \nonumber \\
&& \left. ~~~~~~~~~~~~~~
+ 124416 \zeta_3 N g_3^4 g_4^3 
+ 88584 N g_3^4 g_4^3 
- 41896 N g_3^2 g_4^5 
+ 279936 \zeta_3 g_4^7 
\right. \nonumber \\
&& \left. ~~~~~~~~~~~~~~
+ 184632 g_4^7 \right] ~+~
O(g_i^9) \nonumber \\
\left. \beta_5(g_i) \right|_{m=2} &=& 
\frac{1}{8} \left[ 4 N g_3^3 
- N g_3^2 g_5 + 10 g_4^2 g_5 
- 3 g_5^3 \right] 
\nonumber \\
&& +~ \frac{1}{1152} \left[ 
- 96 N g_1^2 g_3^3 
+ 8 N g_1^2 g_3^2 g_5 
+ 248 N g_1^2 g_4^2 g_5 
- 2592 N g_1 g_3^3 g_4 
- 480 N g_1 g_3^2 g_4 g_5 
\right. \nonumber \\
&& \left. ~~~~~~~~~~~
+ 124 g_2^2 g_4^2 g_5 
- 1248 g_2 g_4^3 g_5 
+ 336 N g_3^5 
+ 344 N g_3^4 g_5 
- 432 N g_3^3 g_4^2 
\right. \nonumber \\
&& \left. ~~~~~~~~~~~
+ 324 N g_3^3 g_5^2 
+ 292 N g_3^2 g_4^2 g_5 
- 126 N g_3^2 g_5^3 
- 2864 g_4^4 g_5 
+ 2340 g_4^2 g_5^3 
\right. \nonumber \\
&& \left. ~~~~~~~~~~~
- 513 g_5^5 \right] \nonumber \\
&& +~ \frac{1}{165888} \left[ 3840 N^2 g_1^4 g_3^3 
+ 136896 N g_1^4 g_3^3 
- 1648 N^2 g_1^4 g_3^2 g_5 
+ 20736 \zeta_3 N g_1^4 g_3^2 g_5 
\right. \nonumber \\
&& \left. ~~~~~~~~~~~~~~
- 34240 N g_1^4 g_3^2 g_5 
- 5920 N^2 g_1^4 g_4^2 g_5 
+ 17056 N g_1^4 g_4^2 g_5 
- 19008 N g_1^3 g_2 g_3^3 
\right. \nonumber \\
&& \left. ~~~~~~~~~~~~~~
+ 9600 N g_1^3 g_2 g_3^2 g_5 
- 56256 N g_1^3 g_2 g_4^2 g_5 
- 130176 N^2 g_1^3 g_3^3 g_4 
\right. \nonumber \\
&& \left. ~~~~~~~~~~~~~~
+ 581760 N g_1^3 g_3^3 g_4 
+ 63360 N^2 g_1^3 g_3^2 g_4 g_5 
- 183552 N g_1^3 g_3^2 g_4 g_5 
\right. \nonumber \\
&& \left. ~~~~~~~~~~~~~~
- 165888 \zeta_3 N g_1^3 g_4^3 g_5 
+ 30528 N g_1^3 g_4^3 g_5 
+ 1920 N g_1^2 g_2^2 g_3^3 
\right. \nonumber \\
&& \left. ~~~~~~~~~~~~~~
- 824 N g_1^2 g_2^2 g_3^2 g_5 
+ 16672 N g_1^2 g_2^2 g_4^2 g_5 
+ 311040 N g_1^2 g_2 g_3^3 g_4 
\right. \nonumber \\
&& \left. ~~~~~~~~~~~~~~
- 207360 \zeta_3 N g_1^2 g_2 g_3^2 g_4 g_5 
+ 161856 N g_1^2 g_2 g_3^2 g_4 g_5 
- 107616 N g_1^2 g_2 g_4^3 g_5 
\right. \nonumber \\
&& \left. ~~~~~~~~~~~~~~
+ 217728 N^2 g_1^2 g_3^5 
- 119616 N g_1^2 g_3^5 
- 98208 N^2 g_1^2 g_3^4 g_5 
- 37952 N g_1^2 g_3^4 g_5 
\right. \nonumber \\
&& \left. ~~~~~~~~~~~~~~
+ 27072 N^2 g_1^2 g_3^3 g_4^2 
+ 995328 \zeta_3 N g_1^2 g_3^3 g_4^2 
+ 21120 N g_1^2 g_3^3 g_4^2 
\right. \nonumber \\
&& \left. ~~~~~~~~~~~~~~
+ 64800 N g_1^2 g_3^3 g_5^2 
- 6704 N^2 g_1^2 g_3^2 g_4^2 g_5 
+ 622080 \zeta_3 N g_1^2 g_3^2 g_4^2 g_5 
\right. \nonumber \\
&& \left. ~~~~~~~~~~~~~~
+ 1466912 N g_1^2 g_3^2 g_4^2 g_5 
- 6984 N g_1^2 g_3^2 g_5^3 
- 114592 N g_1^2 g_4^4 g_5 
\right. \nonumber \\
&& \left. ~~~~~~~~~~~~~~
+ 74448 N g_1^2 g_4^2 g_5^3 
- 65088 N g_1 g_2^2 g_3^3 g_4 
+ 31680 N g_1 g_2^2 g_3^2 g_4 g_5 
\right. \nonumber \\
&& \left. ~~~~~~~~~~~~~~
+ 497664 \zeta_3 N g_1 g_2 g_3^3 g_4^2 
+ 300672 N g_1 g_2 g_3^3 g_4^2 
- 42912 N g_1 g_2 g_3^2 g_4^2 g_5 
\right. \nonumber \\
&& \left. ~~~~~~~~~~~~~~
- 220608 N^2 g_1 g_3^5 g_4 
- 248832 \zeta_3 N g_1 g_3^5 g_4 
- 45504 N g_1 g_3^5 g_4 
\right. \nonumber \\
&& \left. ~~~~~~~~~~~~~~
+ 58752 N^2 g_1 g_3^4 g_4 g_5 
- 497664 \zeta_3 N g_1 g_3^4 g_4 g_5 
- 250944 N g_1 g_3^4 g_4 g_5 
\right. \nonumber \\
&& \left. ~~~~~~~~~~~~~~
+ 248832 \zeta_3 N g_1 g_3^3 g_4^3 
+ 1087488 N g_1 g_3^3 g_4^3 
- 559872 \zeta_3 N g_1 g_3^3 g_4 g_5^2 
\right. \nonumber \\
&& \left. ~~~~~~~~~~~~~~
- 909792 N g_1 g_3^3 g_4 g_5^2 
- 497664 \zeta_3 N g_1 g_3^2 g_4^3 g_5 
- 14592 N g_1 g_3^2 g_4^3 g_5 
\right. \nonumber \\
&& \left. ~~~~~~~~~~~~~~
+ 186624 \zeta_3 N g_1 g_3^2 g_4 g_5^3 
- 72576 N g_1 g_3^2 g_4 g_5^3 
- 4248 g_2^4 g_4^2 g_5 
\right. \nonumber \\
&& \left. ~~~~~~~~~~~~~~
- 82944 \zeta_3 g_2^3 g_4^3 g_5 
- 38544 g_2^3 g_4^3 g_5 
+ 13536 N g_2^2 g_3^3 g_4^2 
- 3352 N g_2^2 g_3^2 g_4^2 g_5 
\right. \nonumber \\
&& \left. ~~~~~~~~~~~~~~
+ 290304 \zeta_3 g_2^2 g_4^4 g_5 
+ 549360 g_2^2 g_4^4 g_5 
+ 37224 g_2^2 g_4^2 g_5^3 
- 82944 \zeta_3 N g_2 g_3^3 g_4^3 
\right. \nonumber \\
&& \left. ~~~~~~~~~~~~~~
- 76032 N g_2 g_3^3 g_4^3 
- 6336 N g_2 g_3^2 g_4^3 g_5 
+ 497664 \zeta_3 g_2 g_4^5 g_5 
\right. \nonumber \\
&& \left. ~~~~~~~~~~~~~~
+ 444768 g_2 g_4^5 g_5 
- 311040 \zeta_3 g_2 g_4^3 g_5^3 
- 387504 g_2 g_4^3 g_5^3 
\right. \nonumber \\
&& \left. ~~~~~~~~~~~~~~
+ 6816 N^2 g_3^7 
- 124416 \zeta_3 N g_3^7 
+ 275712 N g_3^7 
+ 64528 N^2 g_3^6 g_5 
\right. \nonumber \\
&& \left. ~~~~~~~~~~~~~~
+ 207360 \zeta_3 N g_3^6 g_5 
+ 15920 N g_3^6 g_5 
+ 27072 N^2 g_3^5 g_4^2 
\right. \nonumber \\
&& \left. ~~~~~~~~~~~~~~
- 248832 \zeta_3 N g_3^5 g_4^2 
+ 87936 N g_3^5 g_4^2 
- 30456 N^2 g_3^5 g_5^2 
+ 279936 \zeta_3 N g_3^5 g_5^2 
\right. \nonumber \\
&& \left. ~~~~~~~~~~~~~~
- 5184 N g_3^5 g_5^2 
- 4648 N^2 g_3^4 g_4^2 g_5 
- 62208 \zeta_3 N g_3^4 g_4^2 g_5 
- 167936 N g_3^4 g_4^2 g_5 
\right. \nonumber \\
&& \left. ~~~~~~~~~~~~~~
+ 2376 N^2 g_3^4 g_5^3 
+ 139968 \zeta_3 N g_3^4 g_5^3 
+ 199728 N g_3^4 g_5^3 
- 12096 N g_3^3 g_4^4 
\right. \nonumber \\
&& \left. ~~~~~~~~~~~~~~
+ 186624 \zeta_3 N g_3^3 g_4^2 g_5^2 
+ 16848 N g_3^3 g_4^2 g_5^2 
- 77760 \zeta_3 N g_3^3 g_5^4 
\right. \nonumber \\
&& \left. ~~~~~~~~~~~~~~
+ 19116 N g_3^3 g_5^4 
- 96864 N g_3^2 g_4^4 g_5 
+ 90360 N g_3^2 g_4^2 g_5^3 
- 28026 N g_3^2 g_5^5 
\right. \nonumber \\
&& \left. ~~~~~~~~~~~~~~
+ 373248 \zeta_3 g_4^6 g_5 
+ 1293344 g_4^6 g_5 
- 933120 \zeta_3 g_4^4 g_5^3 
- 1361376 g_4^4 g_5^3 
\right. \nonumber \\
&& \left. ~~~~~~~~~~~~~~
+ 629856 \zeta_3 g_4^2 g_5^5 
+ 760428 g_4^2 g_5^5 
- 104976 \zeta_3 g_5^7 
- 137295 g_5^7 \right] \nonumber \\
&& +~ O(g_i^9) ~. 
\end{eqnarray} 
The elements of the mass mixing matrix are 
\begin{eqnarray} 
\left. \gamma_{11}(g_i) \right|_{m=2} &=&
\frac{1}{3} [g_1^2 + g_3^2] \nonumber \\ 
&& + \frac{1}{216} \left[
- 44 N g_1^4 
- 134 g_1^4 
- 30 g_1^3 g_2 
+ 5 g_1^2 g_2^2 
- 268 g_1^2 g_3^2 
+ 10 g_1^2 g_4^2 
- 90 g_1 g_3^2 g_4 
\right. \nonumber \\
&& \left. ~~~~~~~~
- 22 N g_3^4 
+ 4 g_3^4 
+ 10 g_3^2 g_4^2 \right] \nonumber \\
&& + \frac{1}{15552} \left[ 3212 N^2 g_1^6 
+ 31104 \zeta_3 N g_1^6 
- 8032 N g_1^6 
+ 2592 \zeta_3 g_1^6 
+ 31420 g_1^6 
\right. \nonumber \\
&& \left. ~~~~~~~~~~~
- 15552 \zeta_3 N g_1^5 g_2 
+ 4518 N g_1^5 g_2 
+ 15552 \zeta_3 g_1^5 g_2 
- 2964 g_1^5 g_2 
\right. \nonumber \\
&& \left. ~~~~~~~~~~~
+ 7852 N g_1^4 g_2^2 
- 5184 \zeta_3 g_1^4 g_2^2 
+ 18512 g_1^4 g_2^2 
- 9076 N g_1^4 g_3^2 
+ 7776 \zeta_3 g_1^4 g_3^2 
\right. \nonumber \\
&& \left. ~~~~~~~~~~~
+ 94260 g_1^4 g_3^2 
+ 3040 N g_1^4 g_4^2 
- 1964 g_1^4 g_4^2 
- 2859 g_1^3 g_2^3 
+ 15552 \zeta_3 g_1^3 g_2 g_3^2 
\right. \nonumber \\
&& \left. ~~~~~~~~~~~
- 2964 g_1^3 g_2 g_3^2 
- 1578 g_1^3 g_2 g_4^2 
- 15552 \zeta_3 N g_1^3 g_3^2 g_4 
+ 4518 N g_1^3 g_3^2 g_4 
\right. \nonumber \\
&& \left. ~~~~~~~~~~~
+ 46656 \zeta_3 g_1^3 g_3^2 g_4 
- 8892 g_1^3 g_3^2 g_4 
- 4140 g_1^3 g_4^3 
- 51 g_1^2 g_2^4 
- 982 g_1^2 g_2^2 g_3^2 
\right. \nonumber \\
&& \left. ~~~~~~~~~~~
+ 328 g_1^2 g_2^2 g_4^2 
- 10368 \zeta_3 g_1^2 g_2 g_3^2 g_4 
+ 38988 g_1^2 g_2 g_3^2 g_4 
- 1032 g_1^2 g_2 g_4^3 
\right. \nonumber \\
&& \left. ~~~~~~~~~~~
+ 46656 \zeta_3 N g_1^2 g_3^4 
- 2972 N g_1^2 g_3^4 
- 15552 \zeta_3 g_1^2 g_3^4 
+ 27252 g_1^2 g_3^4 
\right. \nonumber \\
&& \left. ~~~~~~~~~~~
+ 12664 N g_1^2 g_3^2 g_4^2 
- 20736 \zeta_3 g_1^2 g_3^2 g_4^2 
+ 74048 g_1^2 g_3^2 g_4^2 
+ 312 g_1^2 g_4^4 
\right. \nonumber \\
&& \left. ~~~~~~~~~~~
- 789 g_1 g_2^2 g_3^2 g_4 
- 6210 g_1 g_2 g_3^2 g_4^2 
- 15552 \zeta_3 N g_1 g_3^4 g_4 
+ 4518 N g_1 g_3^4 g_4 
\right. \nonumber \\
&& \left. ~~~~~~~~~~~
+ 15552 \zeta_3 g_1 g_3^4 g_4 
- 17928 g_1 g_3^4 g_4 
- 10944 g_1 g_3^2 g_4^3 
+ 250 g_2^2 g_3^2 g_4^2 
\right. \nonumber \\
&& \left. ~~~~~~~~~~~
- 516 g_2 g_3^2 g_4^3 
+ 803 N^2 g_3^6 
+ 4018 N g_3^6 
+ 10368 \zeta_3 g_3^6 
+ 9052 g_3^6 
\right. \nonumber \\
&& \left. ~~~~~~~~~~~
+ 4686 N g_3^4 g_4^2 
+ 8854 g_3^4 g_4^2 
+ 312 g_3^2 g_4^4 \right] ~+~ O(g_i^8) \nonumber \\
\left. \gamma_{12}(g_i) \right|_{m=2} &=&
N g_1^2 
+ \frac{N}{12} \left[  
- 2 g_1^4 
- 18 g_1^3 g_2 
- 3 g_1^2 g_2^2 
- 2 g_1^2 g_3^2 
- 18 g_1 g_3^2 g_4 
- 3 g_3^2 g_4^2 \right] \nonumber \\
&& + \frac{N}{864} \left[ 2308 N g_1^6 
+ 1426 g_1^6 
- 1984 N g_1^5 g_2 
+ 1822 g_1^5 g_2 
+ 282 N g_1^4 g_2^2 
+ 864 \zeta_3 g_1^4 g_2^2 
\right. \nonumber \\
&& \left. ~~~~~~~~
+ 1430 g_1^4 g_2^2 
+ 2852 g_1^4 g_3^2 
+ 40 g_1^4 g_4^2 
+ 864 \zeta_3 g_1^3 g_2^3 
+ 1420 g_1^3 g_2^3 
+ 2020 g_1^3 g_2 g_3^2 
\right. \nonumber \\
&& \left. ~~~~~~~~
- 904 g_1^3 g_2 g_4^2 
+ 1426 g_1^3 g_3^2 g_4 
+ 1512 g_1^3 g_4^3 
- 21 g_1^2 g_2^4 
- 300 g_1^2 g_2^2 g_3^2 
\right. \nonumber \\
&& \left. ~~~~~~~~
+ 282 g_1^2 g_2^2 g_4^2 
+ 1728 \zeta_3 g_1^2 g_2 g_3^2 g_4 
+ 1260 g_1^2 g_2 g_3^2 g_4 
- 1080 g_1^2 g_2 g_4^3 
\right. \nonumber \\
&& \left. ~~~~~~~~
+ 1154 N g_1^2 g_3^4 
+ 220 g_1^2 g_3^4 
+ 4060 g_1^2 g_3^2 g_4^2 
+ 756 g_1^2 g_4^4 
- 864 \zeta_3 g_1 g_2^2 g_3^2 g_4 
\right. \nonumber \\
&& \left. ~~~~~~~~
+ 828 g_1 g_2^2 g_3^2 g_4 
+ 3456 \zeta_3 g_1 g_2 g_3^2 g_4^2 
+ 1332 g_1 g_2 g_3^2 g_4^2 
- 992 N g_1 g_3^4 g_4 
\right. \nonumber \\
&& \left. ~~~~~~~~
- 356 g_1 g_3^4 g_4 
+ 1796 g_1 g_3^2 g_4^3 
+ 324 g_2^2 g_3^2 g_4^2 
- 432 g_2 g_3^2 g_4^3 
+ 141 N g_3^4 g_4^2 
\right. \nonumber \\
&& \left. ~~~~~~~~
- 246 g_3^4 g_4^2 
+ 66 g_3^2 g_4^4 \right] ~+~ O(g_i^8) \nonumber \\
\left. \gamma_{13}(g_i) \right|_{m=2} &=&
\frac{1}{2} N g_3^2 
+ \frac{N}{24} \left[ 
- 2 g_1^2 g_3^2 
- 6 g_1^2 g_4^2 
- 36 g_1 g_3^2 g_4 
+ 4 g_3^4 
- 3 g_3^2 g_4^2 \right] \nonumber \\
&& + \frac{N}{1728} \left[ 796 N g_1^4 g_3^2 
+ 1426 g_1^4 g_3^2 
+ 376 N g_1^4 g_4^2 
- 600 g_1^4 g_4^2 
- 198 g_1^3 g_2 g_3^2 
\right. \nonumber \\
&& \left. ~~~~~~~~~~
- 1728 \zeta_3 g_1^3 g_2 g_4^2 
+ 1656 g_1^3 g_2 g_4^2 
- 1984 N g_1^3 g_3^2 g_4 
+ 4040 g_1^3 g_3^2 g_4 
\right. \nonumber \\
&& \left. ~~~~~~~~~~
+ 3456 \zeta_3 g_1^3 g_4^3 
+ 576 g_1^3 g_4^3 
+ 20 g_1^2 g_2^2 g_3^2 
+ 836 g_1^2 g_2^2 g_4^2 
+ 2160 g_1^2 g_2 g_3^2 g_4 
\right. \nonumber \\
&& \left. ~~~~~~~~~~
- 864 g_1^2 g_2 g_4^3 
+ 1512 N g_1^2 g_3^4 
+ 120 g_1^2 g_3^4 
+ 188 N g_1^2 g_3^2 g_4^2 
+ 3456 \zeta_3 g_1^2 g_3^2 g_4^2 
\right. \nonumber \\
&& \left. ~~~~~~~~~~
+ 1660 g_1^2 g_3^2 g_4^2 
+ 132 g_1^2 g_4^4 
- 452 g_1 g_2^2 g_3^2 g_4 
+ 1728 \zeta_3 g_1 g_2 g_3^2 g_4^2 
\right. \nonumber \\
&& \left. ~~~~~~~~~~
+ 1800 g_1 g_2 g_3^2 g_4^2 
- 992 N g_1 g_3^4 g_4 
- 226 g_1 g_3^4 g_4 
+ 3592 g_1 g_3^2 g_4^3 
+ 47 g_2^2 g_3^2 g_4^2 
\right. \nonumber \\
&& \left. ~~~~~~~~~~
- 540 g_2 g_3^2 g_4^3 
+ 432 N g_3^6 
- 864 \zeta_3 g_3^6 
+ 2258 g_3^6 
+ 94 N g_3^4 g_4^2 
\right. \nonumber \\
&& \left. ~~~~~~~~~~
+ 780 g_3^4 g_4^2 
+ 822 g_3^2 g_4^4 \right] ~+~ O(g_i^8) \nonumber \\ 
\left. \gamma_{21}(g_i) \right|_{m=2} &=&
\frac{g_1^2}{2} \nonumber \\
&& + \frac{1}{72} \left[ 14 N g_1^4 
- 20 g_1^4 
- 54 g_1^3 g_2 
- 2 g_1^2 g_2^2 
- 20 g_1^2 g_3^2 
+ 14 g_1^2 g_4^2 
- 54 g_1 g_3^2 g_4 
- 9 g_3^2 g_4^2 \right] \nonumber \\
&& + \frac{1}{10368} \left[ 
- 396 N^2 g_1^6 
- 15552 \zeta_3 N g_1^6 
+ 17596 N g_1^6 
+ 5184 \zeta_3 g_1^6 
+ 3476 g_1^6 
\right. \nonumber \\
&& \left. ~~~~~~~~~~~
- 9792 N g_1^5 g_2 
+ 17532 g_1^5 g_2 
- 500 N g_1^4 g_2^2 
+ 10368 \zeta_3 g_1^4 g_2^2 
+ 10054 g_1^4 g_2^2 
\right. \nonumber \\
&& \left. ~~~~~~~~~~~
+ 848 N g_1^4 g_3^2 
+ 10368 \zeta_3 g_1^4 g_3^2 
+ 6952 g_1^4 g_3^2 
- 792 N g_1^4 g_4^2 
- 676 g_1^4 g_4^2 
\right. \nonumber \\
&& \left. ~~~~~~~~~~~
+ 5184 \zeta_3 g_1^3 g_2^3 
+ 864 g_1^3 g_2^3 
+ 13824 g_1^3 g_2 g_3^2 
- 3888 g_1^3 g_2 g_4^2 
- 2640 N g_1^3 g_3^2 g_4 
\right. \nonumber \\
&& \left. ~~~~~~~~~~~
+ 24948 g_1^3 g_3^2 g_4 
- 10368 \zeta_3 g_1^3 g_4^3 
+ 7344 g_1^3 g_4^3 
- 2592 \zeta_3 g_1^2 g_2^4 
+ 2801 g_1^2 g_2^4 
\right. \nonumber \\
&& \left. ~~~~~~~~~~~
- 696 g_1^2 g_2^2 g_3^2 
- 500 g_1^2 g_2^2 g_4^2 
+ 10368 \zeta_3 g_1^2 g_2 g_3^2 g_4 
+ 12744 g_1^2 g_2 g_3^2 g_4 
\right. \nonumber \\
&& \left. ~~~~~~~~~~~
- 5904 g_1^2 g_2 g_4^3 
- 7776 \zeta_3 N g_1^2 g_3^4 
+ 8374 N g_1^2 g_3^4 
- 3040 g_1^2 g_3^4 
- 704 N g_1^2 g_3^2 g_4^2 
\right. \nonumber \\
&& \left. ~~~~~~~~~~~
+ 20736 \zeta_3 g_1^2 g_3^2 g_4^2 
+ 17600 g_1^2 g_3^2 g_4^2 
- 5184 \zeta_3 g_1^2 g_4^4 
+ 10532 g_1^2 g_4^4 
\right. \nonumber \\
&& \left. ~~~~~~~~~~~
- 2256 g_1 g_2^2 g_3^2 g_4 
+ 15552 \zeta_3 g_1 g_2 g_3^2 g_4^2 
+ 7128 g_1 g_2 g_3^2 g_4^2 
- 3576 N g_1 g_3^4 g_4 
\right. \nonumber \\
&& \left. ~~~~~~~~~~~
- 10368 \zeta_3 g_1 g_3^4 g_4 
+ 3024 g_1 g_3^4 g_4 
+ 5184 \zeta_3 g_1 g_3^2 g_4^3 
- 3672 g_1 g_3^2 g_4^3 
\right. \nonumber \\
&& \left. ~~~~~~~~~~~
+ 984 g_2^2 g_3^2 g_4^2 
- 5184 \zeta_3 g_2 g_3^2 g_4^3 
+ 4968 g_2 g_3^2 g_4^3 
+ 300 N g_3^4 g_4^2 
\right. \nonumber \\
&& \left. ~~~~~~~~~~~
- 5184 \zeta_3 g_3^4 g_4^2 
+ 5856 g_3^4 g_4^2 
- 672 g_3^2 g_4^4 \right] ~+~ O(g_i^8) \nonumber \\
\left. \gamma_{22}(g_i) \right|_{m=2} &=&
\frac{1}{12} \left[ 
- 2 N g_1^2 
+ 5 g_2^2 
- 2 g_4^2 \right] \nonumber \\
&& + \frac{1}{216} \left[ 
- 160 N g_1^4 
- 60 N g_1^3 g_2 
+ 52 N g_1^2 g_2^2 
+ 2 N g_1^2 g_3^2 
+ 48 N g_1 g_3^2 g_4 
- 97 g_2^4 
\right. \nonumber \\
&& \left. ~~~~~~~~
+ 52 g_2^2 g_4^2 
- 60 g_2 g_4^3 
- 11 g_3^2 g_4^2 N 
- 160 g_4^4 \right] \nonumber \\
&& + \frac{1}{62208} \left[ 
- 82472 N^2 g_1^6 
+ 10368 \zeta_3 N g_1^6 
+ 55600 N g_1^6 
+ 45216 N^2 g_1^5 g_2 
\right. \nonumber \\
&& \left. ~~~~~~~~~~~
+ 124416 \zeta_3 N g_1^5 g_2 
- 28128 N g_1^5 g_2 
- 4740 N^2 g_1^4 g_2^2 
+ 57024 \zeta_3 N g_1^4 g_2^2 
\right. \nonumber \\
&& \left. ~~~~~~~~~~~
+ 308076 N g_1^4 g_2^2 
+ 20736 \zeta_3 N g_1^4 g_3^2 
+ 38480 N g_1^4 g_3^2 
- 33368 N g_1^4 g_4^2 
\right. \nonumber \\
&& \left. ~~~~~~~~~~~
- 62208 \zeta_3 N g_1^3 g_2^3 
- 22992 N g_1^3 g_2^3 
- 45888 N g_1^3 g_2 g_3^2 
+ 45216 N g_1^3 g_2 g_4^2 
\right. \nonumber \\
&& \left. ~~~~~~~~~~~
+ 124416 \zeta_3 N g_1^3 g_3^2 g_4 
+ 24672 N g_1^3 g_3^2 g_4 
- 98208 N g_1^3 g_4^3 
- 19768 N g_1^2 g_2^4 
\right. \nonumber \\
&& \left. ~~~~~~~~~~~
+ 6592 N g_1^2 g_2^2 g_3^2 
- 9480 N g_1^2 g_2^2 g_4^2 
- 155520 \zeta_3 N g_1^2 g_2 g_3^2 g_4 
\right. \nonumber \\
&& \left. ~~~~~~~~~~~
+ 174384 N g_1^2 g_2 g_3^2 g_4 
+ 45216 N g_1^2 g_2 g_4^3 
- 5524 N^2 g_1^2 g_3^4 
- 3776 N g_1^2 g_3^4 
\right. \nonumber \\
&& \left. ~~~~~~~~~~~
+ 259200 \zeta_3 N g_1^2 g_3^2 g_4^2 
+ 51776 N g_1^2 g_3^2 g_4^2 
- 33368 N g_1^2 g_4^4 
\right. \nonumber \\
&& \left. ~~~~~~~~~~~
- 24000 N g_1 g_2^2 g_3^2 g_4 
+ 3312 N g_1 g_2 g_3^2 g_4^2 
+ 2304 N^2 g_1 g_3^4 g_4 
- 672 N g_1 g_3^4 g_4 
\right. \nonumber \\
&& \left. ~~~~~~~~~~~
- 62208 \zeta_3 N g_1 g_3^2 g_4^3 
+ 17952 N g_1 g_3^2 g_4^3 
+ 18144 \zeta_3 g_2^6 
+ 52225 g_2^6 
\right. \nonumber \\
&& \left. ~~~~~~~~~~~
- 19768 g_2^4 g_4^2 
- 62208 \zeta_3 g_2^3 g_4^3 
- 22992 g_2^3 g_4^3 
+ 9296 N g_2^2 g_3^2 g_4^2 
\right. \nonumber \\
&& \left. ~~~~~~~~~~~
+ 57024 \zeta_3 g_2^2 g_4^4 
+ 303336 g_2^2 g_4^4 
- 24600 N g_2 g_3^2 g_4^3 
+ 124416 \zeta_3 g_2 g_4^5 
\right. \nonumber \\
&& \left. ~~~~~~~~~~~
+ 17088 g_2 g_4^5 
+ 6 N^2 g_3^4 g_4^2 
- 8408 N g_3^4 g_4^2 
- 1840 N g_3^2 g_4^4 
\right. \nonumber \\
&& \left. ~~~~~~~~~~~
+ 10368 \zeta_3 g_4^6 
- 26872 g_4^6 \right] ~+~ O(g_i^8) \nonumber \\
\left. \gamma_{23}(g_i) \right|_{m=2} &=&
\frac{g_4^2}{2} \nonumber \\
&& + \frac{1}{144} \left[ 
- 54 N g_1^2 g_3^2 
+ 28 N g_1^2 g_4^2 
- 36 N g_1 g_3^2 g_4 
- 4 g_2^2 g_4^2 
- 108 g_2 g_4^3 
+ 7 N g_3^2 g_4^2 
\right. \nonumber \\
&& \left. ~~~~~~~~
- 12 g_4^4 \right] \nonumber \\
&& + \frac{1}{10368} \left[ 
- 2712 N^2 g_1^4 g_3^2 
+ 6060 N g_1^4 g_3^2 
- 396 N^2 g_1^4 g_4^2 
- 5184 \zeta_3 N g_1^4 g_4^2 
\right. \nonumber \\
&& \left. ~~~~~~~~~~~
+ 10928 N g_1^4 g_4^2 
+ 6480 N g_1^3 g_2 g_3^2 
- 5904 N g_1^3 g_2 g_4^2 
+ 2256 N^2 g_1^3 g_3^2 g_4 
\right. \nonumber \\
&& \left. ~~~~~~~~~~~
+ 20736 \zeta_3 N g_1^3 g_3^2 g_4 
- 14400 N g_1^3 g_3^2 g_4 
- 10368 \zeta_3 N g_1^3 g_4^3 
+ 7344 N g_1^3 g_4^3 
\right. \nonumber \\
&& \left. ~~~~~~~~~~~
+ 912 N g_1^2 g_2^2 g_3^2 
- 500 N g_1^2 g_2^2 g_4^2 
+ 15552 \zeta_3 N g_1^2 g_2 g_3^2 g_4 
+ 8424 N g_1^2 g_2 g_3^2 g_4 
\right. \nonumber \\
&& \left. ~~~~~~~~~~~
- 3888 N g_1^2 g_2 g_4^3 
- 1620 N^2 g_1^2 g_3^4 
- 5184 \zeta_3 N g_1^2 g_3^4 
+ 5196 N g_1^2 g_3^4 
\right. \nonumber \\
&& \left. ~~~~~~~~~~~
- 132 N^2 g_1^2 g_3^2 g_4^2 
- 2592 \zeta_3 N g_1^2 g_3^2 g_4^2 
+ 37030 N g_1^2 g_3^2 g_4^2 
- 1468 N g_1^2 g_4^4 
\right. \nonumber \\
&& \left. ~~~~~~~~~~~
- 2112 N g_1 g_2^2 g_3^2 g_4 
- 5184 \zeta_3 N g_1 g_2 g_3^2 g_4^2 
+ 5616 N g_1 g_2 g_3^2 g_4^2 
\right. \nonumber \\
&& \left. ~~~~~~~~~~~
+ 564 N^2 g_1 g_3^4 g_4 
- 10368 \zeta_3 N g_1 g_3^4 g_4 
+ 3744 N g_1 g_3^4 g_4 
- 8064 N g_1 g_3^2 g_4^3 
\right. \nonumber \\
&& \left. ~~~~~~~~~~~
- 2592 \zeta_3 g_2^4 g_4^2 
+ 2801 g_2^4 g_4^2 
+ 5184 \zeta_3 g_2^3 g_4^3 
+ 864 g_2^3 g_4^3 
+ 216 N g_2^2 g_3^2 g_4^2 
\right. \nonumber \\
&& \left. ~~~~~~~~~~~
+ 10368 \zeta_3 g_2^2 g_4^4 
+ 9554 g_2^2 g_4^4 
- 1188 N g_2 g_3^2 g_4^3 
+ 7740 g_2 g_4^5 
- 33 N^2 g_3^4 g_4^2 
\right. \nonumber \\
&& \left. ~~~~~~~~~~~
+ 1256 N g_3^4 g_4^2 
+ 632 N g_3^2 g_4^4 
- 10368 \zeta_3 g_4^6 
+ 20676 g_4^6 \right] ~+~ O(g_i^8) \nonumber \\
\left. \gamma_{31}(g_i) \right|_{m=2} &=&
\frac{g_3^2}{2} 
+ \frac{1}{72} \left[ 
- 20 g_1^2 g_3^2 
- 18 g_1^2 g_4^2 
- 108 g_1 g_3^2 g_4 
+ 7 N g_3^4 
- 2 g_3^4 
+ 5 g_3^2 g_4^2 \right] \nonumber \\
&& + \frac{1}{10368} \left[ 
- 5184 \zeta_3 N g_1^4 g_3^2 
+ 9404 N g_1^4 g_3^2 
+ 5184 \zeta_3 g_1^4 g_3^2 
+ 3476 g_1^4 g_3^2 
+ 2256 N g_1^4 g_4^2 
\right. \nonumber \\
&& \left. ~~~~~~~~~~~
- 1464 g_1^4 g_4^2 
+ 3708 g_1^3 g_2 g_3^2 
- 5184 g_1^3 g_2 g_4^2 
- 7152 N g_1^3 g_3^2 g_4 
\right. \nonumber \\
&& \left. ~~~~~~~~~~~
+ 27648 g_1^3 g_3^2 g_4 
+ 20736 \zeta_3 g_1^3 g_4^3 
+ 3456 g_1^3 g_4^3 
- 374 g_1^2 g_2^2 g_3^2 
+ 5016 g_1^2 g_2^2 g_4^2 
\right. \nonumber \\
&& \left. ~~~~~~~~~~~
+ 10368 \zeta_3 g_1^2 g_2 g_3^2 g_4 
+ 9504 g_1^2 g_2 g_3^2 g_4 
- 10368 \zeta_3 g_1^2 g_2 g_4^3 
+ 9936 g_1^2 g_2 g_4^3 
\right. \nonumber \\
&& \left. ~~~~~~~~~~~
- 10368 \zeta_3 N g_1^2 g_3^4 
+ 7768 N g_1^2 g_3^4 
+ 10368 \zeta_3 g_1^2 g_3^4 
- 4808 g_1^2 g_3^4 
\right. \nonumber \\
&& \left. ~~~~~~~~~~~
- 1760 N g_1^2 g_3^2 g_4^2 
+ 20736 \zeta_3 g_1^2 g_3^2 g_4^2 
+ 22616 g_1^2 g_3^2 g_4^2 
- 1344 g_1^2 g_4^4 
\right. \nonumber \\
&& \left. ~~~~~~~~~~~
- 2280 g_1 g_2^2 g_3^2 g_4 
+ 9072 g_1 g_2 g_3^2 g_4^2 
- 6216 N g_1 g_3^4 g_4 
+ 11556 g_1 g_3^4 g_4 
\right. \nonumber \\
&& \left. ~~~~~~~~~~~
+ 10368 \zeta_3 g_1 g_3^2 g_4^3 
- 7344 g_1 g_3^2 g_4^3 
+ 344 g_2^2 g_3^2 g_4^2 
- 2952 g_2 g_3^2 g_4^3 
- 99 N^2 g_3^6 
\right. \nonumber \\
&& \left. ~~~~~~~~~~~
+ 2654 N g_3^6 
- 5184 \zeta_3 g_3^6 
+ 13820 g_3^6 
- 184 N g_3^4 g_4^2 
\right. \nonumber \\
&& \left. ~~~~~~~~~~~
+ 10368 \zeta_3 g_3^4 g_4^2 
+ 1928 g_3^4 g_4^2 
- 5184 \zeta_3 g_3^2 g_4^4 
+ 9860 g_3^2 g_4^4 \right] ~+~ O(g_i^8) \nonumber \\
\left. \gamma_{32}(g_i) \right|_{m=2} &=&
g_4^2 
+ \frac{1}{24} \left[ 
- 18 N g_1^2 g_3^2 
- 12 N g_1 g_3^2 g_4 
- 6 g_2^2 g_4^2 
- 36 g_2 g_4^3 
+ 7 N g_3^2 g_4^2 
- 4 g_4^4 \right] \nonumber \\
&& 
+ \frac{1}{864} \left[ 1010 N g_1^4 g_3^2 
+ 756 N g_1^4 g_4^2 
+ 1728 \zeta_3 N g_1^3 g_2 g_3^2 
- 900 N g_1^3 g_2 g_3^2 
\right. \nonumber \\
&& \left. ~~~~~~~~~
- 1080 N g_1^3 g_2 g_4^2 
+ 1560 N g_1^3 g_3^2 g_4 
+ 1512 N g_1^3 g_4^3 
- 432 \zeta_3 N g_1^2 g_2^2 g_3^2 
\right. \nonumber \\
&& \left. ~~~~~~~~~
+ 846 N g_1^2 g_2^2 g_3^2 
+ 282 N g_1^2 g_2^2 g_4^2 
+ 2592 \zeta_3 N g_1^2 g_2 g_3^2 g_4 
+ 2268 N g_1^2 g_2 g_3^2 g_4 
\right. \nonumber \\
&& \left. ~~~~~~~~~
- 904 N g_1^2 g_2 g_4^3 
- 496 N^2 g_1^2 g_3^4 
- 178 N g_1^2 g_3^4 
+ 3911 N g_1^2 g_3^2 g_4^2 
+ 40 N g_1^2 g_4^4 
\right. \nonumber \\
&& \left. ~~~~~~~~~
- 216 N g_1 g_2^2 g_3^2 g_4 
- 1728 \zeta_3 N g_1 g_2 g_3^2 g_4^2 
+ 792 N g_1 g_2 g_3^2 g_4^2 
+ 282 N^2 g_1 g_3^4 g_4 
\right. \nonumber \\
&& \left. ~~~~~~~~~
- 492 N g_1 g_3^4 g_4 
- 1344 N g_1 g_3^2 g_4^3 
- 21 g_2^4 g_4^2 
+ 864 \zeta_3 g_2^3 g_4^3 
+ 1420 g_2^3 g_4^3 
\right. \nonumber \\
&& \left. ~~~~~~~~~
- 204 N g_2^2 g_3^2 g_4^2 
+ 864 \zeta_3 g_2^2 g_4^4 
+ 1712 g_2^2 g_4^4 
- 70 N g_2 g_3^2 g_4^3 
- 162 g_2 g_4^5 
\right. \nonumber \\
&& \left. ~~~~~~~~~
- 33 N^2 g_3^4 g_4^2 
+ 1276 N g_3^4 g_4^2 
- 37 N g_3^2 g_4^4 
+ 3734 g_4^6 \right] ~+~ O(g_i^8) \nonumber \\
\left. \gamma_{33}(g_i) \right|_{m=2} &=&
\frac{1}{12} \left[ 
- N g_3^2 
+ 4 g_4^2 \right] \nonumber \\
&& 
+ \frac{1}{432} \left[ 2 N g_1^2 g_3^2 
+ 20 N g_1^2 g_4^2 
- 12 N g_1 g_3^2 g_4 
+ 10 g_2^2 g_4^2 
- 60 g_2 g_4^3 
- 76 N g_3^4 
\right. \nonumber \\
&& \left. ~~~~~~~~
+ 31 N g_3^2 g_4^2 
- 356 g_4^4 \right] \nonumber \\
&& 
+ \frac{1}{31104} \left[
- 206 N^2 g_1^4 g_3^2 
+ 2592 \zeta_3 N g_1^4 g_3^2 
- 4280 N g_1^4 g_3^2 
- 344 N^2 g_1^4 g_4^2 
+ 968 N g_1^4 g_4^2 
\right. \nonumber \\
&& \left. ~~~~~~~~~~~
+ 1200 N g_1^3 g_2 g_3^2 
- 2064 N g_1^3 g_2 g_4^2 
+ 4536 N^2 g_1^3 g_3^2 g_4 
- 12144 N g_1^3 g_3^2 g_4 
\right. \nonumber \\
&& \left. ~~~~~~~~~~~
- 8280 N g_1^3 g_4^3 
- 103 N g_1^2 g_2^2 g_3^2 
+ 656 N g_1^2 g_2^2 g_4^2 
- 10368 \zeta_3 N g_1^2 g_2 g_3^2 g_4 
\right. \nonumber \\
&& \left. ~~~~~~~~~~~
+ 5328 N g_1^2 g_2 g_3^2 g_4 
- 3156 N g_1^2 g_2 g_4^3 
- 7416 N^2 g_1^2 g_3^4 
+ 13436 N g_1^2 g_3^4 
\right. \nonumber \\
&& \left. ~~~~~~~~~~~
- 442 N^2 g_1^2 g_3^2 g_4^2 
+ 113644 N g_1^2 g_3^2 g_4^2 
+ 2152 N g_1^2 g_4^4 
+ 2268 N g_1 g_2^2 g_3^2 g_4 
\right. \nonumber \\
&& \left. ~~~~~~~~~~~
- 7308 N g_1 g_2 g_3^2 g_4^2 
+ 3960 N^2 g_1 g_3^4 g_4 
+ 31104 \zeta_3 N g_1 g_3^4 g_4 
\right. \nonumber \\
&& \left. ~~~~~~~~~~~
- 16680 N g_1 g_3^4 g_4 
- 31104 \zeta_3 N g_1 g_3^2 g_4^3 
- 15216 N g_1 g_3^2 g_4^3 
- 102 g_2^4 g_4^2 
\right. \nonumber \\
&& \left. ~~~~~~~~~~~
- 5718 g_2^3 g_4^3 
- 221 N g_2^2 g_3^2 g_4^2 
- 10368 \zeta_3 g_2^2 g_4^4 
+ 52728 g_2^2 g_4^4 
\right. \nonumber \\
&& \left. ~~~~~~~~~~~
+ 1080 N g_2 g_3^2 g_4^3 
+ 3108 g_2 g_4^5 
- 3292 N^2 g_3^6 
+ 18144 \zeta_3 N g_3^6 
- 1214 N g_3^6 
\right. \nonumber \\
&& \left. ~~~~~~~~~~~
- 284 N^2 g_3^4 g_4^2 
+ 31104 \zeta_3 N g_3^4 g_4^2 
+ 24248 N g_3^4 g_4^2 
- 10644 N g_3^2 g_4^4 
\right. \nonumber \\
&& \left. ~~~~~~~~~~~
+ 67392 \zeta_3 g_4^6 
+ 53200 g_4^6 \right] ~+~ O(g_i^8) ~. 
\end{eqnarray} 

\section{Complex Solutions.}

In this appendix we provide the remaining spectrum of fixed points for 
$N$~$=$~$1000$ and $m$~$=$~$2$ as an example of the mix of purely real, purely 
imaginary or fully complex fixed points. As in the main text we exclude 
solutions related by symmetries. First, there are five sets of complex fixed 
points of the CS type. The first is at
\begin{eqnarray}
x &=& (0.030988i-0.117059) ~+~  (0.225504-0.026447i)\epsilon ~+~
(0.141512+0.0220556i)\epsilon^{2} \nonumber \\
&& +~ O(\epsilon^{3}) \nonumber \\
y &=& (10.616979i+7.410016) ~+~ (-0.306607+5.10124i)\epsilon ~+~
(-3.30970+5.02141i)\epsilon^{2} \nonumber \\
&& +~ O(\epsilon^{3}) \nonumber \\
z &=& (0.041381i+0.950398) ~+~ (0.073839- 0.000086i)\epsilon ~+~
(-0.076593+ 0.240769i)\epsilon^{2} \nonumber \\
&& +~ O(\epsilon^{3}) \nonumber \\
t &=& (- 3.598246i+9.770473) ~+~ (2.741560- 4.671500i)\epsilon ~+~
(17.262953-21.70028i)\epsilon^{2} \nonumber \\
&& +~ O(\epsilon^{3}) \nonumber \\
w &\in& \{ (12.52843i-8.437892) ~+~ (-0.081315+6.488068i)\epsilon ~+~
(79.099646+19.016683i)\epsilon^{2} \nonumber \\
&& ~+~ O(\epsilon^{3}), \nonumber \\
&& ~ (- 8.157796i + 0.215249) ~+~ (10.458651+2.563357i)\epsilon ~+~
(-13.448748+143.623375i)\epsilon^{2} \nonumber \\
&& ~+~ O(\epsilon^{3}), \nonumber \\
&& ~ (- 4.370634i + 8.222643) ~+~ (-1.841641-7.930799i)\epsilon ~+~
(5.191836-47.804171i)\epsilon^{2} \nonumber \\
&& ~+~ O(\epsilon^{3}) \} 
\end{eqnarray}
where we have grouped three solutions together given that the only difference
is in the location of the $w$ coupling. The remainder are
\begin{eqnarray}
x &=& (- 0.003823i+1.005167) ~+~ (-0.054618+0.012486i)\epsilon ~+~
(0.018976+0.143820i)\epsilon^{2} \nonumber \\
&& +~ O(\epsilon^{3}) \nonumber \\
y &=& (2.003185i+18.658117) ~+~ (5.392312+1.086468i)\epsilon ~+~
(14.851433-19.827819i)\epsilon^{2} \nonumber \\
&& +~ O(\epsilon^{3}) \nonumber \\
z &=& ( 0.008575i + 1.025642) ~+~ (-0.085717-0.009151i)\epsilon ~+~
(-0.006651-0.017253i)\epsilon^{2} \nonumber \\
&& +~ O(\epsilon^{3}) \nonumber \\
t &=& (- 4.466842i+8.052554) ~+~ (4.210604+3.034436i)\epsilon ~+~
(8.671652+12.119131i)\epsilon^{2} \nonumber \\
&& +~ O(\epsilon^{3}) \nonumber \\
w &\in& \{ (15.494072i-6.618774) ~+~  (6.586489-2.039527i)\epsilon ~+~
(160.338535-22.392846i)\epsilon^{2} \nonumber \\
&& ~+~ O(\epsilon^{3}), \nonumber \\
&& ~ (- 2.718385i+6.100016) ~+~ (3.357728+1.279947i)\epsilon ~+~
(12.015204+7.083669i)\epsilon^{2} \nonumber \\
&& ~+~ O(\epsilon^{3}), \nonumber \\
&& ~ (- 12.775687i + 0.518757) ~+~  (0.842674+1.030180i)\epsilon ~+~
(56.581857+57.905307i)\epsilon^{2} \nonumber \\
&& ~+~ O(\epsilon^{3}) \} 
\end{eqnarray}
\begin{eqnarray}
x &=& (0.050822i - 0.763449)  ~+~ (-0.342230-0.090513i)\epsilon ~+~
(-1.048864-0.055624i)\epsilon^{2} \nonumber \\
&& ~+~ O(\epsilon^{3}) \nonumber \\
y &=& (7.435673i + 12.90737) ~+~ (11.071087+3.990671i)\epsilon ~+~
(10.916564+32.815621i)\epsilon^{2} \nonumber \\
&& ~+~ O(\epsilon^{3}) \nonumber \\
z &=& (0.0450638i + 0.849646) ~+~ (0.197056-0.055744i)\epsilon ~+~
(0.452791+0.136376i)\epsilon^{2} \nonumber \\
&& ~+~ O(\epsilon^{3}) \nonumber \\
t &=& (- 1.982473i + 12.903491) ~+~ (6.044935-1.720892i)\epsilon ~+~
(6.597369-13.346172i)\epsilon^{2} \nonumber \\
&& ~+~ O(\epsilon^{3}) \nonumber \\
w &\in& \{ (6.421453i - 11.600963) ~+~ (-2.494744+ 2.659224i)\epsilon ~+~
(12.554273+26.208510i)\epsilon^{2} \nonumber \\
&& ~+~ O(\epsilon^{3}), \nonumber \\
&& ~ (-3.154923i + 14.29153) ~+~ (7.238628-2.639826i)\epsilon ~+~
(4.840108-21.122141i)\epsilon^{2} \nonumber \\
&& ~+~ O(\epsilon^{3}), \nonumber \\
&& ~ (- 3.266530i - 2.690567) ~+~ (1.337926+0.955634i)\epsilon ~+~
(33.890419+26.868026i)\epsilon^{2} \nonumber \\
&& ~+~ O(\epsilon^{3}) \} 
\end{eqnarray}
\begin{eqnarray}
x &=& (- 0.555294i - 1.513381)  ~+~ (-23.114452+24.950148i)\epsilon 
\nonumber \\
&& +~ (3858.041480+2762.998578i)\epsilon^{2} ~+~ O(\epsilon^{3}) \nonumber \\
y &=& (34.90363i - 21.842603) ~+~ (890.252852+680.087930i)\epsilon
\nonumber \\
&& +~ (74550.859506-131881.754919i)\epsilon^{2} ~+~ O(\epsilon^{3}) 
\nonumber \\
z &=& (0.549418i + 1.214497) ~+~ (24.255203- 14.435870i)\epsilon
\nonumber \\
&& +~ (-2470.779282-3096.675176i)\epsilon^{2} ~+~ O(\epsilon^{3}) \nonumber \\
t &=& (- 29.55207i + 15.18796) ~+~ (-850.256009-539.584442i)\epsilon 
\nonumber \\
&& +~ (-56438.610729+124039.167626i)\epsilon^{2} ~+~ O(\epsilon^{3}) 
\nonumber \\
w &\in& \{ ( 42.91263i - 23.15191) ~+~ (1127.488695+813.413116i)\epsilon 
\nonumber \\
&& ~+~ (82884.915451-172830.465074i)\epsilon^{2} ~+~ O(\epsilon^{3}), 
\nonumber \\
&& ~ (0.409667i + 1.270019) ~+~ (35.031670+22.655513i)\epsilon 
\nonumber \\
&& ~+~ (1486.251233-4603.586046i)\epsilon^{2} ~+~ O(\epsilon^{3}), \nonumber \\
&& ~ (- 43.32229i + 21.88189) ~+~ (-1155.605341-813.415117i)\epsilon 
\nonumber \\
&& ~+~ (-84174.229119+179594.341857i)\epsilon^{2} ~+~ O(\epsilon^{3}) \} 
\end{eqnarray}
and 
\begin{eqnarray}
x &=& (- 0.006320i - 1.030086) ~+~ (0.107095+0.043240i)\epsilon ~+~
(-0.036554+0.122859i)\epsilon^{2} \nonumber \\
&& ~+~ O(\epsilon^{3}) \nonumber \\
y &=& (2.138785i - 9.592803) ~+~ (-5.961631-11.215238i)\epsilon ~+~
(-22.616924-62.894519i)\epsilon^{2} \nonumber \\
&& ~+~ O(\epsilon^{3}) \nonumber \\
z &=& (0.002767i + 1.0176108) ~+~ (-0.059037-0.021032i)\epsilon ~+~
(0.043635-0.043068i)\epsilon^{2} \nonumber \\
&& ~+~ O(\epsilon^{3}) \nonumber \\
t &=& (1.198860i - 8.255330) ~+~ (-2.187227-7.108184i)\epsilon ~+~
(-8.358400-32.699994i)\epsilon^{2} \nonumber \\
&& ~+~ O(\epsilon^{3}) \nonumber \\
w &\in& \{ (13.36369i - 4.799020) ~+~ (10.845688-1.270978i)\epsilon ~+~
(139.104337-28.027224i)\epsilon^{2} \nonumber \\
&& ~+~ O(\epsilon^{3}), \nonumber \\
&& ~ (- 12.50645i - 2.865408) ~+~ (-1.424629-4.030314i)\epsilon ~+~
(67.742495+17.073470i)\epsilon^{2} \nonumber \\
&& ~+~ O(\epsilon^{3}), \nonumber \\
&& ~ (- 0.857238i + 7.664428) ~+~ (1.116387+5.387238i)\epsilon ~+~
(4.573636+22.217349i)\epsilon^{2} \nonumber \\
&& ~+~ O(\epsilon^{3}) \} ~. 
\end{eqnarray}
There were several sets where some of the fixed points were either real or
imaginary in addition to one being fully complex since we found the solutions
\begin{eqnarray} 
x &=& 0.114419i ~-~ 0.553587i\epsilon ~+~ 6.740110i\epsilon^{2} ~+~ 
O(\epsilon^{3}) \nonumber \\
y &=& 22.486625i ~-~ 31.203475i\epsilon ~+~ 603.274688i\epsilon^{2} ~+~ 
O(\epsilon^{3}) \nonumber \\
z &=& 1.070456 ~-~ 0.504440\epsilon ~+~ 9.518801\epsilon^{2} ~+~ 
O(\epsilon^{3}) \nonumber \\
t &=& -11.601013i ~+~ 36.808088i\epsilon ~-~ 722.589642i\epsilon^{2} ~+~ 
O(\epsilon^{3}) \nonumber \\
w &\in& \{ (22.427674i-1.617314) ~+~ (1.687428 - 38.335633i)\epsilon 
\nonumber \\
&& ~+~ (255.539793+718.517860i)\epsilon^{2} ~+~ O(\epsilon^{3}), \nonumber \\
&& ~~ 3.234627 ~+~ 8.891244\epsilon ~-~ 144.067750\epsilon^{2} ~+~ 
O(\epsilon^{3}) \} 
\end{eqnarray}
and 
\begin{eqnarray} 
x &=& -~ 0.868555 ~-~ 0.203744\epsilon ~-~ 0.915849\epsilon^{2} ~+~ 
O(\epsilon^{3}) \nonumber \\
y &=& 19.811433 ~+~ 7.966436\epsilon ~-~ 15.205442\epsilon^{2} ~+~ 
O(\epsilon^{3}) \nonumber \\
z &=& 1.012581 ~-~ 0.020942\epsilon ~+~ 0.066905\epsilon^{2} ~+~ 
O(\epsilon^{3}) \nonumber \\
t &=& -~ 4.342552 ~-~ 2.231269\epsilon ~-~ 7.303390\epsilon^{2} ~+~ 
O(\epsilon^{3}) \nonumber \\
w &=& (15.014668i-2.955662) ~+~ (4.805741 + 1.917129i)\epsilon \nonumber \\
&& ~+~ (111.835693-52.975405i)\epsilon^{2} ~+~ O(\epsilon^{3}) ~. 
\end{eqnarray}
This completes the set of all CS type solutions in addition to those in section
6.

For the remaining solutions we found at least one of the critical couplings was
zero. First we group those solutions where the couplings are either real or
imaginary. We found
\begin{eqnarray} 
x &=& 0 ~~~,~~~ t ~=~ 0 \nonumber \\
y &=& 14.907120i ~+~ 11.502407i\epsilon ~-~ 10.399304i\epsilon^{2} ~+~ 
O(\epsilon^{3}) \nonumber \\
z &=& 0.998006 ~+~ 0.007339\epsilon ~-~ 0.009793\epsilon^{2} ~+~ 
O(\epsilon^{3}) \nonumber \\
w &\in& \{ (15.559942i-2.659423) ~+~ (4.519271 + 2.244641i)\epsilon 
\nonumber \\
&& ~+~ (110.184708-62.342815i)\epsilon^{2} ~+~ O(\epsilon^{3}), \nonumber \\
&& ~ 5.318846 ~+~ 0.901757\epsilon ~-~ 1.582315\epsilon^{2} ~+~ 
O(\epsilon^{3}) \} 
\end{eqnarray}
and 
\begin{eqnarray}
x &=& 0 ~~~,~~~ z ~=~ 0 ~~~,~~~ t ~=~ 0 \nonumber \\
y &=& 14.907120i ~+~ 11.502407i\epsilon ~-~ 10.399304i\epsilon^{2} ~+~ 
O(\epsilon^{3}) \nonumber \\
w &\in& \{ 10.540926 ~+~ 8.344899\epsilon ~-~ 16.563109\epsilon^{2} ~+~ 
O(\epsilon^{3}), ~ 0 \} ~.
\label{partsoln}
\end{eqnarray}
Included in the first set is a complex $w$ coupling. However, the second 
solution of each set are examples which are similar to pure $\phi^3$ theory 
when its coupling is purely imaginary. That particular $O(N)$ model described 
the Lee-Yang edge singularity problem, \cite{60}. Also the solutions of
(\ref{partsoln}) correspond to the Lagrangian without a $\phi^{ia}$ field. In
the case of the only non-zero coupling $y$ this is the pure cubic theory 
involving only the $\sigma$ field. The remaining solutions with any complex 
roots all have a vanishing critical $z$ coupling and are  
\begin{eqnarray} 
x &=& ( 0.102517i-1.069257) ~+~ (- 0.187674 - 0.485282i)\epsilon \nonumber \\
&& ~+~ (6.388101+3.227418i)\epsilon^{2} ~+~ O(\epsilon^{3}) \nonumber \\
y &=& (8.388316i+4.182323) ~+~ (- 31.690792 - 0.719078i)\epsilon \nonumber \\
&& ~+~ (394.496262-256.273660i)\epsilon^{2} ~+~ O(\epsilon^{3}) \nonumber \\
z &=& 0 \nonumber \\
t &=& (- 12.684990i-5.204227) ~+~ (23.207111 + 4.566306i)\epsilon \nonumber \\
&& ~+~ (-270.175902+154.130691i)\epsilon^{2} ~+~ O(\epsilon^{3}) \nonumber \\
w &\in& \{ ( 13.394284i+8.214396) ~+~ (- 30.663870 - 17.742257i)\epsilon 
\nonumber \\
&& ~+~ (479.083975-270.124833i)\epsilon^{2} ~+~ O(\epsilon^{3}), ~ 0 \} 
\end{eqnarray}
and
\begin{eqnarray}
x &=& 0 ~~~,~~~ z ~=~ 0 \nonumber \\
y &=& (12.918644i+4.311494) ~+~ (2.006636 + 9.961474i)\epsilon \nonumber \\
&& ~+~ (-0.372460-5.181172i)\epsilon^{2} ~+~ O(\epsilon^{3}) \nonumber \\
t &=& ( 9.765251i-1.446829) ~+~ (- 0.575941 + 7.513693i)\epsilon \nonumber \\
&& ~+~ (-1.501827-7.061748i)\epsilon^{2} ~+~ O(\epsilon^{3}) \nonumber \\
w &\in& \{ (7.382858i-3.189516) ~+~ (- 1.150870 + 6.001629i)\epsilon
\nonumber \\
&& ~+~ (-0.334518-6.569977i)\epsilon^{2} ~+~ O(\epsilon^{3}), ~ 0 \} ~.
\end{eqnarray}

The remainder of the solutions are real but interesting patterns emerge in 
several cases. First we record the fixed points where there is no pairing with 
another set. There were four such cases. Of all the real solutions we record we
found only the first two correspond to stable fixed points which are 
\begin{eqnarray} 
x &=& 1.011102 ~-~ 0.026162\epsilon ~+~ 0.022238\epsilon^{2} ~+~ 
O(\epsilon^{3}) \nonumber \\
y &=& 6.637801 ~-~ 1.117476\epsilon ~+~ 0.962982\epsilon^{2} ~+~ 
O(\epsilon^{3}) \nonumber \\
z &=& 0 ~~~,~~~ t ~=~ 0 \nonumber \\
w &=& 10.540926 ~+~ 8.344899\epsilon ~-~ 16.563109\epsilon^{2} ~+~ 
O(\epsilon^{3})  
\end{eqnarray}
and
\begin{eqnarray}
x &=&  0 ~~~,~~~ y ~=~ 0 ~~~,~~~ z ~=~  0 ~~~,~~~ t ~=~ 0 \nonumber \\
w &=& 10.540926 ~+~ 8.344899\epsilon ~-~ 16.563109\epsilon^{2} ~+~ 
O(\epsilon^{3}) ~.  
\end{eqnarray}
The final solution corresponds to the pure $T^{ab}$ theory when $m$~$=$~$2$ but
the $(0,0,0,0,w)$ structure could be analysed in isolation for arbitrary $m$.
However, the stability of these two solutions, in contrast to the remaining
real solutions which are not stable, appears to be driven by the vanishing of
the couplings $g_3$ and $g_4$. In this case there is no interaction whatsoever 
between the pair of fields $\{\phi^{ia},\sigma\}$ and $T^{ab}$ which is 
apparent from (\ref{laglgw6}). In other words one is dealing with a partitioned
Lagrangian and the coupling constant space is also partitioned. So the 
stability here is a reflection of the stability of the two separate 
Lagrangians. In the second of these two solutions the situation is effectively 
trivial since it reflects that one of the two Lagrangians is a free field 
theory which has a zero $\beta$-function. That such solutions representing the 
sum of independent Lagrangians emerge similar to (\ref{partsoln}) ought not to 
come as a surprise and should be regarded as an internal consistency in our 
analysis of solution. The remaining unpaired solutions are 
\begin{eqnarray} 
x &=& -~ 0.854446 ~-~ 0.212078\epsilon ~-~ 0.647751\epsilon^{2} ~+~ 
O(\epsilon^{3}) \nonumber \\
y &=& 18.789145 ~+~ 9.197094\epsilon ~-~ 2.733818\epsilon^{2} ~+~ 
O(\epsilon^{3}) \nonumber \\
z &=& 0 ~~~,~~~ t ~=~  0 \nonumber \\
w &=& 10.540926 ~+~ 8.344899\epsilon ~-~ 16.563109\epsilon^{2} ~+~ 
O(\epsilon^{3})  
\end{eqnarray}
and
\begin{eqnarray} 
x &=& 0.983210 ~-~ 0.011253\epsilon ~+~ 0.063259\epsilon^{2} ~+~ 
O(\epsilon^{3}) \nonumber \\
y &=& 16.345805 ~+~ 10.658027\epsilon ~+~ 21.524495\epsilon^{2} ~+~ 
O(\epsilon^{3}) \nonumber \\
z &=&  0 ~~~,~~~ t ~=~ 0 \nonumber \\
w &=& 10.540926 ~+~ 8.344899\epsilon ~-~ 16.563109\epsilon^{2} ~+~ 
O(\epsilon^{3})  
\end{eqnarray}
which together with the other real solutions correspond to saddle point 
structures. For the paired solutions we have  
\begin{eqnarray}
x &=& 0.986386 ~+~ 0.006824\epsilon ~+~ 0.023890\epsilon^{2} ~+~ 
O(\epsilon^{3}) \nonumber \\
y &=& 3.882413 ~+~ 3.856888\epsilon ~+~ 1.139387\epsilon^{2} ~+~ 
O(\epsilon^{3}) \nonumber \\
z &=& 0 \nonumber \\
t &=& -~ 6.810006 ~+~ 3.229867\epsilon ~+~ 5.755896\epsilon^{2} ~+~ 
O(\epsilon^{3}) \nonumber \\
w &\in& \{ 13.726061 ~+~ 3.729174\epsilon ~-~ 29.707955\epsilon^{2}
~+~ O(\epsilon^{3}), ~ 0 \}   
\end{eqnarray}
and
\begin{eqnarray}
x &=& 0 ~~~,~~~ z ~=~ 0 \nonumber \\
y &=& 17.222217 ~+~ 37.874140\epsilon ~+~ 404.846200\epsilon^{2} ~+~ 
O(\epsilon^{3}) \nonumber \\
t &=& -~ 13.657731 ~-~ 33.103835\epsilon ~-~ 376.517213\epsilon^{2} ~+~ 
O(\epsilon^{3}) \nonumber \\
w &\in& \{ 20.542650 ~+~ 42.608899\epsilon ~+~ 533.442168\epsilon^{2} ~+~ 
O(\epsilon^{3}), ~ 0 \} ~.  
\end{eqnarray}
The final three pairings exhibit a novel feature in that in each set the 
critical $x$ and $t$ couplings are equal. This is clear since we found
\begin{eqnarray} 
x &=& -~ 0.854046 ~-~ 0.211934\epsilon ~-~ 0.647273\epsilon^{2} ~+~ 
O(\epsilon^{3}) \nonumber \\
y &=& 18.789012 ~+~ 9.196991\epsilon ~-~ 2.731101\epsilon^{2} ~+~ 
O(\epsilon^{3}) \nonumber \\
z &=&  0 \nonumber \\
t &=& -~ 0.854046 ~-~ 0.211934\epsilon ~-~ 0.647273\epsilon^{2} ~+~
O(\epsilon^{3}) \nonumber \\
w &\in& \{ -~ 10.598432 ~+~ 8.327096\epsilon ~-~ 16.655991\epsilon^{2} ~+~ 
O(\epsilon^{3}), ~ 0 \} \\  
x &=& 0.982680 ~-~ 0.011165\epsilon ~+~ 0.063307\epsilon^{2} ~+~ 
O(\epsilon^{3}) \nonumber \\
y &=& 16.347740 ~+~ 10.655223\epsilon ~+~ 21.515017\epsilon^{2} ~+~ 
O(\epsilon^{3}) \nonumber \\
z &=&  0 \nonumber \\
t &=& 0.982680 ~-~ 0.011165\epsilon ~+~ 0.063307\epsilon^{2} ~+~ 
O(\epsilon^{3}) \nonumber \\
w &\in& \{ 10.616993 ~+~ 8.279650\epsilon ~-~ 16.867303\epsilon^{2} ~+~ 
O(\epsilon^{3}), ~ 0 \}  
\end{eqnarray}
and 
\begin{eqnarray} 
x &=& 1.010586 ~-~ 0.0261234\epsilon ~+~ 0.022211\epsilon^{2} ~+~ 
O(\epsilon^{3}) \nonumber \\
y &=& 6.633618 ~-~ 1.114926\epsilon ~+~ 0.960725\epsilon^{2} ~+~ 
O(\epsilon^{3}) \nonumber \\
z &=& 0 \nonumber \\
t &=& 1.010586 ~-~ 0.026124\epsilon ~+~ 0.022211\epsilon^{2} ~+~ 
O(\epsilon^{3}) \nonumber \\
w &\in& \{ 10.621358 ~+~ 8.274362\epsilon ~-~ 16.895995\epsilon^{2} ~+~ 
O(\epsilon^{3}), ~ 0 \} ~. 
\end{eqnarray}
While our focus here was on the $O(1000)$~$\times$~$O(2)$ theory it represents
a snapshot of the spectrum of potential solutions for the general symmetry
group. What has emerged are real fixed points in addition to the Heisenberg, AU
and CS type which were motivated by the four dimensional theory.

\end{document}